\documentclass[fleqn,usenatbib]{mnras}

\usepackage{newtxtext,newtxmath}

\usepackage[T1]{fontenc}
\usepackage{ae,aecompl}

\usepackage{graphicx}	
\usepackage{amsmath}	
\usepackage{amssymb}	
\usepackage{acronym}
\usepackage{siunitx}
\usepackage{subfig}
\usepackage{booktabs}
\usepackage[mathcal]{euscript}
\usepackage{array}
\usepackage{tabularx}
\usepackage{rotating}
\usepackage{pdflscape}
\usepackage{afterpage}

\DeclareSIUnit\jansky{Jy}
\DeclareSIUnit\parsec{pc}
\DeclareSIUnit\beam{beam}

\acrodef{AGN}{active galactic nuclei}
\acrodef{ASKAP}{Australian Square Kilometre Array Pathfinder}
\acrodef{ASSOM}{Adaptive Sub-space Self-organising map}
\acrodef{ATLAS}{Australia Telescope Large Area Survey}
\acrodef{BMU}{Best Matching Unit}
\acrodef{BOSS}{Baryon Oscillation Spectroscopic Survey}
\acrodef{BT}{bent-tail}
\acrodef{CNN}{Convolution Neural Network}
\acrodef{CON}[CONNvis]{CONNection VISualisation}
\acrodef{CPU}{Central Processing Unit}
\acrodef{DR}{data release}
\acrodef{EMU}{Evolutionary Map of the Universe}
\acrodef{FIRST}{Faint Images of the Radio-Sky at Twenty centimeters}
\acrodef{FR}{Fanaroff-Riley}
\acrodef{FSC}{FIRST supplemented catalogue}
\acrodef{GID}{group identification}
\acrodef{GPU}{Graphical Processing Unit}
\acrodef{GRC}{group reference catalogue}
\acrodef{GRG}{Giant Radio Galaxy}
\acrodefplural{GRG}{Giant Radio Galaxies}
\acrodef{GRS}{Giant Radio Source}
\acrodefplural{GRS}{Giant Radio Sources}
\acrodef{HYMORS}[HyMoRS]{Hybrid morphology radio sources}
\acrodef{IGM}{inter-galactic medium}
\acrodef{JVLA}{ Karl G. Jansky Very Large Array}
\acrodef{LOFAR}{Low Frequency Array}
\acrodef{LOTS}[LoTSS]{LOFAR Two-metre Sky Survey}
\acrodef{ML}{Machine Learning}
\acrodef{MWA}{Murchison Widefield Array}
\acrodef{NAN}[\texttt{NaN}]{not-a-number}
\acrodef{NAT}{narrow-angle tail}
\acrodef{NVSS}{NRAO VLA Sky Survey}
\acrodef{PDF}{probability density function}
\acrodef{PINK}[\texttt{PINK}]{Parallelized rotation and flipping INvariant Kohonen maps}
\acrodef{PSF}{point spread function}
\acrodef{RGZ}{Radio Galaxy Zoo}
\acrodef{SDSS}{Sloan digital sky survey}
\acrodef{SED}{spectral energy distribution}
\acrodef{SFG}{star forming galaxy}
\acrodefplural{SFG}{star forming galaxies}
\acrodef{SKA}{Square Kilometre Array}
\acrodef{SMBH}{super massive black hole}
\acrodef{SOM}{self-organising map}
\acrodef{UGID}{unique group identifier}
\acrodef{VLA}{Very Large Array}
\acrodef{WAT}{wide-angle tail}
\acrodef{WCS}{world coordinate system}
\acrodef{WISE}[\textit{WISE}]{Wide-field Infrared Survey Explorer}

\newcommand{\revision}[1]{#1}
\newcommand{\revisiontwo}[1]{#1}

\newcommand{\colstyle}[1]{\texttt{#1}}

\title[Cataloging with unsupervised ML]{Cataloging the radio-sky with unsupervised machine learning: a new approach for the SKA era}

\author[T. J. Galvin et al.]{
T. J. Galvin,$^{1}$\thanks{E-mail: tim.galvin@csiro.au}
M. Huynh,$^{1,2}$\thanks{E-mail: minh.huynh@csiro.au}
R. P. Norris,$^{3,4}$
X. R. Wang,$^{6}$
E. Hopkins,$^{7}$
K. Polsterer,$^{7}$\newauthor
N. O. Ralph,$^{3}$
A. N. O'Brien,$^{3,4,8}$
G. H. Heald$^{1}$\\
$^{1}$CSIRO Astronomy and Space Science, PO Box 1130, Bentley WA 6102, Australia\\
$^{2}$International Centre for Radio Astronomy Research (ICRAR), M468, The University of Western Australia, 35 Stirling Highway, Crawley, WA 6009, Australia\\
$^{3}$Western Sydney University, Penrith Campus, Locked Bag 1797, Penrith NSW 2751\\
$^{4}$CSIRO Astronomy and Space Science, PO Box 76, Epping, NSW 1710, Australia\\
$^{5}$CSIRO Data61, Australia, PO Box 76, Epping, NSW 1710, Australia\\
$^{6}$Western Sydney University, Parramatta South Campus, NSW Australia\\
$^{7}$Astroinformatics, HITS gGmbH, Schloss-Wolfsbrunnenweg 35, 69118 Heidelberg, Germany\\
$^{8}$Department of Physics, University of Wisconsin - Milwaukee, Milwaukee, WI 53201, USA\\
}

\date{Accepted XXX. Received YYY; in original form ZZZ}

\pubyear{2019}

\begin{document}
\label{firstpage}
\pagerange{\pageref{firstpage}--\pageref{lastpage}}
\maketitle

\begin{abstract}
We develop a new analysis approach towards identifying related radio components and their corresponding infrared host galaxy based on unsupervised machine learning methods. By exploiting \ac{PINK}, a self-organising map algorithm, we are able to associate radio and infrared sources without the \textit{a priori} requirement of training labels. We present an example of this method using $894,415$ images from the \ac{FIRST} and \ac{WISE} surveys centred towards positions described by the \ac{FIRST} catalogue. We produce a set of catalogues that complement \ac{FIRST} and describe $802,646$ objects, including their radio components and their corresponding All\ac{WISE} infrared host galaxy. Using these data products we (i) demonstrate the ability to identify objects with rare and unique radio morphologies (e.g. `X'-shaped galaxies, hybrid FR-I/FR-II morphologies), (ii) can identify the potentially resolved radio components  that are associated with a single infrared host and (iii) introduce a ``curliness'' statistic to search for bent and disturbed radio morphologies, and (iv) extract a set of 17 giant radio galaxies between $700-1100$\,kpc. As we require no training labels, our method can be applied to any radio-continuum survey, provided a sufficiently representative \ac{SOM} can be trained. 
\end{abstract}




\section{Introduction}
\acresetall

Radio astronomy will enter its `golden age' as the \ac{SKA} and its pathfinder instruments begin to operate at full capacity  over the next decade. This new suite of radio interferometers, including the \ac{LOFAR} \citep{2013A&A...556A...2V}, \ac{MWA} \citep{2013PASA...30....7T,2018PASA...35...33W}, \ac{ASKAP} \citep{2008ExA....22..151J}, MeerKAT \citep{2016mks..confE...1J} and the \ac{JVLA} \citep{2011ApJ...739L...1P}, offer exceptional gains in sensitivity and survey speeds when compared to previous generations of instruments \citep{2017NatAs...1..671N}. They will be capable of routinely producing deep continuum surveys containing tens of thousands of radio sources on timescales as short as hours, a feat that would have taken months of dedicated telescope time some years ago. With this gain in sensitivity and survey speed comes the need to redesign how data analysis and interpretation is performed. Existing efforts that are primarily powered by human `intelligence' are expensive in both time and effort and will not be able to scale to these exceptionally high data volumes. 

Among these non-trivial tasks is the process of applying a morphological classification to each of the detected radio sources within some image. This involves describing all of the potentially resolved structure of a single intrinsic object, including situations where some of its components may be separated by some distance (i.e. many units of the resolution element of the instrument). 

Tools have been developed to help distribute this classification problem. Perhaps the most successful to date is the Galaxy Zoo online portal \citep{2008MNRAS.389.1179L}, a publicly accessible website that allows `citizen scientists' (members of the general public who may not formally be trained in the field) to participate in a broad set of astronomical classification problems alongside domain experts. Especially relevant to this work is \ac{RGZ} \citep{2015MNRAS.453.2326B}, a project being operated on the Galaxy Zoo platform. They ask the public to identify related radio components and their infrared host galaxy of complex sources using primarily images from the 1.4\,GHz \ac{VLA} \ac{FIRST} \citep{1995ApJ...450..559B} and the all-sky \ac{WISE} \citep{2010AJ....140.1868W} surveys. \ac{RGZ} has been successful in producing a dataset that can be used as training sets, allowing for \ac{ML} approaches to contribute to this source classification problem \citep{2018MNRAS.476..246L,2018MNRAS.478.5547A,2019MNRAS.482.1211W}. 

These studies each apply a \textit{supervised} \ac{ML} method, where an algorithm needs examples of data with known or trusted attributes (e.g. galaxy/morphology type, redshifts) in order to optimise a cost function and produce a usable model. Without training labels (produced by \ac{RGZ} or a similar project) these methods can not be used. As radio interferometers each have their own  performance characteristics (e.g. observing frequency, angular scale sensitivities), the types of objects and physical processes that each survey can be sensitive to may be vastly different. Some recent work has begun to evaluate how well some supervised \ac{ML} models can be transferred to datasets they were not trained against \citep{2019MNRAS.488.3358T},  but it remains to be seen how successful these approaches will be and how far this extrapolation of models and labels can be stretched.

An alternate approach is an \textit{unsupervised} \ac{ML} framework. This is a broad set of algorithms that are executed without \emph{a priori} knowledge about the dataset. Instead the focus is on recognising the \textit{structure} within a dataset \citep{Ghahramani2004}. Examples include $k$-means clustering \citep{Lloyd82leastsquares}, Gaussian mixture models \citep{10.1093/biomet/56.3.463}, principal component analysis \citep{doi:10.1080/14786440109462720} and \ac{SOM} \citep{Kohonen1982}, the latter being of focus in this study.  As there is no error function conditioned against existing labels, the structure that an unsupervised method infers may not correspond to existing classification schemes described by a set of labels. Depending on the \ac{ML} method chosen, the relationships may be transparently examined, or may be hidden within several layers of complexity, often making it difficult to know exactly what the method has discovered. However, if the structure of the data is understood, then such models can be applied to a variety of problems that may not have been foreseen when they were initially trained.

SOMs have been used in the astronomical literature for a variety of tasks, including the classification of light curves \citep{2004MNRAS.353..369B}, clustering and analysis of gigahertz-peaked spectrum sources \citep{2008A&A...482..483T}, detecting structure within point data \citep{2011ApJ...727...48W} and object classification and photometric redshift estimation \citep{2012MNRAS.419.2633G}. They operate by constructing a set of prototypes fixed on some regular lattice like structure that are representative of the types of structures seen within a training dataset. Proximity of prototypes upon this lattice represent the level of similarity of the structures those prototypes are describing. 

Applying the \ac{SOM} method onto image datasets may require introducing a redefinition on how similarity is measured. Often we want image processing methods to be invariant to affine transformations (i.e. scaling, flipping, rotation, and translation). For example, the orientation of an \ac{AGN} and its radio lobes is not a significant characteristic.

\citet{2019PASP..131j8011R} approach the problem by using a convolutional auto-encoder to first construct a latent vector (a compressed space that encodes defining characteristics within an image) to minimise the effect of affine transformations, and then train a \ac{SOM} based on these latent vectors. Although this offers orders of magnitude improvement with regards to training time, the representative features within the neurons are features within the compressed latent vector space and are not directly interpretable without an additional decoding step. 

An alternate solution is to attack the affine transform problem directly. \ac{PINK} is an algorithm that incorporates flipping and rotational invariance into the \ac{SOM} algorithm \citep{polsterer2016}. By exploiting products determined during the training process, the \ac{SOM} itself can be used as a tool to transfer knowledge \citep{2019PASP..131j8009G}. Given that the 
\ac{SOM} requires no training labels and is able to construct prototypes representative of source morphologies, it may prove to be an ideal tool for distinguishing between alternative explanations. For example, it may be able to distinguish the two lobes of a radio \ac{AGN} from a pair of  \acp{SFG}.

\citet{2019PASP..131j8009G} used \ac{PINK} to create prototypes capable of breaking this degeneracy by including infrared image information, thereby producing prototypes with estimated host galaxy positions. 

In our study we investigate how \ac{PINK} can be used to identify related radio components and their corresponding host galaxy. We present a proof-of-concept method that demonstrates how we can accomplish this without the requirement of training labels. This manuscript is structured as follows. In \S\ref{sec:som} we describe the construction of a \ac{SOM}, and in \S\ref{sec:approach} briefly describe our component grouping procedure. The datasets and their preprocessing steps used outlined in \S\ref{sec:dataset}, and our approach to source collation and catalogue creation is described in \S\ref{sec:cata_create}. We present our results in \S\ref{sec:results}, and discuss improvements and future outlooks in \S\ref{sec:discussion}. We adopted the cosmology of \citet{2013ApJS..208...19H} where $\Omega_{m} = 0.287$, $\Omega_\lambda = 0.713$ and $H_0 = \SI{69.3}{\kilo\meter\per\s\per\mega\parsec}$.

\section{Self Organising Maps}
\label{sec:som}

\acp{SOM} are a type of neural network that constructs a set of features that are representative of those in a training dataset \citep{Kohonen1982}. They can be thought of as a way of projecting high dimensional data onto a lower dimensional lattice or manifold. Data can be compared to these constructed feature sets using a similarity measure, where the terms `closer' and `distant' refer to data items being similar or dissimilar to items on the SOM lattice, respectively. For a properly trained \ac{SOM}, each of the major features within the training dataset should be represented by at least one constructed feature without over-representing any single class. 

The \ac{SOM} consists of a set of neurons $n \in N$, each with a position $p$ on the \ac{SOM} lattice, and a set of prototype weights, $w$. \revision{These prototype weights can be initialised in various ways, including random noise following some  distribution}. For a single training iteration, $i$, an item $t \in T$ is selected, where $T$ is a training dataset, and a similarity measure $\Delta\left(t, w_p\right)$ \revision{is computed} (see \S\ref{sec:rot_invariance}). The neuron most similar to $t$ is the \ac{BMU}. Once the position of the \ac{BMU}, $j$, is identified, all prototype weights are updated 

\begin{equation}
\label{eq:update}
	w^\prime_p = w_p + \left(\phi\left(t\right)-w_p\right) \cdot d\left(p,j\right) \cdot l\left(i\right),
\end{equation}

\noindent where $\phi$ may be necessary to spatially align $t$ onto $w_p$. 
$d$ is the distance component, or neighbourhood function, and $l$ is the time component, or learning rate. 
\revision{ $d$  down-weights the updates by an amount that increases with the separation between the position coordinates $p$ and $j$. The exact functional form of this radial decay is arbitrary, but is often parameterised to take the form of a Gaussian,}

\begin{equation}
\label{eq:neighbour}
	d\left(p, j\right) = \frac{1}{\sigma_u\sqrt{2\pi}} \exp\left[{-\frac{1}{2}\left(\frac{s\left(p, j\right)}{\sigma_u}\right)}\right],
\end{equation}

\noindent \revision{where $s\left(p, j\right)$ is the separation between $p$ and $j$ lattice coordinate positions and $\sigma_u$ is a term used to control the distance updates are propagated (which may also evolve with $i$). 
The time component $l$ of Equation~\ref{eq:update} is a term that further dampens the significance of the weighting updates applied for each $i$. An exponential decay for $l$ could be adopted,}

\begin{equation}
	l\left(i\right) = \exp\left[-i/\tau\right]
\end{equation}

 \noindent\revision{ where $\tau$ is a constant to further dampen the evolution of $l$. In the \ac{PINK} implementation both $\sigma_u$ and $l$ are user defined values that do not automatically evolve with increasing $i$. }

\subsection{Rotation and Flipping Invariance}
\label{sec:rot_invariance}

When dealing with image data, we need methods that are invariant to rotation and flipping transforms. \ac{PINK} implements a modified Euclidean distance metric that will search for the optimal set of rotation and flipping parameters to align an example image to prototype weights on a \ac{SOM}. Their measure is formalised as 

\begin{equation}
\label{eq:ed}
\begin{aligned}
\Delta(t,w_j) =& \\
\underset{\forall \phi \in \Phi}{minimize(\phi)}& \sqrt{\sum_{c=0}^{C}\sum_{x=0}^{X}\sum_{y=0}^{Y}\left(w_{j\left(c,x,y\right)} - \phi (t_{c,x,y}) \right)^2},
\end{aligned}
\end{equation}

\noindent where $c$ are image channels within $t$ and $w_j$, $x$ and $y$ pixel coordinates, and $\phi$ corresponds to an affine image transform taken from a set of transforms $\Phi$. The optimal $\phi$ will describe whether $t$ has to be flipped and the rotation that should be applied to align it with features contained in $w_j$. The best matching realisation of $t$  is carried forward to update the entire set of prototypes. 

Because the technique is  rotationally invariant, we need to avoid the blank corner regions that rotating an image would introduce. The prototypes are therefore a factor of $\sqrt{2}$ times smaller than the input images when a quadrilateral shape is used. \ac{PINK} performs rotation in a clockwise direction around the central pixel and flips images across the vertical axis.

Computationally generating and comparing all  transforms in $\Phi$ is an expensive process. \ac{PINK} leverages the massive parallel processing capabilities of modern \acp{GPU} to accelerate the brute force optimisation of Equation~\ref{eq:ed}. With the performance optimisation implemented by \ac{PINK}, the algorithm scales in linear proportion to the number of images, neurons and transforms. Throughout this study we use \ac{PINK} \texttt{v1.1}. 

\section{Overview of Our Approach}
\label{sec:approach}
Supervised morphological classification algorithms require data to be labelled (e.g. single, double, bent-tail, etc), and the cost of doing so can be prohibitive for all-sky surveys expected from \ac{SKA} and its pathfinder projects. Furthermore, such labels can often not be transferred to other projects, because of varying resolution, sensitivity, etc. Instead, here we present an unsupervised framework for classifying and identifying  objects.

Additionally, by understanding the underlying \textit{structure} of the training dataset, we can create a framework that is versatile without requiring the training of a new model, unlike typical supervised methods.

For our approach to identifying and cataloguing (potentially resolved) galaxy morphologies across wavebands we train a \ac{SOM} whose role is to project the high order image data onto a manifold in a lower dimensional space. This space is made up of prototypes with meaningful features constructed by \ac{PINK} that represent the dominant structures in the training dataset. Since there are only a small number of prototypes compared to the number in   the training dataset, we can annotate important features and create  descriptive labels for each prototype. The manifold constructed by the \ac{SOM} is used as the basis to transfer knowledge from the annotated prototypes onto items in the training dataset. We do this by  reversing the direction of the mapping carried out against the \ac{SOM}. Essentially, we use an unsupervised \ac{ML} algorithm to create a set of classes that are representative of the types of objects in the image data which are then classified by a domain expert. These labels are then transferred back to the original training data. If the training data were representative of a larger set, this transfer of knowledge can be made to previously unseen data.

By using \ac{PINK}, the solution required to `undo' the affine transformation of an image and map it onto a corresponding neuron is explicitly derived. Therefore, as the \ac{WCS} of each input image is known, features annotated at the pixel level can be converted to absolute sky positions. 

Our approach has three main stages,

\begin{enumerate}
	\item training a representative \ac{SOM} that exhibits physically meaningful object morphologies,
	\item annotating the constructed prototype weights, and
	\item transferring these labeled features back onto the training images and collating them into a useable catalogue.
\end{enumerate} 

We apply our approach to a combination of radio and infrared images and catalogue data from large surveys. With this combination there is sufficient information to distinguish the host galaxy producing the radio emission and separate unrelated nearby galaxies. The radio catalogue describes individual components, and multiple radio components can share the same host galaxy. Ideally, for any radio component a single infrared source is identified as the host galaxy. Practically though this may not be possible for all radio components, particularly in regions where the infrared image is confused, saturated or simply too faint to be detected in the infrared image. 

To maximise the usefulness of this approach, we  minimise the initial upfront cost that is required before training can commence. We require a set of catalogued source components and access to their corresponding images. We do not need supplementary information such as redshifts, magnitudes, flux densities. 

\section{Datasets, Pre-Processing and SOM creation}
\label{sec:dataset}

Distinguishing between  morphologies (e.g. \acp{AGN} and \acp{SFG}) requires multi-wavelength information. We therefore use a combination of radio and infrared data. The infrared emission is used to localise the host galaxy, enabling us to  distinguish between radio components that are related to a single galaxy versus radio components that happen to be near one another by random chance.

\begin{figure*}
  \includegraphics[width=\linewidth]{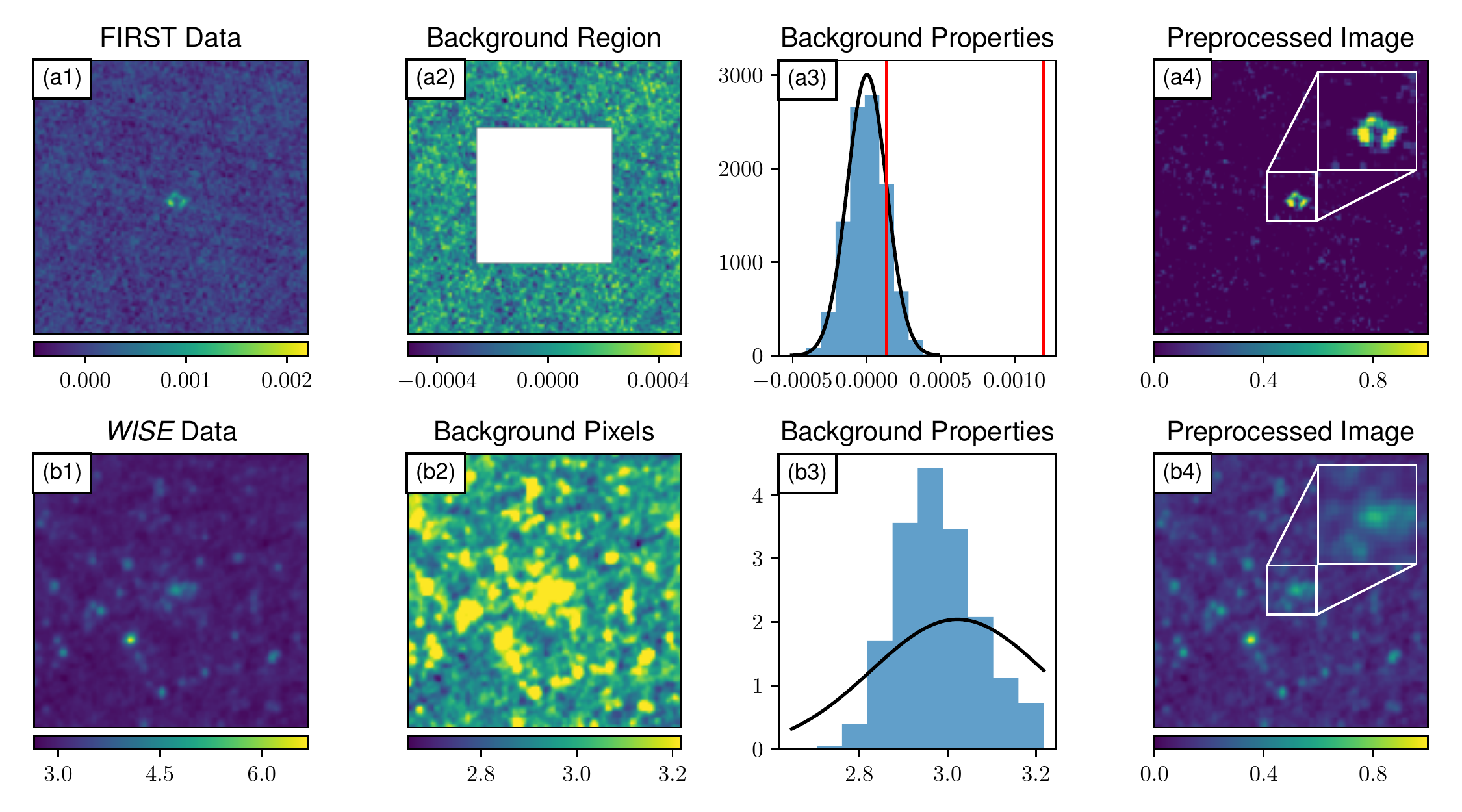}
  \caption{The pre-processing stages applied to the FIRST (top row) and \textsc{WISE} W1 band data (bottom row)  for the sky position towards (J2000) RA, DEC = \texttt{11:37:34 30:00:10} that contains a known wide angle tailed 
  \ac{AGN} \citep{2004cgpc.sympE...4B}. 
Panels (a1) and (b1) show the raw \ac{FIRST} and \ac{WISE} input images  respectively. 
Panel (a2) shows the masking region used to  estimate  the background noise statistics of the FIRST radio continuum image of each source.
Pixels from this background region make up the histogram in panel (a3). 
The overlaid Gaussian on this histogram shows the model we construct to replace empty and missing pixel values from the \ac{FIRST} input images based on the mean and standard deviation of pixels in the background region. 
The left and right red lines represent the $1\sigma$ lower limit and the $9\sigma$ upper limit used to clip the input \ac{FIRST} images.
The final FIRST pre-processed image, shown in (a4), is produced by applying a MinMax normalisation to the clipped \ac{FIRST} data.
To better emphasize the noise characteristics of the \ac{WISE} image, we created panel (b2) by applying stretch to hide the brightest pixels of panel (b1). 
These stretched pixels were used to create the distribution in panel (b3), and their mean and standard deviation were used to establish the overlaid Gaussian profile.
The \textsc{WISE} data were then placed onto a logarithmic scale before being normalised onto a zero to one intensity scale, shown as the final pre-processed image in panel (b4). 
We show the pixel intensity range under each appropriate panel as an accompanying colour bar.
The inset figure within (a4) and (b4) show the zoomed region of the centre of each preprocessed image.
Pixel values in the original FIRST and \ac{WISE} images are Jy/beam and Digital Numbers (DN) respectively. \label{fig:preprocessing}}
\end{figure*}

\subsection{Radio and Infrared Data}

As our base catalogue we used positions of radio components detected from  \ac{FIRST} \citep{1995ApJ...450..559B,2015ApJ...801...26H}, a $\SI{1.4}{\giga\hertz}$ survey covering over 10,000 deg$^2$ of sky conducted with the \ac{VLA}in the  B array configuration.  \ac{FIRST} achieves a resolution of \ang{;;5} with an rms sensitivity of \SI{0.15}{\milli\jansky\per\beam}. Its positional accuracy is  \ang{;;0.05} for radio components whose flux density is $\sim$\SI{0.75}{\milli\jansky}. FIRST detected 946,432 radio components with approximately 35\% showing resolved structure \citep{1995ApJ...450..559B}. Throughout this study we use the most recent version of the \ac{FIRST} catalogue\footnote{\url{http://sundog.stsci.edu/first/catalogs/first_14dec17.fits.gz}}. \revision{The term radio component refers to a single two-dimensional Gaussian that describes a region of radio emission in an image, parameterised by its on-sky position, brightness, orientation angle and angular size.  Extended or highly resolved sources may be described by a collection of radio components}.

Complementing this radio survey is the \ac{WISE} all sky infrared  survey \citep{2010AJ....140.1868W}. It is made up of four wavelength bands corresponding to 3.4, 4.6, 12 and 22\,$\mu$m (labelled as $W1, W2, W3$ and $W4$) reaching 5$\sigma$ point source sensitivities of 0.08, 0.11, 1. and 6.0\,mJy, respectively. The corresponding resolutions for these bands are \ang{;;6.1}, \ang{;;6.4}, \ang{;;6.5} and \ang{;;12}. This study uses only the W1 band.
 
\subsection{Data Pre-Processing}
\label{sec:data_preprocess}

To build our training dataset, we downloaded \ang{;5;}$\times$\ang{;5;} images for each \ac{FIRST} catalogue position 
from the appropriate \ac{FIRST}\footnote{\url{https://third.ucllnl.org/cgi-bin/firstcutout}} and \ac{WISE}\footnote{\url{https://irsa.ipac.caltech.edu/ibe/search/wise/allsky/4band_p3am_cdd}} image cutout services. Using \ac{WCS} meta-data   that accompanied each image \citep{greisen2002representations}, the \ac{WISE} $W1$ band data were \revision{were placed on the same pixel grid as the \ac{FIRST} images using bilinear re-interpolation (the same interpolation algorithm used by \ac{PINK})}. In practise this corresponded to a small rotation correction, generally less than \ang{15;;}. \ac{PINK} is not aware of the \ac{WCS} and works strictly on pixel intensities when searching for the \ac{BMU}. Although this reprojection was not strictly required, it was carried out to ensure that the prototype weights could easily be compared across wavelength channels. A small positional error may be introduced from this reprojection, but this would be most drastic for infrared structures towards the border of the images and are  minimal.  

We then preprocessed the images
to enhance feature characteristics. We expand the procedure outlined by \citet{2019PASP..131j8009G}, which we describe below. 

\ac{FIRST} radio continuum images suffer from correlated Gaussian noise and imaging artefacts, both of which are produced by the \ac{VLA} \ac{PSF}. Although iterative image deconvolution methods \citep[and its modern derivates]{1974A&AS...15..417H} have been carried out on these images \citep{1995ApJ...450..559B}, for many bright sources there still remains imaging artefacts that can contaminate the similarity measure and consequently degrade the set of representative features that \ac{PINK} can construct. To counter both of these issues, a background region was used to estimate the noise properties for each \ac{FIRST} image. If there were any pixels with a \ac{NAN} value (as was the case for images near the edge of a mosaic region)  they were replaced by  values drawn from a pixel distribution with a mean and standard deviation established by the background region (panel a3 of Figure~\ref{fig:preprocessing}). The $1\sigma$ value of this noise distribution was then subtracted from each pixel to set it as the new zero point. Next, to place more emphasis on the morphological shapes of objects within an image, an upper ceiling of 9$\sigma$ was applied.  Finally these images were normalised to a zero to one pixel intensity scale. This procedure is shown in the top row of Figure~\ref{fig:preprocessing}.

\ac{WISE} images followed a similar set of preprocessing stages. However, neighbouring pixels of \ac{WISE} images are not correlated  in the same way as radio-interferometric images. Because of the weight update process of a \ac{SOM}, the lack of correlated noise features means that over many iterations the confused infrared images tend to cancel out all but the consistent features. Therefore, no background masking was attempted for these \ac{WISE} images. We did attempt to characterise the pixel intensity profile only to replace pixels with non-finite \ac{NAN} values. However, as the noise characteristics of the \ac{WISE} instrument are not entirely Gaussian and images are approaching the source confusion limit, our approach is only a first order approximation. To avoid any bias potentially introduced by this incomplete profile, sources whose image contained more than twenty pixels to replace were dropped. This often happened when an image was towards the edge of the field of view of the co-added \ac{WISE} mosaics. If too many pixels are blanked then the potential of clipping out distinguishing structure increases. Finally, \ac{WISE} images were placed onto a logarithmic scale and a MinMax normalisation was performed following
\begin{equation}
    I_{normalized} = \frac{I - \mathrm{min}\left(I\right)}{\mathrm{max}\left(I\right) - \mathrm{min}\left(I\right)},
\end{equation}
where $I$ is the image data to be normalised. This process is presented as the bottom row in Figure~\ref{fig:preprocessing}.

The remaining set of successfully preprocessed images were then formed into two-channel images, with 95\% and 5\% channel weights being applied to the \ac{FIRST} and \ac{WISE} $W1$ data respectively. These weights make the Euclidean distance more sensitive to inconsistencies in the radio channel when searching for the \ac{BMU}. We show the effects of the preprocessing stages in Figure~\ref{fig:preprocessing}.

In Table~\ref{tab:overview} we show the number of images that failed throughout the data acquisition and preprocessing stages. Note that that there were no images that could not be retrieved from the \ac{FIRST} server. A total of $894,415$ of the total $946,432$ \revision{images centred towards} \ac{FIRST} catalogued radio components were successfully preprocessed and placed into our training dataset. 

In principal, we could download images that are centred towards the positions described by an infrared catalogue instead of those from \ac{FIRST}. This may require  some level of curation or subsetting of input images as the source density of the infrared sky is many times higher than the source density of current radio-continuum surveys, and therefore increases the computational requirements of the entire process.  

\begin{table}
\centering
\begin{tabular}{lr}
\hline\hline
Step & Number \\
\hline	
\ac{FIRST} Components & 946,432 \\
\midrule
\ac{WISE} Download Failed & 268 \\
\ac{WISE} Reprojection Failed & 44 \\
\ac{WISE} \ac{NAN} Check Failed & 51,705 \\
\midrule
Training Dataset & 894,415 \\
\midrule
\end{tabular}
\caption{An overview of the number of images that were successfully downloaded and preprocessed into our training dataset from the base \ac{FIRST} radio component catalogue. We include the number of components whose postage stamp image could not be downloaded from the either the \ac{FIRST} or \ac{WISE} server, were unable to be reprojected onto a common pixel grid, or did not pass preprocessing steps due to to many missing or invalid \ac{WISE} pixels (\ac{NAN} Check Failed).  \label{tab:overview}}
\end{table}

\subsection{Training the SOM}
\label{sec:som_training}

It is difficult to construct a \ac{SOM} that represents all the features from a complex training dataset, as the small number of  neurons tends to be  dominated by the most common features in the training dataset. This could be countered by  increasing the number of neurons, but this is very computationally expensive. We therefore adopt a training procedure which uses two hierarchical \ac{SOM} layers, as described below.

\subsubsection{Layer One}
\label{sec:som_layer_1}

We trained an initial \ac{SOM} layer using \ac{PINK} across five simple stages, which are outlined in Table~\ref{tab:training}. Within the nomenclature of \ac{PINK} a single iteration refers to each item in the training dataset being used once to update the prototypes on the \ac{SOM}. A \ac{SOM} of $10\times10$ neurons in a quadrilateral layout was initialised using the \ac{PINK} options `random noise with  `preferred direction.' This will \revision{initialise weights as uniformly distributed noise between 0-1, and will set pixel intensities across the top-left to bottom-right diagonal to values of one}, which attempts to seed the orientation of resolved radio structures in this direction. Throughout all training stages no periodic boundary conditions (i.e. updates that wrap around the edges of the \ac{SOM} lattice) nor update radius cutoffs (i.e. weight updates only applied to prototypes if they are less than some distance to the \ac{BMU}) were used. We used a Gaussian neighbourhood function\revision{ (implemented in \ac{PINK})} to weigh the updates made to prototype weights. \revision{Each training stage has a user defined learning rate that remains constant throughout all iterations. The specifications of each training stage are provided in Table~\ref{tab:training}.}

The goal of the early training stages is to establish the broad layout of features among neurons. This requires only a minimal set of rotations. Subsequent training stages concentrate on the small scale structure. Our first training stage had three iterations and only 48 rotations with increments of $7.5^{\circ}$. Although this introduces the possibility of a small misalignment between an image and prototype ($\sim6$ pixels in the worst case for our training conditions), the $7.5$ factor of improvement in training time was substantial and justified during these earliest training epochs. The size of the neighbourhood function and learning rate are also the largest, allowing for many prototypes to be modified with each update. 

\begin{table}
  \centering
  \begin{tabular}{rrrrr}
  \hline\hline
  Training & $\sigma_u$ & $l$ & Rotations & Iterations \\
  Stage & &  & & \\
  \hline
  1 & 1.5 & 0.10 & 48 & 3 \\
  2 & 1.0 & 0.05 & 92 & 5 \\
  3 & 0.7 & 0.05 & 92 & 5 \\
  4 & 0.7 & 0.05 & 360 & 5 \\
  5 & 0.3 & 0.01 & 360 & 10 \\
  \hline
  \end{tabular}
  \caption{\revision{Parameters used when training \acp{SOM} with \ac{PINK} throughout this study. The column `$\sigma_u$' denotes the one sigma width of the neighbourhood function used to apply prototype updates (Equation~\ref{eq:neighbour}). The learning rate used for each stage during the weighting update process (Equation~\ref{eq:update}) is shown in column `$l$'. } \label{tab:training}} 
  \end{table}

Training stages two and three begin to slowly reduce the region of influence of each prototype update by shrinking the neighbourhood function and lowering the learning rate. A larger set of rotations are also used to capture some of the more refined details. Finally, stages four and five focus on the smallest level of details by allowing 360 rotations with a small region of influence. 

\revision{Hyper-parameter selection is important as there are no formal convergence criteria for the \ac{SOM} algorithm.  If the neighbourhood functions is too large, then all prototype weights are modified, whereas a neighbourhood function that is too small will decouple neurons from each other. Similarly if the learning rate is too small the \ac{SOM} will require a large computation time to train, whereas too large produces abrupt and unstable prototype changes. We converged on these five training stages largely by experimentation across several trials where we qualitatively examined the \ac{SOM} searching neurons that carried physically meaningful morphologies or other distinguishing characteristics. In practise we found the \ac{SOM} was most sensitive to the background clipping level applied to the radio images. By virtue of the weight averaging process, we found that \ac{PINK} was remarkably efficient at extracting the uncleaned remnants of the \ac{VLA} \ac{PSF}  from below the noise, which consequently became the dominating feature across the \ac{SOM}.}

\begin{figure}
\centering
	\includegraphics[width=\linewidth]{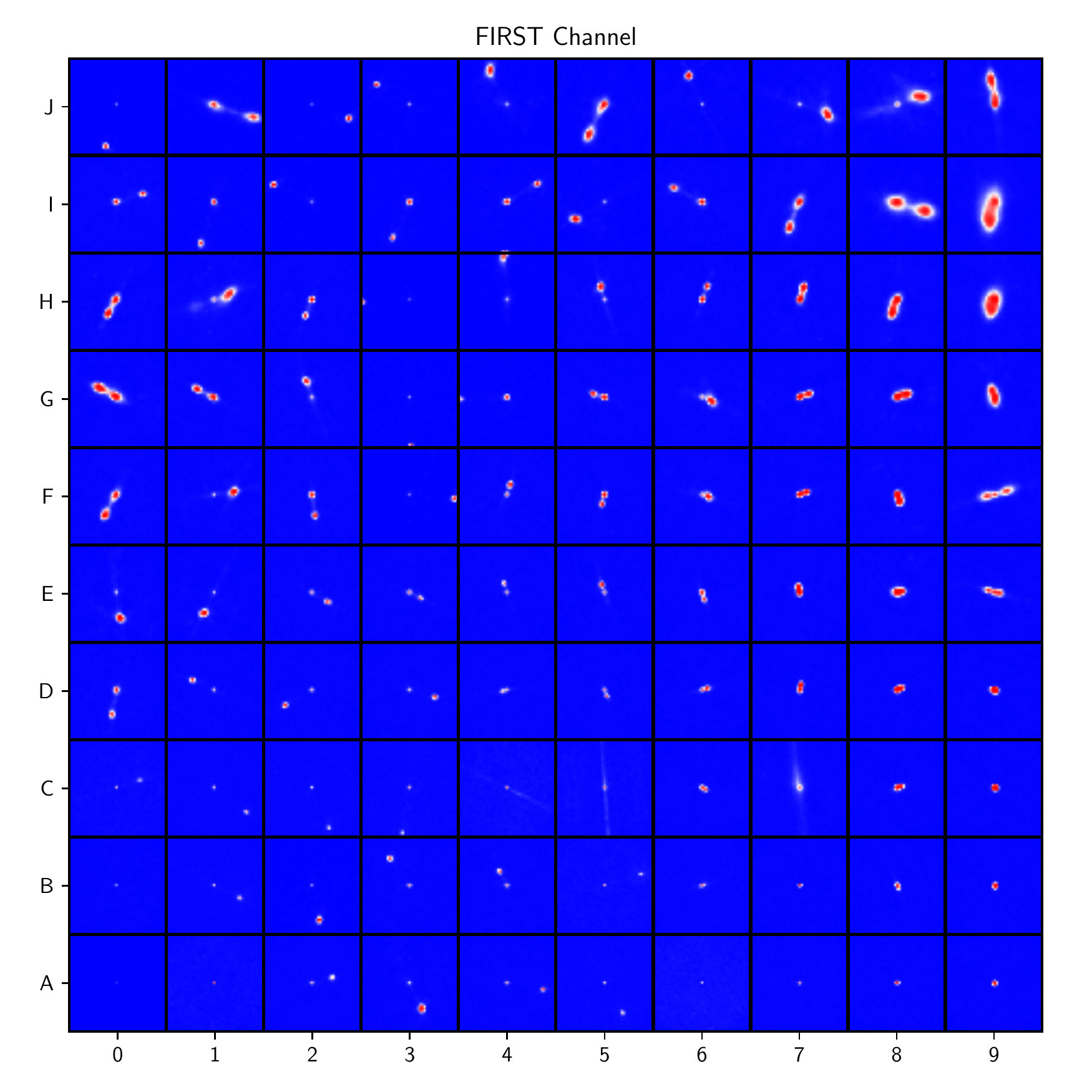}
	\includegraphics[width=\linewidth]{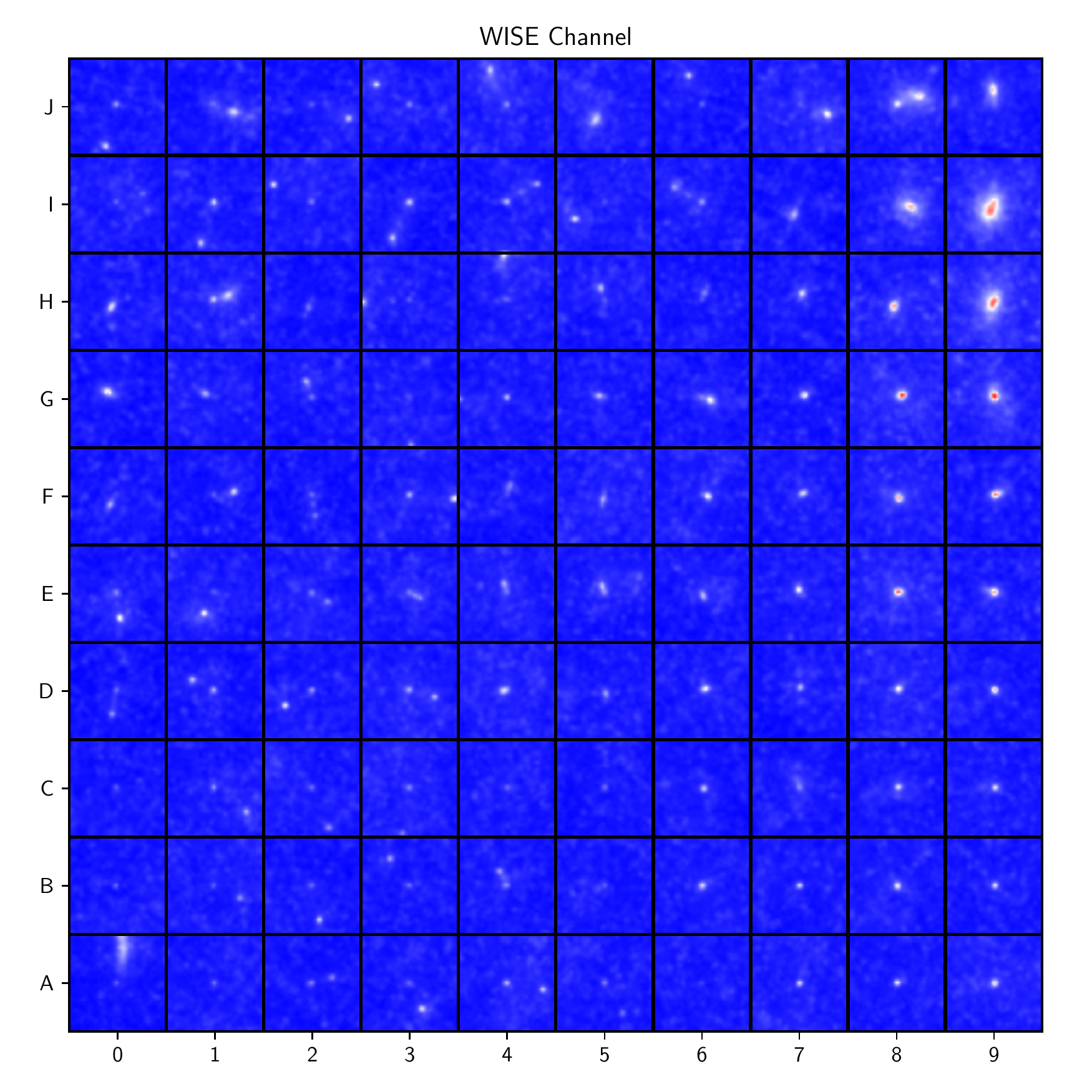}
	\caption{The \ac{FIRST} (top) and \ac{WISE} (bottom) channels of the  layer one \ac{SOM} using the preprocessing and training steps described throughout \S\ref{sec:data_preprocess}-\ref{sec:som_layer_1}.\label{fig:som}}
\end{figure}

The final Layer One \ac{SOM} is shown in Figure~\ref{fig:som}. Across its lattice there is strong evolution of morphologies. Towards the bottom right corner of the \ac{FIRST} channel near the (B, 8) neuron is a concentration of point source objects without companions. When moving towards (B, 1), these single radio features evolve into two distinct components. The neurons around (E, 4) also possess two separate unresolved radio components without signs of extended morphologies or low level diffuse structure. Towards (H, 1) these features slowly become more pronounced and diffuse. The top right corner near (I, 8) contains the largest set of radio features, with most containing two individual radio components with a clear bridge of diffuse emission connecting them. 

The constructed prototypes of the \ac{WISE} channels show less variety. Most neurons have a single unresolved source. However, some of these unresolved objects are offset from the centre of the neuron (see Figure~\ref{fig:ex_proto}, panel (b2) for an example). For neurons in columns 0 to 4 there is often a secondary infrared component offset from the centre of the neuron. Near (I, 8) there are extended structures in the recovered infrared prototypes.  

We emphasise that for the \ac{WISE} images we applied no background clipping or region masking to images in the training dataset. The structures learnt within the \ac{WISE} channel have been recovered through what is essentially an image stacking procedure performed by \ac{PINK} during its prototype weighting update step. The consistent features within individual images re-enforce one another throughout the training stages, and since the instrumental \ac{PSF} of \ac{WISE} if far less significant than the synthesised \ac{PSF} of the \ac{VLA}, the predominate features in the infrared prototype weights are genuine source morphologies. Without background clipping radio images \ac{PINK} does a remarkable job of recovering the residual un-cleaned sidelobe artefacts of the synthesised \ac{PSF} that remain below the noise level and treats these as distinguishing features in the radio prototype weights. 

\subsubsection{Layer Two}

A domain expert may notice that some expected morphological features are missing from Figure~\ref{fig:som}. Because the  \ac{SOM} makes data compete for representation, relatively rare and complex morphological shapes in the training dataset may not be properly represented in the lattice. This is especially true when $\sim900,000$ training images are being projected onto a $10\times10$ lattice embedding. Simply enlarging the lattice structure would give these items sufficient space to grow at the expense of increasing training time, which can become intractable. 
 
Instead we employ a `divide-and-conquer' approach. Using the Layer One \ac{SOM} (Figure~\ref{fig:som}) we divide our training dataset into 100 subsets based on the individual \acp{BMU} of each training item. Next, a $4\times4$ \ac{SOM} is independently trained using \ac{PINK} for each of the segmented data subsets. While training each of these smaller \acp{SOM}, \revision{we use the training procedure used to produce the initial Layer One \ac{SOM} (Table~\ref{tab:training})}. By both segmenting the full training dataset into many smaller subsets and reducing the size of each \ac{SOM}, the training time was significantly accelerated without compromising feature representation. This process is easily parallelisable across many \ac{GPU} compute nodes as each subset is independently trained.
 
After each of the 100 $4\times4$ \acp{SOM} were individually trained, they were concatenated together to form a single $40\times40$ \ac{SOM}, shown in Figure~\ref{fig:big_som}. Each of the individual $4\times4$ \acp{SOM} were placed in the same location as the corresponding Layer One \ac{BMU} neuron in order to preserve the broad morphological structures established by this initial layer. As a consequence there is no longer a smooth evolution of morphological features across the \ac{SOM} lattice. Subsequent training stages could be carried out against this concatenated Layer Two \ac{SOM}, but the effect on the neurons and subsequent processing steps is expected to be minimal. 
  
Training was carried out on a cluster of workstations equipped with 28 \ac{CPU} cores, 64\,GB of main memory and four \acp{GPU}. The initial Layer One \ac{SOM} required 36 hours to train using only a single \ac{GPU} on a single workstation. After segmenting the dataset into 100 subsets, training was significantly accelerated taking only four hours to complete with upwards of 20 tasks running in parallel. All subsequent processing and analysis was carried out on a 2017 Apple Macbook with 16\,GB of system memory and a 8-core \ac{CPU} clocked at \SI{3.1}{\giga\hertz}.

\section{Collating Sources and Catalogue Creation}
\label{sec:cata_create}

In this section we describe how we use data products from \ac{PINK} to identify radio components and their corresponding infrared host galaxies. We also describe a set of statistics to extract sets of important or interesting objects and assess measures of reliability. 

\subsection{Annotating Prototypes}
\label{sec:ant_scheme}

As the training images are centred on known radio source positions, there will always be a radio feature at the centre of each radio channel. When coupled with infrared images there is sufficient information to physically interpret the constructed prototypes.  Examining each set of prototype weights there are four main scenarios consistently encountered, which were:  

\begin{enumerate}
	\item a single radio feature that coincides with a single infrared feature,\label{item:single}
	\item two bright radio features that are separated with an infrared feature located between them, \label{item:double}
	\item three bright radio features with an infrared coincident with the centred radio feature, and\label{item:triple}
 	\item combinations of (i) and (ii) within a single set of prototype weights.\label{item:combo}

\end{enumerate}

\begin{figure*}
\includegraphics[width=0.9\linewidth]{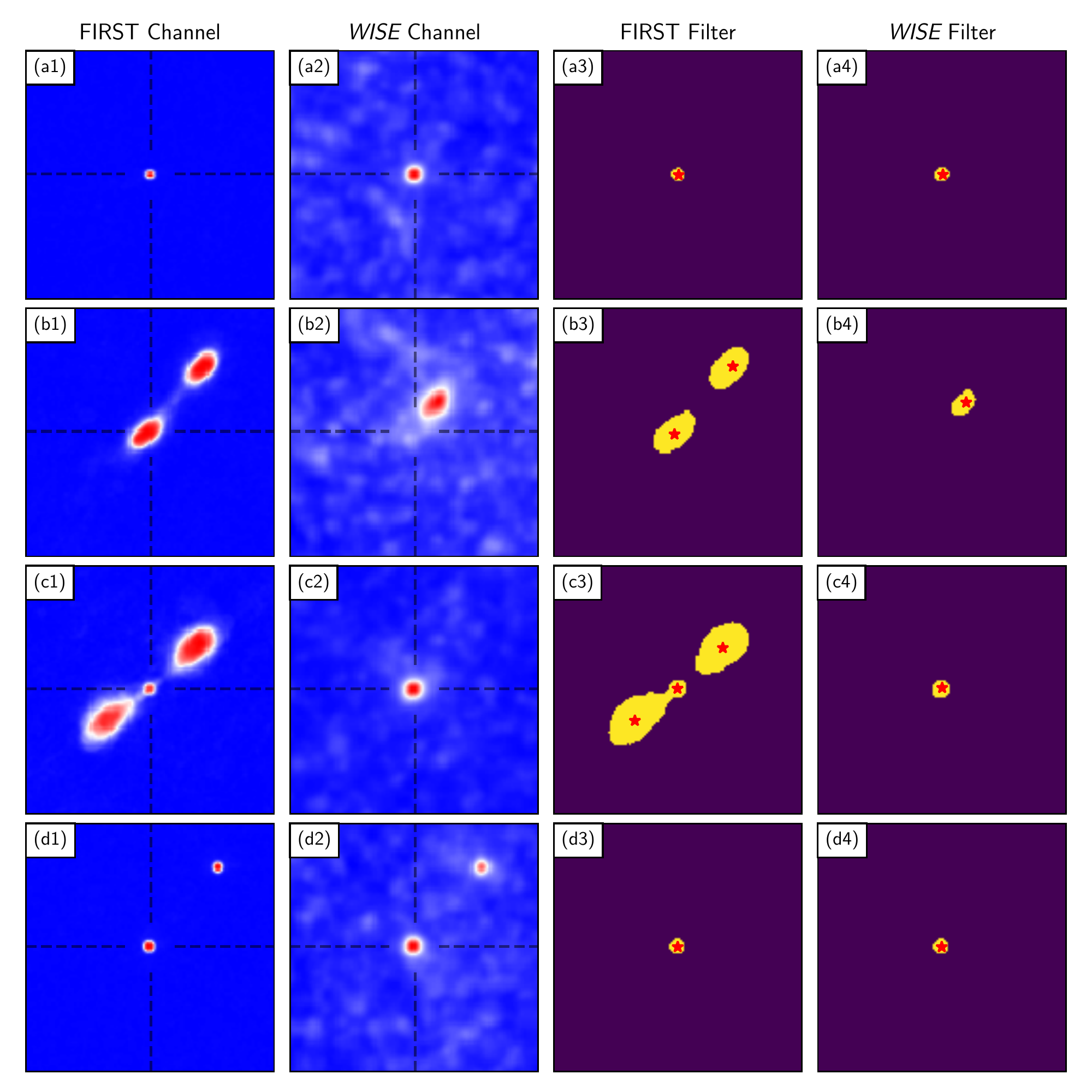}
\caption{Example prototypes and their corresponding filters constructed by \ac{PINK} belonging to neurons from Figure~\ref{fig:big_som} that have been selected to demonstrate the broad set of morphologies described in \S\ref{sec:ant_scheme}. The `\ac{FIRST} Channel' (\textit{first}) and `\ac{WISE} Channel' (\textit{second}) columns are the FIRST and \ac{WISE} channels of prototype weights selected from Figure~\ref{fig:big_som}. Overlaid are  vertical and horizontal dash lines that denote the centre of the image. These prototypes are \ang{;3.5;} in angular size at the \ac{FIRST} resolution. The `\ac{FIRST} Filter' (\textit{third}) and `\ac{WISE} Filter' (\textit{fourth}) columns show corresponding filters produced using the segmentation method described in \S\ref{sec:filter_construction}. Denoted as red stars on each of these filters are the corresponding pixel positions obtained by the annotation process described in \S\ref{sec:ant_scheme}. The yellow islands represent the inclusion region of each filter.  \label{fig:ex_proto}}	
\end{figure*}

\revision{These scenarios were repeated across multiple neurons because \ac{PINK} is not invariant to difference in angular scales or pixel separations between image features. } We interpret images matching item~\ref{item:single} as being an unresolved galaxy that has both a radio and infrared component. Resolved \acp{AGN} with clearly distinguished radio lobes but no detectable radio core would belong to item~\ref{item:double}. \acp{AGN} with a radio core coinciding with the infrared feature would be represented by item~\ref{item:triple}. The key to distinguishing these unresolved galaxies from radio lobes of a resolved \ac{AGN} is the location of the presumed infrared host. Finally, item~\ref{item:combo}  describes scenarios where there are unrelated objects, either resolved or unresolved, that are  close  to each another by random chance. We show examples of these scenarios in Figure~\ref{fig:ex_proto}. 

A properly trained \ac{SOM}  should contain a representative neuron for each item in the training dataset. Therefore,  labels attached to a single neuron can be transferred to any items that share this neuron as their \ac{BMU}.  We therefore annotated each of the $1,600$ neurons on our Layer Two \ac{SOM}. We maintain a simple labelling scheme where we recorded the pixel positions of any radio and infrared features that were related to the centred radio component. Across all prototypes only a single infrared feature was annotated, since we expect that each host galaxy will have only one infrared component. We place no limit to the number of radio features that were  recorded, but in practice found that three was sufficient. We tried to place annotated pixel positions at the approximate centre of each island of pixels. These only needed to be approximate positions as they are subsequently used as seeds for an island segmentation method (\S\ref{sec:filter_construction}).

\subsection{Constructing filters}
\label{sec:filter_construction}

The aim of this study was to identify sets of related radio components and their corresponding host galaxy within an unsupervised \ac{ML} framework. To do this we first mapped each of the individual training images onto the Layer Two \ac{SOM} (Figure~\ref{fig:big_som}) to retrieve their \ac{BMU} and the corresponding $\phi$ (angle of rotation and flip flag) used to minimise Equation~\ref{eq:ed}.

Next we used the pixel positions of annotated features for each neuron to craft an equivalent set of binary filters that represent arbitrary shapes \revision{based on the prototype weights constructed by \ac{PINK}}. We created these filters by first applying a thresholding operation to \revision{each of the prototype weights to} produce a set of candidate islands. The threshold intensity was defined to be $N_t\times\left(I_{\mathrm{max}} - I_{\mathrm{min}}\right) + I_{\mathrm{min}}$, where $I_{\mathrm{max}}$ and $I_{\mathrm{min}}$ were the maximum and minimum pixel intensities \revision{within a channel of the prototype weights}, respectively, and $N_t$ was initially set to $0.9$. If an island contained an annotated pixel position, it was retained, otherwise it was discarded. To the remaining islands we applied filling and dilation steps to ensure they contained no empty pixels \citep{scikit-image}. These remaining islands correspond to an inclusion zone. If there were fewer than $10$ pixels in the inclusion zone we repeated this procedure with $N_t = 0.8$. In Figure~\ref{fig:ex_proto} we show four examples of these filters, including the annotated pixel positions made against the neuron. \revision{We retain these filters as simple images that are programmatically treated as multi-dimensional arrays in subsequent processing (\S~\ref{sec:filter_components}). }  

\subsection{Mapping and filtering source components}
\label{sec:filter_components}

\revision{Source catalogues represent a model of the sky as observed within some image. In a similar manner, \ac{PINK} also attempts to construct prototypes that model features within an image dataset. Sharing annotated knowledge from annotated \ac{PINK} neurons to a source catalogue requires a transformation between the Cartesian coordinate system of a prototype/filter to an absolute on-sky position. }

To do this, consider a single  radio component \revision{at some on-sky position. We first identify all nearby radio components within a \ang{;6;} radius.} This large radius is adopted to select all components potentially within the field of view of that  component's corresponding filter \revision{after a transform has been applied}. The angular offsets on the celestial sphere between the  radio component position and the positions of nearby components are calculated and then converted to Cartesian pixel offsets $\left(x_0, y_0\right)$ using the known \ac{FIRST} resolution (\ang{;;1.8}/pix). \revision{Next, the \ac{BMU} of the subject radio component image is identified and the appropriate \ac{PINK} derived transform $\phi$ is extracted.} Based on the angle of rotation ($\theta$) and flipping flag described by $\phi$, these pixel offsets are then transformed following
\begin{align}
	x_{0}^{\prime}  & = y_0\times\mathrm{cos}\left(\theta\right) - x_0\times\mathrm{sin}\left(\theta\right) \\
	y_{0}^{\prime} & = y_0\times\mathrm{sin}\left(\theta\right) + x_0\times\mathrm{cos}\left(\theta\right),
\end{align}
and in the case where a flipping action was performed $x_{0}^{\prime}$ is negated. The positions $\left(x_{0}^{\prime}, y_{0}^{\prime}\right)$ now describe nearby components in the reference frame of the \ac{BMU} filter. We can therefore directly evaluate whether \revision{the components in a Cartesian frame} fall within the bounds determined by the inclusion region \revision{of a filter}. The corresponding nearby components whose transformed positions fell within the inclusion region were then noted as being related to the centred subject radio component. This process is carried out for both the \ac{FIRST} and \ac{WISE} channels, where for the \ac{WISE} channel we instead search for nearby infrared sources from the All\ac{WISE} catalogue \citep{2013yCat.2328....0C}. The use of carefully constructed shaped filters offers a large degree of flexibility to capture unique morphologies. We refer to this approach of grouping components as a \textit{`cookie-cutter'} method. For convenience $M_c$ will be used to describe sets of candidate radio components and infrared sources that passed through a subject's filter. 

We applied this cookie-cutter to all $894,415$ radio component positions in our training dataset, which produced as many as $M_c$ sets. For the infrared neuron we also transformed the annotated pixel position to an absolute sky position for each source. We will refer to these as `infrared predicted positions'. For consistency we only filter \ac{FIRST} components whose images were in our training set. 

\subsection{Collating related source components}
\label{sec:collation}

As each of the $894,415$ radio components in our training dataset now carry a $M_c$ set after being mapped to Figure~\ref{fig:big_som} in isolation, we now collate these results into a  catalogue of groups that describe related radio components and their corresponding infrared host. \revision{For heavily resolved radio objects represented by two or more catalogued radio components (e.g. \ac{AGN} with two distinct radio lobes) there is potential for redundant information to be present among the multiple $M_c$ sets. Due to field of view effects and unique morphological features, often these $M_c$ sets may not be entirely self-consistent.} This is especially true for resolved \ac{AGN} with large angular separations between their radio lobes. 

We adopt a straightforward method of collating this information together using graph theory \citep{SciPyProceedings_11}. First, each of the $894,415$ radio components within our training dataset were placed as nodes on a graph. Next we sorted neurons \revision{based on the maximum distance between any two annotated click positions within the radio channel of each neuron. Sorting in this manner often placed neurons with \acp{AGN} centred on their core towards the beginning of the sorted list}, and placed point sources towards the end. For neurons with a single radio position annotated, they were simply ordered by their position on the \ac{SOM} (top-left to bottom-right). 

Our procedure then iterated across the ordered list of neurons. For each neuron we selected all $M_c$ sets produced by this neuron's filter. Given a $M_c$, we would work out each combination of listed radio component pairs and add an edge between their nodes on the graph. Once a node had an edge, the $M_c$ set produced by that component would be excluded from subsequent processing. This essentially was a greedy process, where we aimed to connect together as many nodes (i.e. radio components) together with as few $M_c$ sets as possible \revision{while giving preference to $M_c$ sets whose images were resolved radio \ac{AGN} that were centred towards their host (as these neurons were processed at the earliest stages of this greedy process)}. When this was finished, isolated trees of nodes on this graph correspond to unique groups of related radio components within the \ac{FIRST} catalogue. Each group of components was assigned a \ac{GID}. 

Potentially, many $M_c$ sets may be used for a heavily resolved radio object with many radio components. To reduce the number of All\ac{WISE} sources included in each group as potential hosts, we only select the All\ac{WISE} sources from the $M_c$ that contained the largest set of \ac{FIRST} components. If this $M_c$ contained more than ten All\ac{WISE} sources (which occurred in confused fields or with an erroneously constructed filter based on a neuron with ambiguous or poorly defined infrared features) we drop all infrared host information from the collated set. 

Ideally each $M_c$ for each collected object would contain a single infrared All\ac{WISE} source. In practice, this was not always true, as the filters may not be restrictive enough to select a single All\ac{WISE} source in all situations. We therefore assign a numerical flag (\colstyle{grp\_flag}) to each collated group, where:

\begin{enumerate}
	\item[Flag 1 -] a single All\ac{WISE} source passes through the infrared filter and it is less than \ang{;;3.4} from a \ac{FIRST} component in the same group,
	\item[Flag 2 -] a single All\ac{WISE} source passes through the infrared filter and is further than \ang{;;3.4} from any \ac{FIRST} component in the same group,
	\item[Flag 3 -] more than one All\ac{WISE} source passes through the infrared filter with at least one source being less than \ang{;;3.4} from a \ac{FIRST} component in the same group,
	\item[Flag 4 -] more than one All\ac{WISE} source passes through the infrared filter and none were less than \ang{;;3.4} from a \ac{FIRST} component in the same group,
	\item[Flag 5 -] no All\ac{WISE} sources passed through the infrared filter. 
\end{enumerate} 

These flags are not meant to correspond to specific physical scenarios, but are instead intended to act as a potential queryable catalogue property. As we are promoting an unsupervised framework based purely on morphological information, it is foreseeable that subsequent steps could further refine infrared host selection or use this flag as a criterion during a later sample section. The $\ang{;;3.4}$ limit was based on results later described in Figure~\ref{fig:wise_dist} and \S\ref{sec:cross_reliability}. 

For this study our primary aim is to demonstrate a new \textit{unsupervised} \ac{ML} based approach to this difficult cross-matching problem. In the future a more advanced filtering and collating method could be developed that tightly integrates these two stages together. We discuss this improvement in \S\ref{sec:future}. This method of multi-wavelength cross-cataloging (and subsequent processing described below) may also be tested with appropriate simulated image data to further refine the approach and quantify its effectiveness, \revisiontwo{which may be insightful alongside measures of \ac{SOM} connectivity \citep[e.g.][]{connvis} and subsequent analysis steps (\S~\ref{sec:cross_reliability})}. 

\subsection{Size and Curvature properties}
\label{sec:size_curve}

After collating together \ac{FIRST} components we derive a set of additional group properties. Given a group, $G$, that contains the \ac{FIRST} radio components $F = \left[f_0, f_1, ..., f_n \right]$, we calculate the maximum distance, $D$, of each group as 
\begin{equation}
\label{eq:max_dist}
	D  = \mathrm{max}\left[d\left(f_i, f_j\right)\right] : f_i, f_j \in F,
\end{equation}
where $d$ is a function to calculate the on-sky separation between components $f_i$ and $f_j$. This was performed as an exhaustive search between all combinations of two components and was only possible for groups with two or more \ac{FIRST} components. This gives a robust estimation of the projected angular size of all groups that we have collated together. As prototypes are generalised representations of predominate morphologies, simply transferring a an angular distance from an annotated neuron to an individual object may not be ideal. 

With the set of components for each group and the corresponding pair of components with the largest distance ($f_i$ and $f_j$) in that $G$, we can calculate a ratio to identify curved or morphologically disturbed objects. We measure the shortest path, $SP$, by finding the optimal order to $F$ such that 
\begin{equation}
	\label{eq:shortest_path}
	SP = \mathrm{min}\left(\sum_{k=0}^{n} d(f_k, f_{k+1}) \right) : f_i = f_0, f_j = f_n.
\end{equation}
	 
These two competing quantities can be used to defined a `curliness factor' ($Q$), which is simply,
\begin{equation}
	\label{eq:curliness}
	Q = \frac{SP}{D}.
\end{equation}

For groups with only two \ac{FIRST} components, $Q$ will evaluated to be 1. However, for groups with three or more components, the $Q$ will be high for instances where components are not in a straight line (e.g. bent-tail \ac{AGN}). This $Q$ statistic would be a lower limit for extremely bent \ac{AGN} where the radio lobes are closer together than they are to the host position. 

\section{Results}
\label{sec:results}

\subsection{Assessing All\ac{WISE} Source Matches}
\label{sec:wise_match}

Using a dimensionality reduction technique, we have created a set of $1,600$ representative galaxy morphologies at radio and infrared wavelengths. Combining this with a simple annotation scheme and pixel intensity thresholding, a corresponding set of filters were created. These filters were then projected through catalogue space to identify sets of related components. As we have no training labels, it is difficult to establish and quote accuracy, reliability and purity measures that are often used in supervised learning contexts for the same problem \citep{2018MNRAS.476..246L,2018MNRAS.478.5547A,2019MNRAS.482.1211W}. 

As an alternative, the approach we adopt is to use the predicted positions (absolute sky positions based on transformed annotated pixel positions) and perform nearest-neighbour searches against genuine and randomised catalogues. Although this may lose spatial information maintained by the filters, conceptually it is similar to other metrics already used through the literature \citep[e.g.][]{1986MNRAS.218...31D}. We pursue this approach throughout this study to assess reliability. Practically it is straight forward to implement with minimal computation cost and should be accurate as annotated positions were recorded at the centre of important features. 

\subsubsection{Cross-matching infrared predicted positions}
\label{sec:cross_reliability}

We now assess the usability of the infrared host identified for each collated group. The source sky-density of \ac{WISE} is $\sim\SI{15000}{\deg^{-2}}$ and is higher than \ac{FIRST}'s by a factor $\sim200$, making this aspect of the collation problem more difficult. We first cross-referenced our infrared predicted positions to the All\ac{WISE} catalogue \citep{2013yCat.2328....0C}. We searched for all All\ac{WISE} sources $<\ang{;;10}$ of these positions. To establish the by-chance coincidence rate of a foreground or background All\ac{WISE} source at any arbitrary position we performed a Monte-Carlo test where we created a `shifted' predicted position catalogue by adding a constant \ang{;3;} offset to the declination to the genuine set of infrared predicted positions.

\begin{figure}
\includegraphics[width=\linewidth]{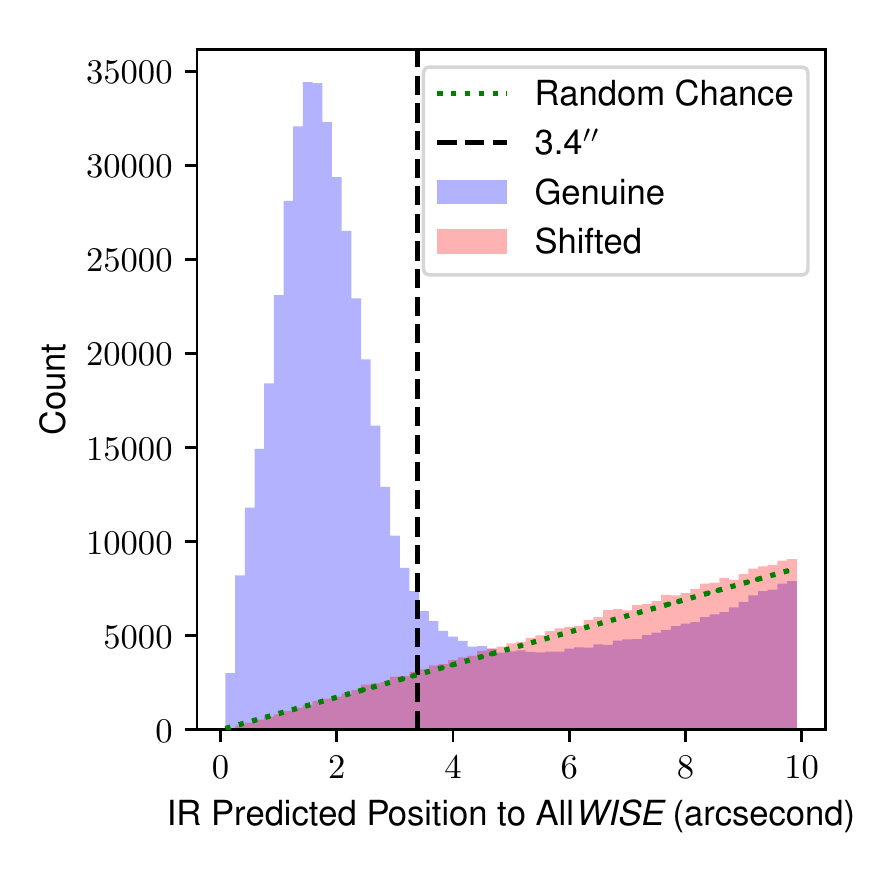}
\caption{The number of All\ac{WISE} objects found in 60 annuli evenly spaced between $0-\ang{;;10}$ for positions described in a genuine and shifted catalogues. This was performed only for groups with a single \ac{FIRST} component (i.e. point sources). Overlaid as the green dotted curve is the chance alignment rate described by Equation~\ref{eq:by_chance}. The vertical dash-dash line shows the point where the genuine catalogue has double the number of sources than the shifted catalogue, which we calculate to be at \ang{;;3.4}.\label{fig:wise_dist}}	
\end{figure}

Next we construct 60 annuli evenly spaced between  $0-\ang{;;10}$ around each position in the genuine and shifted infrared host catalogues and count the number of returned matches. These counts for groups with a single collated \ac{FIRST} radio component are presented in Figure~\ref{fig:wise_dist}. For the shifted catalogue positions, the number of returned All\ac{WISE} sources is proportional to the area  within each annulus. This establishes the chance coincidence rate of foreground or background objects with respect to some arbitrary position. \revision{The genuine catalogue has  $351,310$ more All\ac{WISE} counterparts $<\ang{;;3}$ than the shifted catalogue.} By assuming the source density of All\ac{WISE} to be $\rho_{0}\sim$\SI{15000}{\deg^{-2}} and uniform across the segmented area, the chance count, $B$, of a foreground or background object being present between offsets of $r$ and $r+dr$ can be formulated as,
\begin{equation}
	\label{eq:by_chance}
	B = N\rho_{0}\left(2\pi r\right)dr,
\end{equation}
where $N$  corresponds to the number of positions searched around. Overlaying Equation~\ref{eq:by_chance} onto Figure~\ref{fig:wise_dist} shows that it agrees with the object counts of the shifted catalogue. 

We find that, at an offset of \ang{;;3.4}, the number of sources in the genuine infrared predicted position catalogue is twice the number of the shifted infrared predicted position catalogue. At this radius the likelihood of there being a genuine match for a predicted host position is about the same as random chance. 

Following \citet{2017MNRAS.464.1306C}, a measure of reliability, $R$, of a cross-matched catalogue can be established by, 
\begin{equation}
	\label{eq:reliability}
	R = 1 - N_{\mathrm{shift}}^{\mathrm{match}} /  N_{\mathrm{genuine}}^{\mathrm{match}}, 
\end{equation}
where $N_{\mathrm{shift}}^{\mathrm{match}}$ and $N_{\mathrm{genuine}}^{\mathrm{match}}$ are the number of matches found between some target catalogue and the set of shifted (i.e., random) and genuine catalogues, respectively. Given Figure~\ref{fig:wise_dist}, at \ang{;;3.4} the $R$ statistic is $\sim92.5$\%, and improves as the offset distance becomes closer to zero. \revision{\citet{2017MNRAS.464.1306C} provided optical identifications using data from \ac{SDSS} of $19,179$ of radio sources from \ac{FIRST} using over an area of \SI{800}{\deg^{2}}. They obtain a reliability of $93.5$\% using a rule-based methodology with visualisation inspection of complex \ac{FIRST} sources. } 

\begin{figure}
	\includegraphics[width=0.98\linewidth]{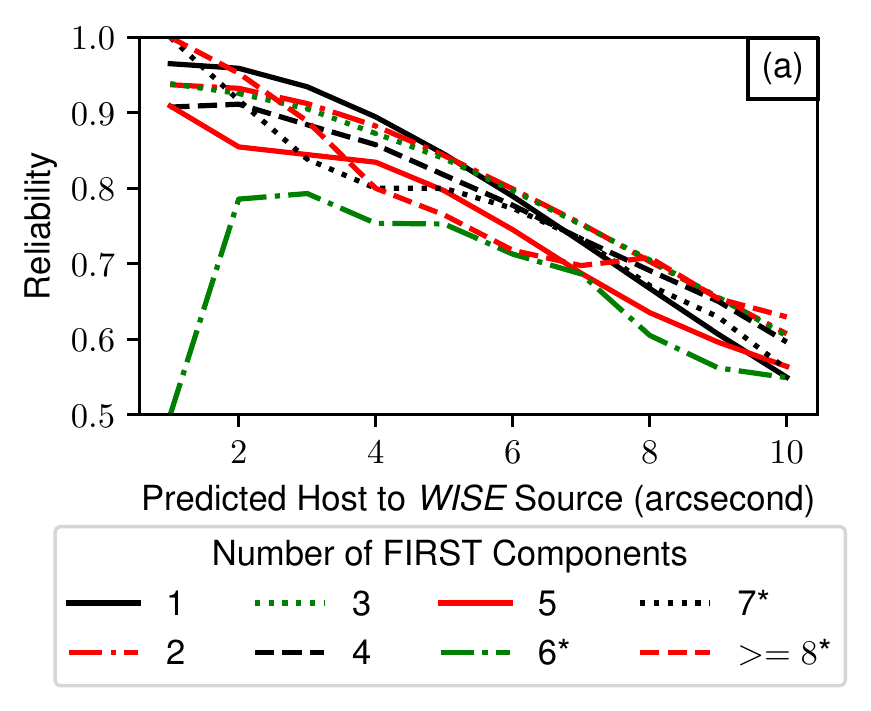}
	\includegraphics[width=0.98\linewidth]{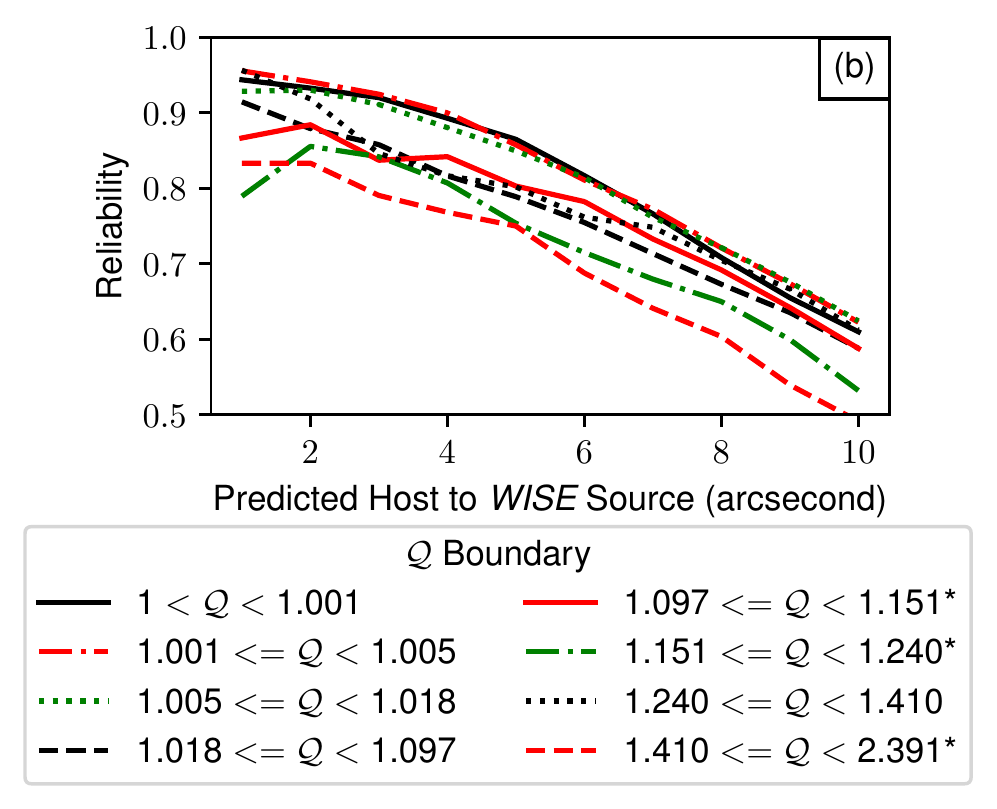}
	\includegraphics[width=0.98\linewidth]{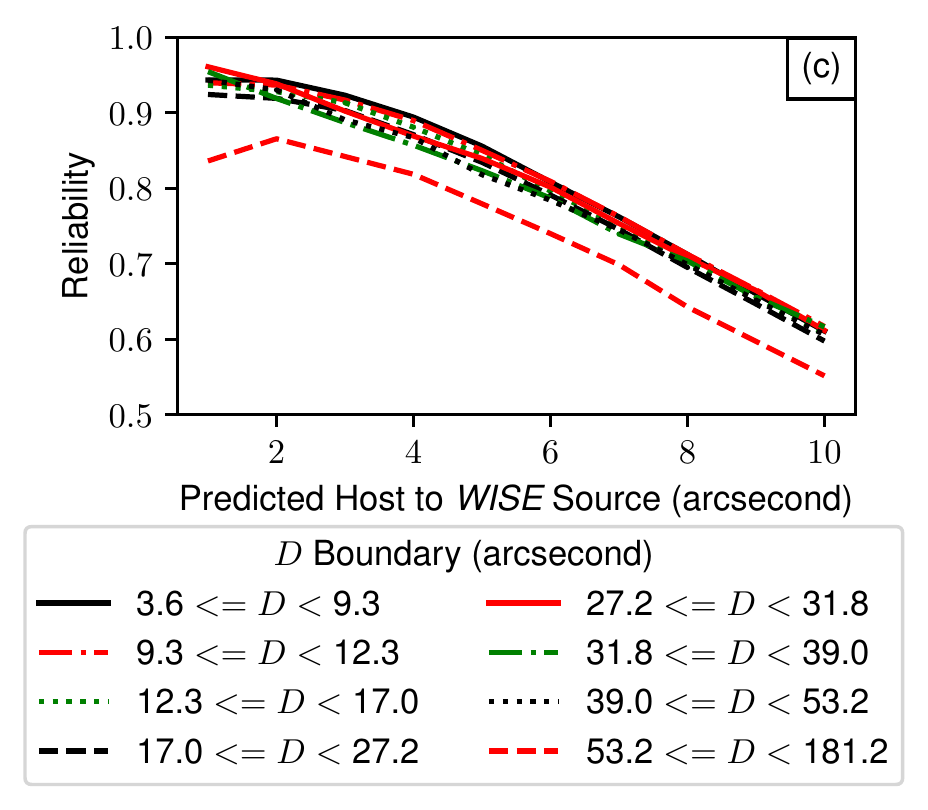}
	\caption{\textit{(a)} The reliability curves of the cross-matched infrared predicted positions against All\ac{WISE} as a function of number of collated \ac{FIRST} components. \textit{(b)} The reliability curves of the cross-matched infrared predicted positions made against All\ac{WISE} as a function of the $\mathcal{Q}$ parameter. Boundaries of the $\mathcal{Q}$ were spaced to contain roughly the same number of groups to maximise the clarity of the different reliability trends. All groups included in this figure contain three or more FIRST components. \textit{(c)} The same as \textit{(b)} except as a function of $D$. All groups included in this figure contain two or more FIRST components. \revision{Labels with a `*'-suffix denote categories with fewer than 15 genuine matches with separations $<\ang{;;1}$}. \label{fig:wise_rel} }
\end{figure}

We generalise this analysis to all collated groups and investigate any second order effects as functions of collated number of \ac{FIRST} components, maximum distance ($D$) and curliness ($Q$). The $R$ was evaluated across a range of separations (spaced from $0-\ang{;;10}$ in steps of $\ang{;;1}$) after dividing our collated groups into meaningful subsets. For the subsets in the $D$ and $Q$ dimensions we established boundaries at the 20$^{\mathrm{th}}$, 40$^{\mathrm{th}}$, 60$^{\mathrm{th}}$, 80$^{\mathrm{th}}$, 85$^{\mathrm{th}}$, 90$^{\mathrm{th}}$ and  95$^{\mathrm{th}}$ percentiles, where finer bins were used in regions of the parameter space with rapid variations identified after inspection the distribution. Otherwise we  created boundaries based on the number of associated \ac{FIRST} components in each collated group. We interpolate between $R$ measures across these dimensions to create a set of reliability curves, which we present as Figure~\ref{fig:wise_rel}. Broadly speaking these curves show that as measures of complexity increase, judged either by the number of collated \ac{FIRST} components or $Q$, the infrared predicted position becomes less reliable. At a threshold of $\ang{;;3.4}$ these $R$ are above 80\% with significant spreads of upwards of $\sim12$\% between the simple and complex groups. However, the reliability curves in Figure~\ref{fig:wise_rel}c suggest groups across all $D$ have about the same degree of reliability. We use these interpolated reliability curves to provide a measure of reliability for all collated groups in our catalogue data products for each of these dimensions. For groups outside the range of these reliability curves we set the $R=0.5$ across all dimensions. \revision{Six of the 24 reliability curves in Figure~\ref{fig:wise_rel} have fewer than 20 genuine matches with separations $<\ang{;;1}$ and are the most susceptible to small number statistical effects. Since increasing separations accumulate counts from smaller separations (i.e. bins are not exclusive), the number of genuine matches very quickly grows so that all other curves and separation bins have at least twenty genuine matches.}

We highlight that assessing reliability in this manner may be incomplete as it is only testing for the presence of an infrared source near a specific position, and not necessarily testing the relationship between the radio and infrared characteristics of a subject galaxy. Given that the neuron prototype weights are constructed to represent features within an image, it is not surprising that there will often be a infrared source at our predicted positions. \revisiontwo{It would be interesting to repeat this procedure after removing the infrared channel from the pre-processed images and Figure~\ref{fig:big_som} to assess its influence. However, we suspect that without the infrared information available the determination of the host position given just radio information will be less reliable because the nature of the radio emission becomes ambiguous}. Also, our infrared predicted position are ultimately tied to the selection of a \ac{BMU}, which in-of-itself may be perturbed by a spurious infrared source in a subject image. In such cases using multiple neurons to recognise conflicting information or a more robust collation procedure may be required to produce a more physically reliable classification. \revisiontwo{Invoking alternative \ac{SOM} statistics \citep[e.g.][]{connvis} could also be useful in detecting and handling ambiguity within the cookie-cutter and subsequent collation process. }

\subsection{Description of Catalogues}
\label{sec:catalogue}

We present two catalogues that have been produced using our `cookie-cutter' approach to grouping and host identification. 

The first catalogue we present is based on the \ac{FIRST} catalogue, but is supplemented with additional columns of information. It details the $894,390$ radio components whose \ac{FIRST} and \ac{WISE} images were successfully downloaded, preprocessed and included in our training dataset (Table~\ref{tab:overview}). We provide all information from the \ac{FIRST} catalogue related to the properties of the radio components themselves (not value added information from earlier cross-matching). As additional columns we include \ac{PINK} provided data products, including the \ac{BMU} position upon the \ac{SOM} lattice, Euclidean distance to the \ac{BMU}, and transformation parameters derived by Equation~\ref{eq:ed}. These \ac{BMU} and transform information could be applied to other image datasets directly if the layout of our Layer Two \ac{SOM} is accepted. For each radio component we also include a \ac{GID} to distinguish unique collated groups of related radio components. We will refer to this as the \ac{FSC}. 

As a second separate catalogue we describe properties of each of the $802,646$  collated groups we have identified. For each group we supply an appropriate \ac{GID} to identify the related \ac{FIRST} radio components from our \ac{FSC}. Included are the solutions to Equations~\ref{eq:max_dist}, \ref{eq:shortest_path} and \ref{eq:curliness}. In column \colstyle{D\_idx} we present a comma-separated tuple detailing the indices of the \ac{FIRST} components in the \ac{FSC} that solved Equation~\ref{eq:max_dist}. A similar column titled \colstyle{SP\_idx} contains comma separated tuples that highlight the order of \ac{FIRST} components visited when solving Equation~\ref{eq:shortest_path}. Note here that the path will start and end on the sources described in the \colstyle{D\_idx} components. Reliability measures for each collated group as functions of $Q$, number of collated \ac{FIRST} components and maximum component separations are also included. As the column \colstyle{comp\_host} we provide a comma-separated tuple that includes the names of any All\ac{WISE} sources that passed through the infrared filter. Of these sources, our \colstyle{prob\_host} contains the All\ac{WISE} source name of the source closest to any \ac{FIRST} component in the same group if the on-sky separation is less than $\ang{;;3.4}$. The infrared predicted position, that were used for the reliability analysis, are also included. We refer to this as the \ac{GRC} and provide a concise summary of \revision{columns in both the \ac{FSC} and \ac{GRC}} in Table~\ref{tab:grp_cat} \revision{and also include a five-row extract of each in Appendix~\ref{sec:cata_extract}}. 
 
 \begin{table}
  \begin{tabular}{lp{6cm}}
  \hline
  Column & Description \\
  \hline 
  \multicolumn{2}{l}{\textbf{\ac{FIRST} Supplemented Catalogue}} \\
  \hline
    \colstyle{GID} &  The unique identifier for each group \\
    \colstyle{idx} & A unique identifier for each \ac{FIRST} radio component \\
    \colstyle{neuron\_index} & A three element tuple containing the index of the \ac{BMU} for the image corresponding to this \ac{FIRST} component position \\
    \colstyle{ED} & The Euclidean distance between the transformed image and the \ac{BMU} \\
   \colstyle{flip} & Whether the image was flipped to align with the \ac{BMU} \\
   \colstyle{rotation} & The angle of rotation that the input image  was rotated to align with the \ac{BMU}, in radians \\

  \hline
  \multicolumn{2}{l}{\textbf{Group Reference Catalogue}} \\
    \hline
    \colstyle{GID} &  The unique identifier for each group \\
  \colstyle{prob\_host} & All\ac{WISE} source name of the probable host \\
  \colstyle{comp\_host} & Names of any All\ac{WISE} sources that also passed through the infrared filter \\
  \colstyle{grp\_flag} & The set of flags described in \S\ref{sec:collation} \\
  \colstyle{D\_idx} & The indices of the \ac{FIRST} components with the largest angular separation\\
  \colstyle{D} & The distance $D$ in arcseconds\\
  \colstyle{SP\_idx} & List of indices of \ac{FIRST} components that produce $SP$\\
  \colstyle{SP}  & Length of the shortest path in arcseconds \\
  \colstyle{Q} & The $Q$ factor \\
    \colstyle{rel\_g} & Approximate reliability as a function of number of collated \ac{FIRST} components \\
  \colstyle{rel\_Q} & Approximate reliability as a function of $Q$ \\
 \colstyle{rel\_D} & Approximate reliability as a function of $D$ \\
 \colstyle{ir\_pp\_ra} & The RA of the infrared predicted position in degrees \\
 
 \colstyle{ir\_pp\_dec} & The Dec of the infrared predicted position in degrees \\
  \hline
  \end{tabular} 
  \caption{A description of the columns that make up the catalogues our method of source collation have produced. The units for each column are dimensionless unless specified otherwise. \label{tab:grp_cat}}
\end{table}

\begin{figure}
	\includegraphics[width=\linewidth]{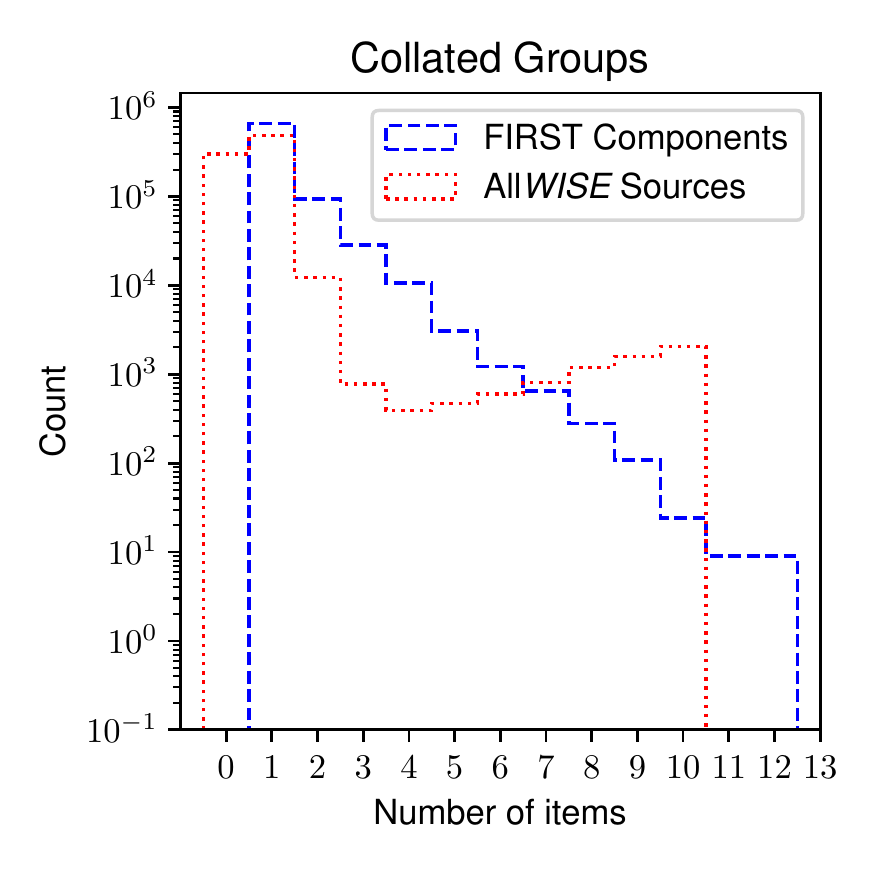}
	\caption{Distribution of the number of \ac{FIRST} components and potential All\ac{WISE} host sources associated with each collated group.\label{fig:group_no}}
\end{figure}

We present in Figure~\ref{fig:group_no} the distribution of the number of \ac{FIRST} components collated and their occurrence. As expected, the overwhelming majority (about 732,000) are groups with a size of one. These correspond to \revision{either point source objects, which remain unresolved at the \ac{FIRST} resolution, or slightly resolved objects that are characterised well by a single two-dimensional Gaussian}. Cross-referencing these point sources to other surveys is largely a simple task that can be achieved with a straight-forward nearest neighbour search  \citep{2011PASA...28..215N}. 
Resolved  objects with many radio components represent the situations where the simple nearest neighbour approach to cross-matching often fails, as these resolved radio components are potentially separated from one another and their host galaxy \citep{2015MNRAS.453.2326B,2018MNRAS.478.5547A}. 
In 62\% 
of these groups, at least one All\ac{WISE} source has been listed as a competing host.

\subsection{Outlying sources}
\label{sec:outliers}

The data deluge from future all-sky radio continuum surveys requires an approach of intelligently assigning meaningful objects to expert users for inspection, whose time and effort should be considered a scarce resource. To maximise the scientific impact, objects presented to human classifiers should be exceptionally rare, interesting and unexpected \citep{2017ASPC..512..109C,2017PASA...34....7N}. Further, it will likely be on these exceptional objects that most classification algorithms, no matter their sophistication, will struggle to perform reliably. 

Assuming an adequately trained map, predominant features in the training dataset should have a representative neuron upon the \ac{SOM} lattice. We examined the distribution of Euclidean distances (Equation~\ref{eq:ed}) of all $894,415$ objects in our training dataset and found it to follow a log-normal distribution with a mean and standard deviation of 15.3 and 12.9, respectively. We could consider a mapping to be `good' if its euclidean distance is less than $+1\sigma$ from the mean (i.e. $<28.2$).
  
\begin{figure*}
\includegraphics[width=0.95\linewidth]{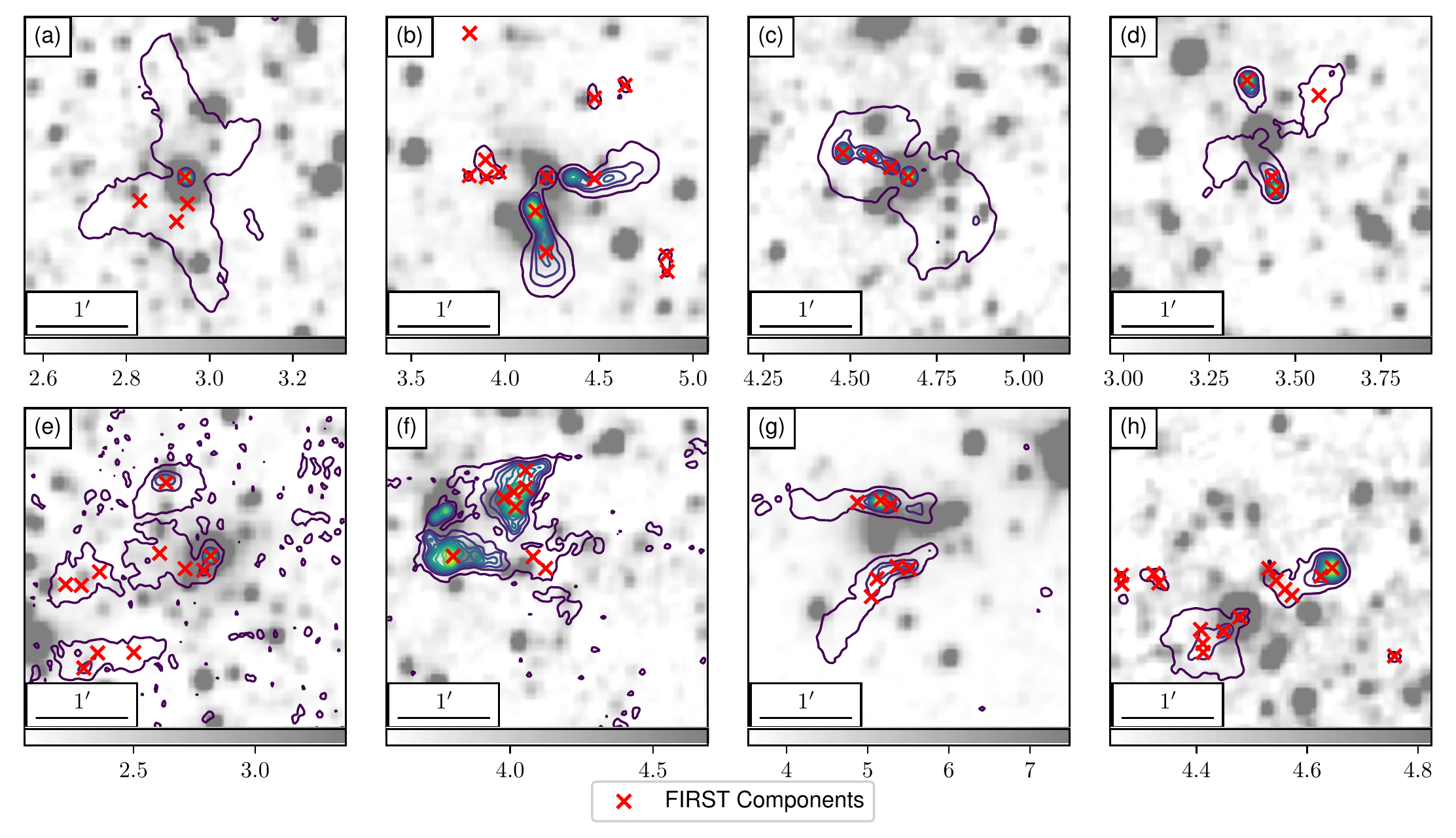}	
\caption{Positions centred towards the \ac{FIRST} radio components with the worst Euclidean distance calculated by \ac{PINK} out of the entire training dataset. The grayscale images and corresponding colourbar are the \ac{WISE} $W1$ band image towards these positions, with pixel intensities in units of DN. Colour contours represented 12 levels of intensity that are evenly spaced between $0.25\sigma$ to the maximum intensity of the \ac{FIRST} image. Note that we are not highlighting the results of our collation procedure and are simply overlaying \ac{FIRST} components.  \label{fig:outliers}}
\end{figure*}

Selecting sources based on their distance from the \ac{PINK} produced prototypes would be an obvious method of segmenting small sets of objects from a larger dataset. To highlight how this measure can be mined for rare and exceptional objects, we show in Figure~\ref{fig:outliers} the eight positions whose mapping to our \ac{SOM} produced the largest set of Euclidean distances (Equation~\ref{eq:ed}), and provide a brief overview of their significance. These objects were not intentionally selected or curated, and are presented to demonstrate that segmenting outliers in this manner is an effective tool capable of isolating physically interesting objects that are in active areas of research. 

From the eight images we highlight in Figure~\ref{fig:outliers} as outliers, panels \textit{(a)} and \textit{(d)} have clear X-shaped radio features. So called `X-shaped' resolved radio galaxies are an ongoing and active area of research \citep{1984MNRAS.210..929L,2002MNRAS.330..609D,2002A&A...394...39C,2007AJ....133.2097C,2009ApJS..181..548C,2009ApJ...695..156S,2018ApJ...852...48S,2019arXiv190506356Y}. These are a set of objects in which there is an additional pair of low surface brightness wings of emission at an angle to the active set of radio lobes  \citep{2007AJ....133.2097C}. The origin of this secondary set of wings is undecided, but common scenarios describe a blackflow of plasma from the set of active lobes, or the rapid realignment of jets, possibly after the merger of two black holes. \citet{2019arXiv190506356Y} recently examined a recent version of the \ac{FIRST} catalogue and identified 290 X-shaped radio sources, 184 of which being labeled as `probable', making them a fairly rare morphological feature. Our two X-shaped radio galaxies are among their sample. 

\ac{HYMORS} are a class of objects that exhibit different \ac{FR} classes on opposite sides of their nuclei \citep{1974MNRAS.167P..31F}. There is ongoing discussion about the mechanism that causes these different morphological structures and whether their manifestation is a symptom of an environmental (nurture) or intrinsic (nature) condition \citep{2017AJ....154..253K}. Building statistically significant samples presents an opportunity towards understanding conditions of this nurture versus nature debate. The radio sources in Panels \textit{(c)} and \textit{(h)} of Figure~\ref{fig:outliers} exhibit \ac{HYMORS}-like morphologies.

Panels \textit{(b), (f)} and \textit{(g)} of Figure~\ref{fig:outliers} exhibit \ac{BT} radio morphologies, sometimes referred to as \acp{WAT} and \acp{NAT}, with their radio lobes appearing to bend away from their typical linear trajectory toward a common direction \citep{2010MNRAS.406.2578M,2018MNRAS.481.5247O}. \acp{BT} have often been used as tracers of cluster and dense \ac{IGM} regions, as their jets are bent by ram pressure applied from the relative motion of its host galaxy moving through a dense medium. 

Finally, panel \textit{(e)} shows a set of radio contours that are highly disturbed and irregular. Even with the inclusion of All\ac{WISE} information it is unclear whether these radio components are all related to a single intrinsic object or many distinct objects in close proximity. This region was also presented as figure~7 of \citet{2015MNRAS.453.2326B} as an example of unusual structures that have been identified by citizen scientists participating in \ac{RGZ}. We emphasis that Figure~\ref{fig:outliers} was based on information from an entirely unsupervised \ac{ML} algorithm without curation. This coincidence further highlights the capability of segmenting interesting objects from a larger dataset using this parameter space.

\subsection{Groups of Radio Sources}
\label{sec:groups}

\begin{figure*}
	\includegraphics[width=0.85\linewidth]{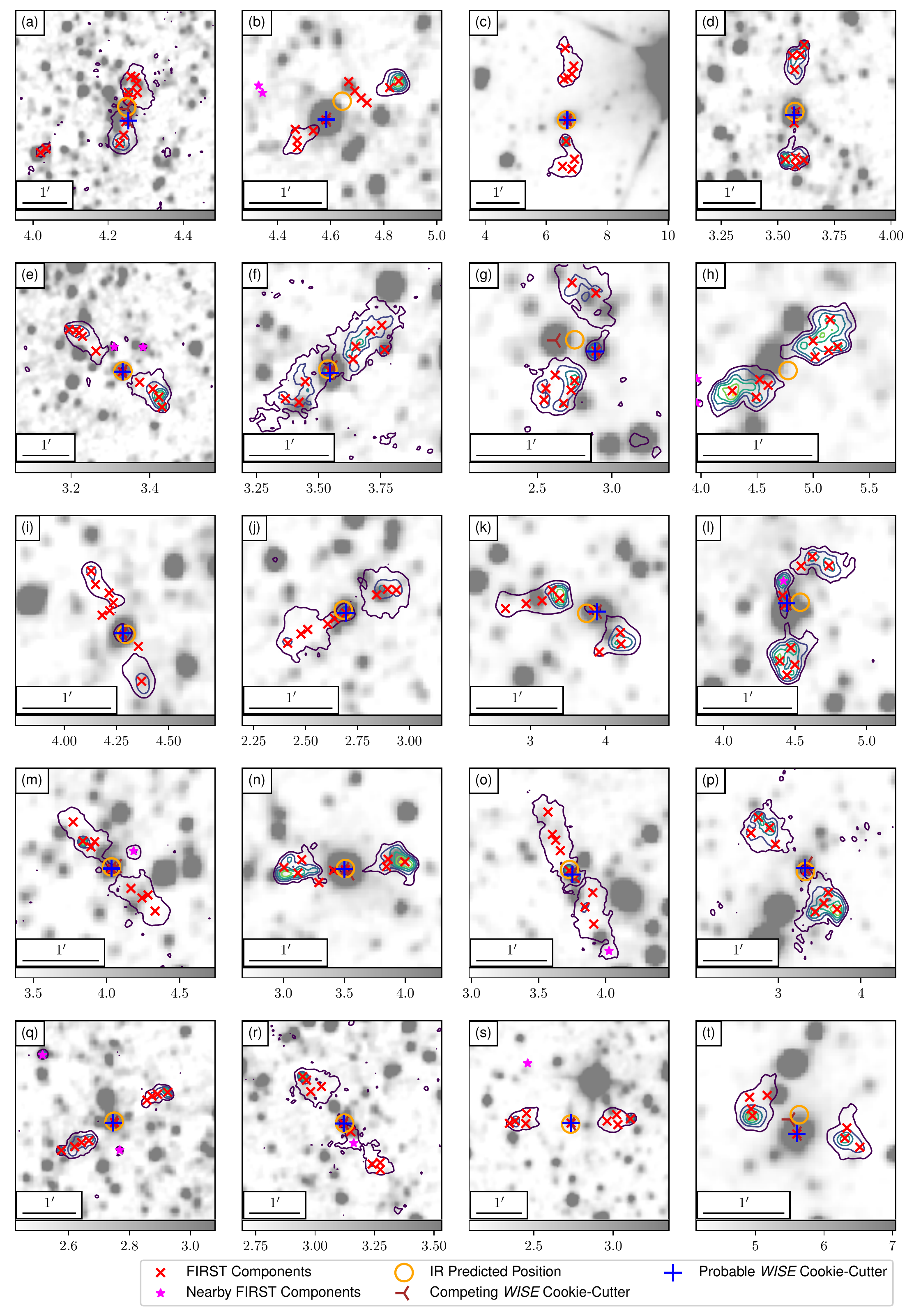}
	\caption{The twenty collated groups with the largest collection of \ac{FIRST} components. For each panel the grayscale image and corresponding colour bar represent the \ac{WISE} $W1$ image, with pixel intensities in unit of DN. \ac{FIRST} radio components with the target \ac{GID} are labelled as `\ac{FIRST} components' whereas components with an unrelated \ac{GID} are `Nearby \ac{FIRST} Components'. Any All\ac{WISE} source that passed through the projected filter are labelled as `Competing \ac{WISE} Cookie-Cutter', and of these sources (if any), we label the one closest to a \ac{FIRST} component as `Probable \ac{WISE} Cookie-Cutter' if it was within $\ang{;;3.4}$ (\S\ref{sec:collation}). The `IR Predicted Position' describes the annotated infrared pixel position after being transformed to an absolute sky position. Colour contours are six levels evenly spaced between  $0.5\sigma$ to the maximum intensity and are based on the \ac{FIRST} image. \label{fig:example_srcs1}}
\end{figure*}

To highlight the structures recovered from our method, we show in Figure~\ref{fig:example_srcs1} the twenty groups with the largest set of collated \ac{FIRST} components. From these examples a number of points can be made. The most obvious is that applying the \ac{BMU} filters to the \ac{FIRST} components as a selection tool has recovered meaningful relationships that extend beyond simple islands of contiguous pixels. \revision{Seven} of the example groups have radio lobes that are separated from their presumed host by angular distances $>\ang{;;45}$. Immediately this extends the capabilities of modern source finders which can include island information for decomposed radio components \citep{2015ascl.soft02007M,2018PASA...35...11H,2018MNRAS.476.3137R}. 

Secondly, projecting meaningfully constructed filters through the catalogue space (the `cookie-cutter' procedure) has performed  remarkably well at separating unrelated source features from one another. Images in panels \textit{(b)}, \textit{(e)}, \textit{(h)}, \textit{(m)} and \textit{(q)} of Figure~\ref{fig:example_srcs1} demonstrate instances where nearby \ac{FIRST} components have not been associated with the target group despite being in relatively close proximity \revision{($<\ang{;;30}$)} to some of their resolved components. In contrast, panels \textit{(l)} and \textit{(r)} show cases where there are unassociated \ac{FIRST} components that could reasonably be grouped with the target collated group. Including island information from modern source finding codes process may help for these cases, as there is an additional queryable parameter linking (presumably) related radio components together. 

We draw attention to panel \textit{(a)} of Figure~\ref{fig:example_srcs1}, which shows an \ac{AGN} roughly \ang{;2;} in angular scale that has had two \ac{FIRST} components (located towards the South-West region of the panel) incorrectly associated with it. This incorrect grouping is caused by the \ang{;3.5;} field of view of our neurons. When images centred on these two \ac{FIRST} components were mapped onto our \ac{SOM}, \ac{PINK} only had visibility of the southern \ac{AGN} lobe (and not the northern lobe). By chance there was also an infrared source located roughly between these two radio components and the southern lobe. Hence, \ac{PINK}'s similarity measure responded well to prototypes corresponding to \ac{AGN} radio morphologies. Our present approach of collating results together only considers the \ac{BMU} mapping of each component in isolation and does not attempt to verify a self-consistent sky-model. We discuss potential improvements to address these scenarios in \S\ref{sec:future}.

The infrared sky as surveyed by \ac{WISE} has a high source density (\SI{15000}{\deg^{-2}}) and is beginning to approach its source confusion limit. Our approach of projecting an appropriately derived filter through catalogue space does a reasonable job of selecting candidate host objects from the All\ac{WISE} catalogue. In some cases, such as panels \textit{(a), (f), (g), (m), (n), (p)} and \textit{(t)}, these filters have captured more than one All\ac{WISE} source. These tend to be in especially confused fields. Where possible, we break this degeneracy by relying upon one of these host candidates being coincident with a \ac{FIRST} radio component. Otherwise, for the remaining panels, with the exception of \textit{(h)}, an All\ac{WISE} source has been identified by passing unaccompanied through the projected filter. 

During this stage we also highlight the sky position obtained by transforming the annotated pixel position made against the infrared channel of the prototype weights for panels in Figure~\ref{fig:example_srcs1}. These are denoted as `IR Predicted Position' and often are aligned closely to the captured All\ac{WISE} hosts. We include these transformed feature locations to argue that our reliability assessment (\S\ref{sec:cross_reliability}) is an appropriate first order approximation of our overall `cookie-cutter' methodology. For cases where no All\ac{WISE} sources passed through the appropriate filter, searching around this transformed infrared feature location in a nearest neighbour type fashion may itself be an effective method if locating a probable infrared host galaxy. 

We manually inspected a set of 200 groups (sorted in a descending order by the number of \ac{FIRST} components) and find that these presented examples are representative. 
\revision{However,  $\sim20\%$ of heavily resolved objects with extremely bent radio lobes had an incorrect of misaligned predicted infrared host.}

\subsection{Curved Sources}

\begin{figure*}
	\includegraphics[width=0.95\linewidth]{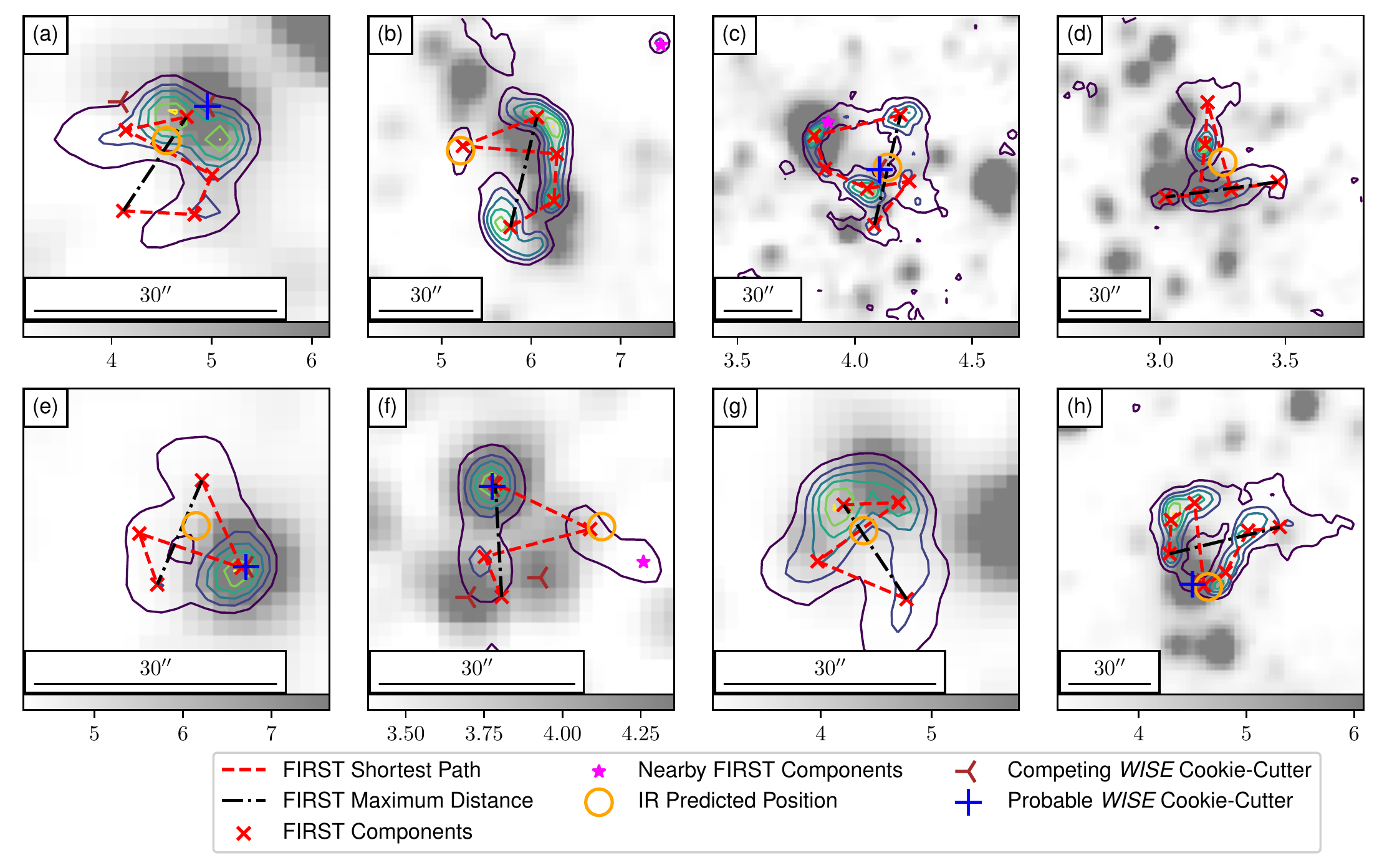}
	\caption{A set of eight groups with the largest $Q$ factors. Legend items carry the same meaning as Figure~\ref{fig:example_srcs1}, except we include as the dot-dash black line we show the pair of \ac{FIRST} components with the maximum separation (Equation~\ref{eq:max_dist}), and as the dash-dash red line we show the corresponding shortest path (Equation~\ref{eq:shortest_path}). Colour contours represent six intensity levels evenly spaced between $2\sigma$ and the maximum intensity of the \ac{FIRST} image. } 
\end{figure*}

We derived the $Q$ statistic (Equation~\ref{eq:curliness}) after examining the prototypes constructed by \ac{PINK} and noticing that there were few neurons that exhibit clear radio lobes with disturbed and bent morphologies. Instead, prototypes were constructed to have very circular lobes with the practical effect of matching to \textit{any} radio morphology with bent lobes. As a consequence, individual images of actual \ac{BT} radio objects had unreasonable infrared sources presented as likely hosts. Therefore, crafting some measure of `curliness' would be a useful quantity to (i) identify these interesting objects with bent morphologies, and (ii) include a metric that would suggest that our potential infrared hosts and their associated \ac{FIRST} components are not as reliable as a more typical object. 

We show the eight collated groups with the largest $Q$ statistic from our \ac{GRC} in Figure~\ref{fig:curl}. A set of general impressions can be drawn from these example groups. The first is this $Q$ statistic is successful in extracting a set of groups that would be considered to be \ac{BT} objects, panels \textit{(a), (c), (d), (f), (g)} and \textit{(h)}, depending on the definition adopted. Compared to the objects from Figure~\ref{fig:outliers}, the groups with high $Q$ values are generally much smaller in angular scales, with all being $<\ang{;1.5;}$. 

Visually inspecting these eight (and other) high $Q$ groups reveals that  our cookie-cutter method has sometimes allowed more than a single All\ac{WISE} source to pass through the projected infrared filter and present as a potential host. Many of these are not believable candidates and are an artefact of a misapplied filter made against an under-represented galaxy morphology. To some degree, searching for coincident \ac{FIRST} radio components has helped to reduce this competing set of infrared sources to a more compelling host - a criteria we describe in \S\ref{sec:collation}. Panels \textit{(a)} and \textit{(f)} illustrate this, as both have at least two All\ac{WISE} sources nominated as potential hosts. However, for these examples a reasonable All\ac{WISE} source has been selected based on its proximity to a corresponding collated \ac{FIRST} component. 

Included in each panel of Figure~\ref{fig:curl} are the infrared predicted positions. Comparing these to the set of All\ac{WISE} sources nominated as potential hosts the projection of a filter through catalogue space highlights how there can be significant disagreement between the two approaches. In cases where no All\ac{WISE} sources passed through a filter, the infrared predicted positions generally lies towards the geometric mean of all grouped \ac{FIRST} components, not towards the `root' of both jets. This could be thought of as a natural consequence of these types of objects being under-represented with appropriate neurons on the lattice, and an over-generalisation of the few neurons that attempt to characterise their unique shapes. We discuss potential avenues of improvement in \S\ref{sec:future} for these class of objects.

\subsection{Identification of GRGs}

\begin{figure*}
	\includegraphics[width=0.925\linewidth]{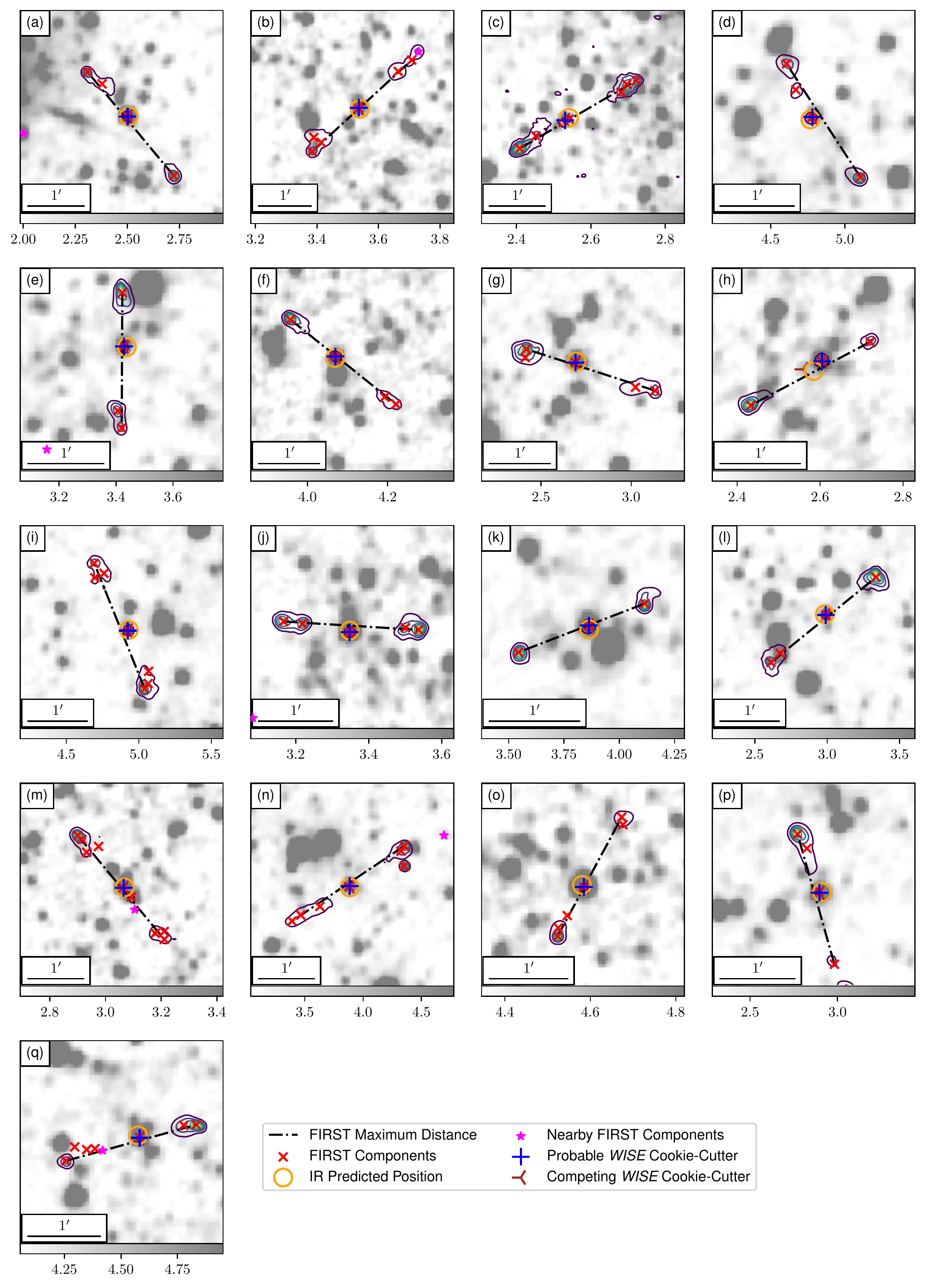}
	\caption{The 17 GRGs we have found from the \ac{FIRST} survey using our method of component collation.  Legend items, images and contours carry the same meaning as Figure~\ref{fig:example_srcs1}, except we include as the dot-dash black line we show the pair of \ac{FIRST} components with the maximum separation (Equation~\ref{eq:max_dist}) that were used to obtain a physical size. \label{fig:grg_example}}

\end{figure*}

Another scientific outcome from our compiled \ac{GRC} is the search for \acp{GRG}. \acp{GRG} are radio sources with physical sizes exceeding \SI{0.7}{\mega\parsec} \citep{2017MNRAS.469.2886D}, making them among the largest individual astronomical objects in the Universe. It is thought that their immense radio lobes are powered by an accreting \ac{SMBH} of mass between $10^8$-$10^{10}$\,M$_\odot$ \citep{1969Natur.223..690L,1984RvMP...56..255B}. However, the mechanism that allows their radio jets grow to these extreme physical scales is debatable. One scenario is that the matter surrounding the \acp{GRG}, the  \ac{IGM}, has a low density, thereby allowing the jets to grow without being confined or frustrated \citep{2008MNRAS.385.1286J,2008ApJ...677...63S,2009MNRAS.393....2S,2015MNRAS.449..955M}. Alternatively, \acp{GRG} could simply be an older population of typical radio galaxies that have grown to their current size over long time-scales \citep{1997MNRAS.292..723K}. Building sufficiently large samples of \acp{GRG} that can begin to distinguish these scenarios is difficult due to their rarity. Mining all-sky radio-continuum surveys for \acp{GRG} are one of the most efficient methods of locating them \citep{2017MNRAS.469.2886D,2019arXiv190400409D}. The task of identifying related radio lobes is only a single component of searching for \acp{GRG}, with the more difficult being host galaxy identification. Our collation method and resulting catalogue set contains this information. 

We search for \acp{GRG} in our \ac{GRC} by first selecting groups with
\begin{enumerate}
	\item a \colstyle{grp\_flag} of either `1' or `3', and \label{item:grg_1}
	\item a \colstyle{D} greater than zero. \label{item:grg_2}
\end{enumerate}

Item \ref{item:grg_1} describes sources where an All\ac{WISE} passed through the projected filter and was coincident with to within $<\ang{;;3.4}$ of a \ac{FIRST} component in the same collated set. This criteria selects radio objects where our collation method has identified a likely infrared host. The secondary criteria \ref{item:grg_2} ensures only resolved objects with a measurable angular extent are returned. 

 Applying these criteria reduced the set of $802,646$ groups in our \ac{GRC} to $19,539$. To obtain a distance measure to the host object we search for spectroscopic redshifts of the $19,539$ All\ac{WISE} sources from two \ac{SDSS} catalogues. Our primary catalogue was from \citet{2017A&A...597A..79P}, who describe spectra of $297,301$ visually confirmed quasars from the \ac{BOSS} from \ac{SDSS}. Our secondary catalogue was \ac{SDSS} data release 12 \citep{2015ApJS..219...12A}, which contained upwards of 470 million spectra. We place preferences towards matches made against \citet{2017A&A...597A..79P} as their pipeline was tailored towards the broad-line emission and absorption features that are typical of quasar optical spectra that the \ac{SDSS} pipeline has difficulty with. Matches for $13,931$ of the $19,539$ sources were found within a \ang{;;3.25} radius, $4,337$ of which possessed a spectroscopic redshift. Cross-matching was performed using the \texttt{X-Match} remote service from CDS\footnote{\url{www.cds.u-strasbg.fr}}. Although there are newer \ac{SDSS} data releases \citep{2017ApJS..233...25A,2018ApJS..235...42A,2019ApJS..240...23A}, they are not yet available through the \texttt{X-Match} interface. 
 
 We then calculated the proper physical size of these $4,337$ sources, accounting for the apparent change in angular scale of objects at cosmologically significant distances \citep{1999astro.ph..5116H}. Adopting a \ac{GRG} definition of $>\SI{0.7}{\mega\parsec}$ we identify a set of 17 \acp{GRG}. 
  
We present as Figure~\ref{fig:grg_example} images of these 17 \acp{GRG} and outline their properties in Table~\ref{tab:grp_cat}. Included are their redshifts from \ac{SDSS}, their derived physical sizes and the separation between the All\ac{WISE} source we label as the host and the cross-matched \ac{SDSS} source. All of these separations are $<\ang{;;0.5}$. As we use the maximum distance between grouped radio components and not the maximum distance between resolved extended emission, their prescribed angular and physical sizes depend on the reliability of the source finding software used by \ac{FIRST}, and in some cases could be considered as lower limits.

\begin{table}
\setlength{\tabcolsep}{3.5pt}
	\resizebox{\columnwidth}{!}{%
\begin{tabular}{lrlrrrr}
\toprule
 Panel & GID &             SDSS & Offset & Angular &    $z_{\mathrm{spec}}$ & Physical \\
&     &   	DR12 & 		 & Size	   & & Size	  \\
\midrule
     &	&				& ($''$)	 & ($''$)	   & & (kpc)	   \\ 
\midrule
a &  7  &  J160953.42+433411.4 &    0.06 &    142.7 & 0.76 &        1069.2 \\
b &  10 &  J085743.54+394528.7 &    0.49 &    151.3 & 0.53 &         964.7 \\
c &   16 &  J151903.65+315008.6 &    0.33 &    143.5 & 0.56 &         941.3 \\
d &  181 &  J111215.45+112919.2 &    0.33 &    108.5 & 1.13 &         907.5 \\
e &  259 &  J031413.72-075023.7 &    0.04 &    106.0 & 1.25 &         901.1 \\
f &   25 &  J100550.63+211652.7 &    0.32 &    131.8 & 0.56 &         862.2 \\
g &    4 &  J125142.03+503424.6 &    0.22 &    131.3 & 0.55 &         853.7 \\
h &   62 &  J103050.90+531028.7 &    0.03 &    100.3 & 1.20 &         847.1 \\
i &   31 &  J100751.16+165243.3 &    0.31 &    135.2 & 0.49 &         824.5 \\
j &  373 &  J120821.99+221958.1 &    0.09 &    108.9 & 0.74 &         809.9 \\
k &  463 &  J121431.15+182815.0 &    0.13 &     89.8 & 1.59 &         776.3 \\
l &   19 &  J161502.40+285819.0 &    0.36 &    134.8 & 0.43 &         769.4 \\
m &   12 &  J105224.06+373004.5 &    0.10 &    143.9 & 0.37 &         751.7 \\
n &  183 &  J155140.30+103548.6 &    0.41 &    143.9 & 0.37 &         741.8 \\
o &  185 &  J123604.51+103449.2 &    0.11 &    102.1 & 0.67 &         726.2 \\
p &  453 &  J131524.33+383044.1 &    0.29 &    107.1 & 0.59 &         719.1 \\
q &  465 &  J102215.04+174648.9 &    0.09 &    111.0 & 0.53 &         706.0 \\

\bottomrule
\end{tabular}
}
\caption{The 17 \acp{GRG} we have identified after cross-matching \ac{WISE} sources to \citet{2017A&A...597A..79P} and \citet{2015ApJS..219...12A}, ordered by their proper physical size. The `SDSS DR12' and `Offset' columns contain the name of the nearest \ac{SDSS} object name and the corresponding separation when cross-matching the infrared host objects of selected groups. The angular size is the maximum distance found following Equation~\ref{eq:max_dist} for each group. We include the spectroscopic redshift ($z_{\mathrm{spec}}$) provided by \ac{SDSS}. Note that by chance all these objects contain redshifts are from \citet{2015ApJS..219...12A}. The physical size has been calculated using \texttt{astropy.cosmology} machinery and is in units of kpc. \label{tab:grg} }
\end{table}

Provided that the redshifts from \citet{2015ApJS..219...12A} and \citet{2017A&A...597A..79P} are reliable, these 17 \acp{GRG} are likely genuine. Each of the matched All\ac{WISE} sources have an angular offset to their \ac{SDSS} source that is $<\ang{;;0.5}$, making them fairly robust. To estimate the false match rate of these All\ac{WISE} sources we shifted all $19,539$ positions submitted to \texttt{X-Match} by \ang{;3;} in declination. Of these submitted sources there were only $1,301$ matches, only 7 of which possessed a spectroscopic redshift. We can therefore estimate that there is about a 1 in 200 rate ($7/1,301 = 0.5$\%) of a spurious object entering our sample \textit{before} our \ac{GRG} criteria is applied. Of course, erroneous collated groups could be produced from mis-applied filters. Visually inspecting these 17 \acp{GRG} suggests many are legitimate and reliable groupings, although the host galaxy selected for panel \textit{(m)} may be slightly ambiguous.

We compared our catalogue to other studies \citep{2017MNRAS.469.2886D,2019arXiv190400409D}, and to our knowledge 16 of these 17 \acp{GRG} are previously unidentified. \citet{2019arXiv190400409D} use data from the \ac{LOTS} data release one \citep{2019A&A...622A...1S} to identify 240 \acp{GRG} in a \SI{400}{\deg^2} region, and also identified the \ac{GRG} we label as panel \textit{(g)}. In principle we could perhaps identify a larger sample of \acp{GRG} by expanding our criteria to include (i) photometric redshifts available from \ac{SDSS}, and (ii) groups where a single All\ac{WISE} source passed through the infrared filter but did not coincide with a collated \ac{FIRST} radio component. For this study we wish to primarily publish our collation method. Instead we will leave a more thorough extraction and analysis of a \ac{GRG} sample as future work. 

\section{Discussion and Future Outlook}
\label{sec:future}
\label{sec:discussion}

Our primary focus throughout this study was to present a new approach to the source collation problem with an emphasis towards exploiting unsupervised \ac{ML} algorithms. Our preliminary results based on our proof-of-concept design have produced a set of catalogues with value added data products which we have used to identify (i) rare and interesting sources, (ii) complex \ac{AGN} (including the introduction of a statistic to search for \ac{AGN} with curved or disturbed morphologies) and (iii) identification of 17 \acp{GRG}. Below we discuss current limitations of our preliminary framework and future directions to improve its known shortcomings. 

\subsection{Constructing the `Sky-Model'}

Any image or catalogue is a model of the sky-brightness observed at a particular wavelength. For simplicity when producing a catalogue describing unique groups, we treat the cookie-cutter segmentation and the collation of their results as two distinctly separate process. However, revising our implementation to tightly integrate these stages together could help produce a more self-consistent sky-model. Consider the pair of \ac{FIRST} components that were incorrectly associated with the larger \ac{AGN} in Figure~\ref{fig:example_srcs1}a. This scenario ultimately failed as our the field of view of our neurons were too small. Presently we only consider the \ac{BMU} of each mapped image. The notion of a \ac{BMU} in this context can be expanded to also consider the validity of information obtained from the cookie-cutter and its consistency with the current sky-model. Sets of individual mappings could be treated as an ensemble voting upon associating sets of components together. This would likely reveal inconsistencies for situations like Figure~\ref{fig:example_srcs1}a, as the many genuine components of the \ac{AGN} would consistently discriminate against the two outlying \ac{FIRST} components that were incorrectly associated. By detecting this disagreement the problematic voting members could have their mapped \ac{BMU} replaced with another neuron with a comparable (but larger) Euclidean distance.

Alternatively, all the neurons could be used to form a crude ensemble type method, similar to a random forest classifier. Mapping a single image to each prototype and the corresponding filter would produce a set of $M_c$. These could collectively be weighted by the Euclidean distance that accompanies each mapping. As each neuron has the same field of view, the individual components that make up each $M_c$ would be the same. Therefore, the position of each component within all filters could be considered, and when weighted by the Euclidean distance of all mappings a regressed like probability score could be constructed. Placing a minimum threshold on this value would either include or exclude a component or source from membership of a group.

Our procedure attempts to expose the underlying distribution of galaxy morphologies in an accessible manner. Leveraging this allows for each step of our methodology to be understood and improved upon outside of a blackbox, which is in stark contrast to some supervised \ac{ML} approaches. This may be more desirable for users who can begin to introduce domain knowledge in a more tractable way. For instance, criteria of the expected \ac{WISE} colour-colour properties of \ac{AGN} cores could be supplied as a property to consider when applying the cookie-cutter for filters with \ac{AGN} morphologies. Such steps can be introduced in a modular fashion without the need to retrain previous steps. 

\subsection{Probability Treatment of Neuron Features}
\label{sec:probability}

\begin{figure}
	\includegraphics[ width=\linewidth]{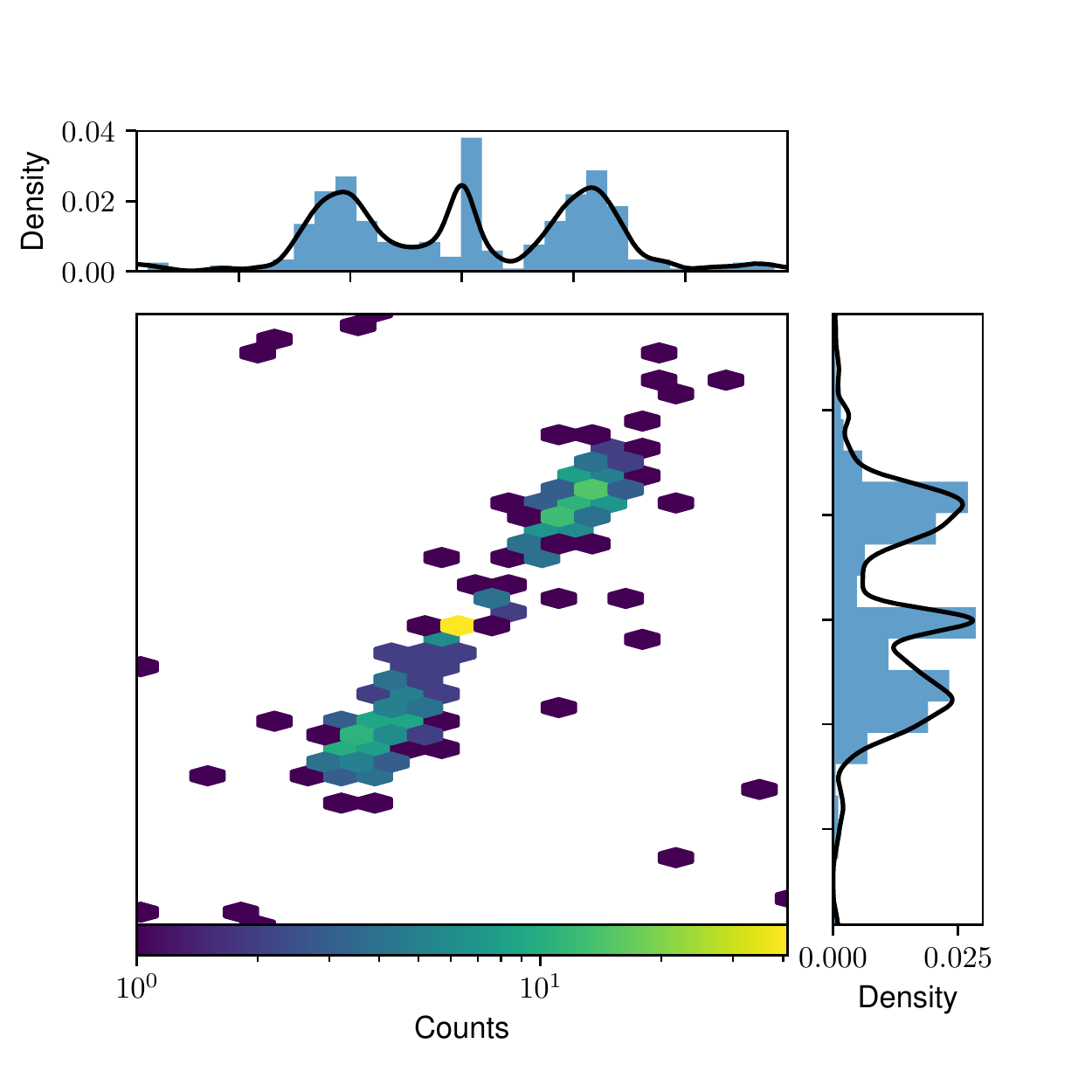}
	\caption{The two dimensional distribution of \ac{FIRST} sources accumulated as they passed through the radio filter of a single neuron. The radio neuron was selected to contain a remanent of the \ac{VLA} \ac{PSF}. Empty hexagonal bins have been masked out. The vertical and horizontal histograms show the normalised density distribution after marginalising across the corresponding axis. Overlaid onto each the result of a kernel density estimator. \label{fig:first_cluster}}
\end{figure}

Neurons and their prototype weights could be likened as something that approximates some underlying two-dimensional \ac{PDF} of predominant object morphologies.  Prototypes are constructed in a manner similar to a weighted averaging process, so the distribution of pixel intensities across the prototype weights could be thought of as the mean to a distribution of intensities observed across each pixel. 

To illustrate this we maintained a two-dimensional histogram count of \ac{FIRST} component locations that passed through the Figure~\ref{fig:ex_proto}c filter.  We show an example of these accumulated counts in Figure~\ref{fig:first_cluster}. Over a sufficient number of matches, structures that have been created by \ac{PINK} in the prototypes can be recovered from catalogue space data, where the relative count of individual bins corresponds approximately to the pixel intensities of the prototypes. Comparing the marginal histograms and their relative densities to the pixel intensities of Figure~\ref{fig:ex_proto}e shows that they are roughly equivalent. Therefore, rather than adopting a intensity cutoff to craft corresponding binary filters, instead a minimum likelihood could be used to test the hypothesis of components and/or infrared sources being related to one another. In situations where conflicting information are detected when constructing the sky-model, this likelihood proxy could also be considered to break degeneracies.

\subsection{GRG limiting size}
\label{sec:grg_size}

The prototype weights we constructed using \ac{PINK} are \ang{;3.5;} in angular scale, a factor $\sqrt{2}$ smaller than our training images. As a consequence the potential maximum angular scale of features within a neuron would be $\ang{;5;}$ orientated across the diagonal of its prototype weight. In practise though this is unlikely, as most radio features constructed by \ac{PINK} neither extend completely to the prototype weights boundary nor are perfectly aligned along its diagonal. Instead we estimate that the maximum angular scale of the largest radio feature across all neurons to be no more than \ang{;3;} in size.

Based on our maximum angular scale sensitivity of \ang{;3;}, the maximum physical size recoverable would be about \SI{1.5}{\mega\parsec} at $z=1.61$. From Table~\ref{tab:grg} we find a single \ac{GRG} over \SI{1}{\mega\parsec} in size, which has one of the larger angular scales in our sample of 17. Similarly, an object with an angular scale of \ang{;3;} would need a $z>0.25$ for it to be considered a \ac{GRG}. As \ac{SDSS} is a shallow survey focusing on the local Universe up to $z\sim0.1$ \citep{2015ApJS..219...12A}, our \ac{GRG} sample is insensitive to local \ac{GRG} populations.

\citet{2017MNRAS.469.2886D} used \ac{NVSS} to identify 25 \acp{GRG}, none of which are in common with our sample of 17. This is almost exclusively because we are limited to detecting related structures that are $<\ang{;3;}$ with our current dataset. The \ac{NVSS} image resolution is about 8 times larger than \ac{FIRST}. Objects that are clearly resolved in the high resolution \ac{FIRST} survey maybe difficult to distinguish as resolved in the lower resolution \ac{NVSS} survey. Similarly, objects that are clearly resolved in \ac{NVSS} would probably have angular scales larger than what our current collection of neurons are capable of characterising. This is especially true if there are no radio components coinciding with the host galaxy, meaning that input images to our method would be centred on radio lobes. Indeed, examining the population of \acp{GRG} described by \citet{2017MNRAS.469.2886D} shows that their smallest angular size is \ang{;3.1;} and extends upwards to \ang{;16;}. Similarly, \citet{2016ApJS..224...18P} compile a catalogue of $1,614$ \acp{GRS} found in \ac{NVSS}. Only 26 of these \acp{GRS} have angular sizes $<\ang{;4;}$, the smallest of which is $\ang{;3;}$. These are all larger than the estimated largest angular scale in our current selection of neurons. 

In the future we could revise our \acp{SOM} to be trained against images with larger fields of view, allowing for larger radio structures to be constructed by \ac{PINK}. We could also explore introducing additional channels to our input images that are at lower resolution but cover larger fields of view.

\subsection{Distinguishing sidelobe artefacts in catalogue space}

Due to the short integration lengths adopted by the \ac{FIRST} survey, there are strong imaging artefacts surrounding bright sources. To a source finding set of codes these artefacts can incorrectly be interpreted as a legitimate radio component and included in a radio component catalogue. To combat this, the original \ac{FIRST} catalogue attempted to distinguish genuine sources from sidelobe artefacts around bright sources by assigning a sidelobe probability for each radio component. Described by \citet{2015ApJ...801...26H}, this statistic was derived by an ensemble of oblique decision tree classifiers trained on a curated set of catalogue information from sources with clear \ac{VLA} \ac{PSF} remnants. A user of the \ac{FIRST} catalogue may define a criterion based on this sidelobe probability to filter out components to a level appropriate for their science requirements. 

\begin{figure}
\includegraphics[width=\linewidth]{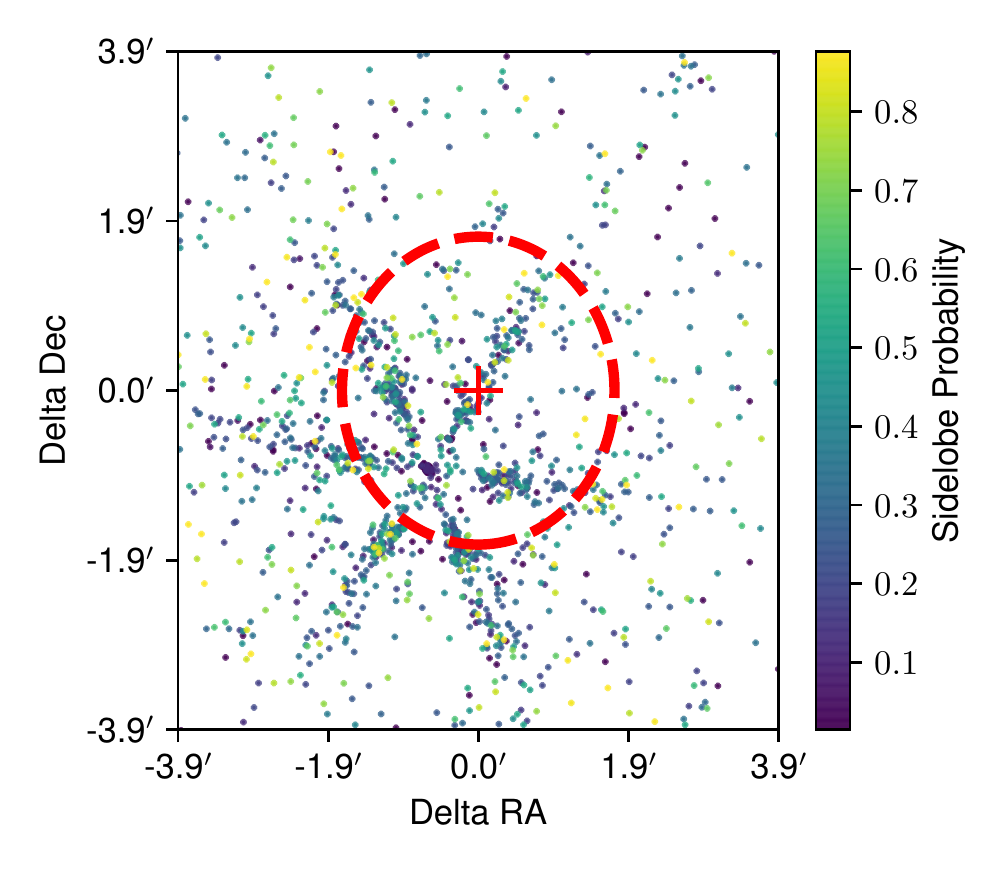}
\caption{The accumulated radio components from the \ac{FIRST} catalogue surrounding sources whose images  shared a single \ac{BMU} that exhibit \ac{PSF} features in the radio channel. These components have been transformed following rotation and flipping information provided as outputs from \ac{PINK}. The red circle denotes the \ang{;3.5;} angular size of the prototype weights with the red cross being position of the centre. The colour of each marker corresponds to the sidelobe probability of each component included as part of the \ac{FIRST} catalogue calculated by the ensemble of decision trees.\label{fig:scatter_psf}}	
\end{figure}

We are capable of performing a similar procedure using exclusively the data products from our trained \ac{SOM} with no \textit{a priori} knowledge about the sidelobe characteristics. For example, we located one neuron that exhibited remnants of the \ac{VLA} \ac{PSF} around a bright source offset from the centre. From mapping all images in our training dataset we identified 126 with this neuron as their \ac{BMU}. We searched $\ang{;6;}$ around the 126 positions these images are centred towards and found a total of $2,025$ radio components in our \ac{FSC}. The 126 transformations derived by \ac{PINK} were applied to the corresponding  set of returned components. We show these transformed component offsets in Figure~\ref{fig:scatter_psf}, where the \ac{VLA} \ac{PSF} can be seen as an over-densities of radio components. For illustration we also show how these features can be extended beyond the $\ang{;3.5;}\times\ang{;3.5;}$ field of view of the prototype weights. 

\begin{figure}
	\includegraphics[width=\linewidth]{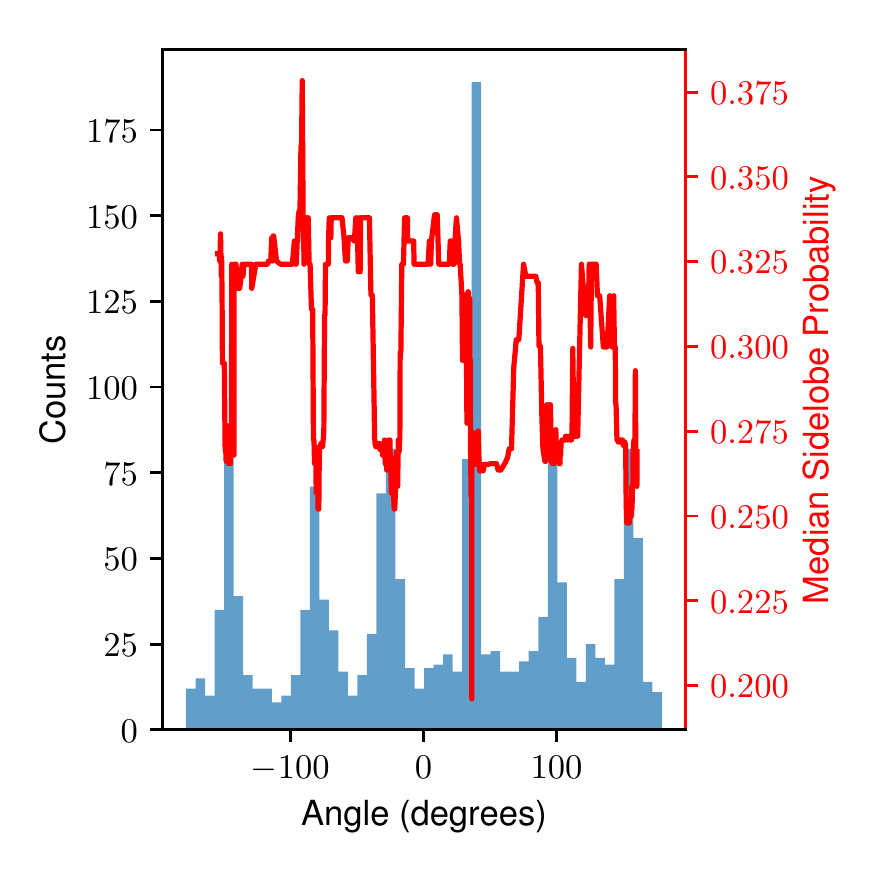}
	\caption{The distribution of $1,008$ angles from components after being transformed to a polar coordinate system. Each peak corresponds to a spoke of the \ac{VLA} \ac{PSF}. The largest count corresponds to a position towards the 126 position centred on a sidelobe artefact. Overlaid as a red line is the running median across 50 components (after sorting by their angle) of the sidelobe probability from the \ac{FIRST} catalogue. For the running median we blank out edge effects from the trend. \label{fig:seg_sideprob}}
\end{figure}

Arguably the sidelobe probabilities from \citet{2015ApJ...801...26H} may be lower than what could qualitatively be expected judging by the density of components from Figure~\ref{fig:scatter_psf}. To illustrate we first shifted the origin of Figure~\ref{fig:scatter_psf} to the approximate centre of the \ac{VLA} \ac{PSF} feature and converted the delta-offset Cartesian positions to a polar coordinate system. We mask both the inner most $\ang{;0.2;}$ region to remove legitimate, bright \ac{FIRST} components and the region beyond $\ang{;2;}$ to remain near the bounds of the field of view of the neuron. After applying this mask there were $1,008$ valid \ac{FIRST} components. Examining the collection of angles from the polar projection in Figure~\ref{fig:seg_sideprob} shows a highly non-uniform distribution with peaks spaced by \SI{\pi/3}{\radian}, each corresponding to a spoke along the \ac{VLA} \ac{PSF}. We then computed the running median value of the corresponding sidelobe probabilities for 50 components at a time (after sorting them by their angle). Comparing the distribution of angles to the computed running median shows that, surprisingly, the components have lower sidelobe probability estimates when there is an excess of components. Likely, this is a manifestation of the poor invariance to affine transforms of the ensemble of decision tree classifiers. These results suggests that \ac{PINK} would be an excellent tool towards identifying and characterising artefact components.

For future catalogue releases (either of the \ac{FIRST} catalogue we present in this work or other surveys) this can be combined with a probability treatment of the neurons (\S\ref{sec:probability}) to obtain a likelihood measure of a radio component being a genuine source component or simply an imaging artefact.  

\subsection{Extending the Annotation Labelling Scheme}

We explicitly elected for a minimal annotation scheme to demonstrate the immediate potential of our overall collation framework. For this reason we restricted our annotations to describe positions of features in Cartesian space. No additional properties were recorded (e.g. core label, presence of sidelobes, optimal threshold level for individual neurons, Fanaroff-Riley type). Subsequent processing steps use these positions in combination with simple, generalised stages to extract knowledge about objects and their, potentially resolved, morphologies. In many ways this transfer of knowledge from the lower dimensional embedding (set of \ac{SOM} neurons) could further be refined by also recording these higher order products during the annotation stage. 

As an example, a user defined threshold value would be of particular importance. We used a simple method that would attempt to identify islands of contiguous pixels that encompassed the user specified pixel positions (Figure~\ref{fig:ex_proto}). This method sometimes failed for neurons with weak or absent features, which was particularly common for the infrared channel of neurons which learnt to identify radio sidelobe artefacts present around bright \ac{FIRST} source components. As a consequence there are cases of artificial groups with ill-defined infrared host positions. A more sophisticated annotation scheme would be able to largely avoid this by carrying forward strict labels or flagging operations with a more reasonable thresholding level defined explicitly during the annotation stage. If necessary user-defined regions could also be interactively defined and used in place of a simple pixel-threshold value. 

Developing this idea further, regions within the prototype weights could be defined to carry further information or actions. For instance, the filters we characterise in Figure~\ref{fig:ex_proto} will only be used to group together components and sources. Expanding these to include regions that will ensure components remain disassociated under all conditions may help to ensure our greedy collation process considers all available information when crafting component groups. 

\section{Conclusions}

We have demonstrated a new approach towards identifying related radio components and isolating their corresponding infrared host galaxies using \ac{PINK}, an unsupervised machine learning algorithm. Unlike supervised methods, which can be likened to attempting to model the operation of a generalised function, an unsupervised machine learning algorithm attempts to model the \textit{structure} of an assumed dataset. \ac{PINK} implements a similarity measure that is rotationally and flipping invariant \revision{by searching for a transform that best aligns an image with structures that are either predefined or learnt through an iterative training process}. In this study we have used \ac{PINK} to

\begin{enumerate}
	\item applied an unsupervised machine learning method against a training set of radio and infrared images to produce a set of physically meaningful prototype morphologies without the need of training labels, and
	\item constructed a framework to meaningfully explore the previously unstructured complex image data, including the introduction of a statistic to identify bent radio morphologies.
\end{enumerate}

\revision{Using the \ac{PINK} products and outputs from our collation process we have }
\begin{enumerate}
	\item used data products from \ac{PINK} to collate together related \ac{FIRST} radio components and predict their corresponding infrared host, 
	\item demonstrated an ability to efficiently extract rare and unusual object morphologies, and  
	\item compiled a list of 17 \acp{GRG} that were identified after isolating their resolved lobes with our collation method.
\end{enumerate}

Throughout we maintained a simplified workflow to demonstrate a new methodology towards this difficult problem. In the future we will further develop our framework to better exploit the data products produced by \ac{PINK} with with emphasis on \ac{SKA} surveys. 

\section*{Acknowledgements}

We thank our reviewer, Mike Walmsley, for constructive feedback that improved the presentation and clarity of the manuscipt. 

TG would like to thanks Ivy Wong, Chen Wu, Kieran Luken and Matthew Alger for enlightening discussions throughout this work. KP and EH gratefully acknowledge the support of the Klaus Tschira Foundation. 

This research made use of the cross-match service provided by CDS, Strasbourg. This research has made use of NASA's Astrophysics Data System Bibliographic Services.

This research made use of Scikit-learn \citep{scikit-learn}, SciPy \citep{jones_scipy_2001}, NumPy \citep{van2011numpy} and Scikit-image \citep{scikit-image}. This research made use of matplotlib, a Python library for publication quality graphics \citep{Hunter:2007}. This research made use of Astropy, a community-developed core Python package for Astronomy \citep{2013A&A...558A..33A}.

This research has made use of the VizieR catalogue access tool, CDS, Strasbourg, France. This research has made use of the NASA/IPAC Infrared Science Archive, which is operated by the Jet Propulsion Laboratory, California Institute of Technology, under contract with the National Aeronautics and Space Administration. This publication makes use of data products from the Wide-field Infrared Survey Explorer, which is a joint project of the University of California, Los Angeles, and the Jet Propulsion Laboratory/California Institute of Technology, and NEOWISE, which is a project of the Jet Propulsion Laboratory/California Institute of Technology. WISE and NEOWISE are funded by the National Aeronautics and Space Administration. This publication makes use of data products from the Wide-field Infrared Survey Explorer\citep{2010AJ....140.1868W}, which is a joint project of the University of California, Los Angeles, and the Jet Propulsion Laboratory/California Institute of Technology, funded by the National Aeronautics and Space Administration. 

\section*{Data Availability}

Image and catalogue data that were used throughout this study are publicly accessible from their respective survey data portals. Output source catalogues will be made publicly available alongside this article and through data access portals (e.g. \texttt{ViZieR}). The corresponding authors may also be contacted for further details if required.




\bibliographystyle{mnras}
\bibliography{cata_references.bib} 



\appendix

\section{Layer Two SOM}

In Figure~\ref{fig:big_som} we present the Layer Two \ac{SOM} constructed in \S\ref{sec:som_training}. We have divided it into 25 $8\times8$ regions to allow the \ac{PINK} prototype weights to be visualised. The coordinate labels are consistent among each of the 25 regions.  

\begin{figure*}
	\includegraphics[width=0.85\linewidth]{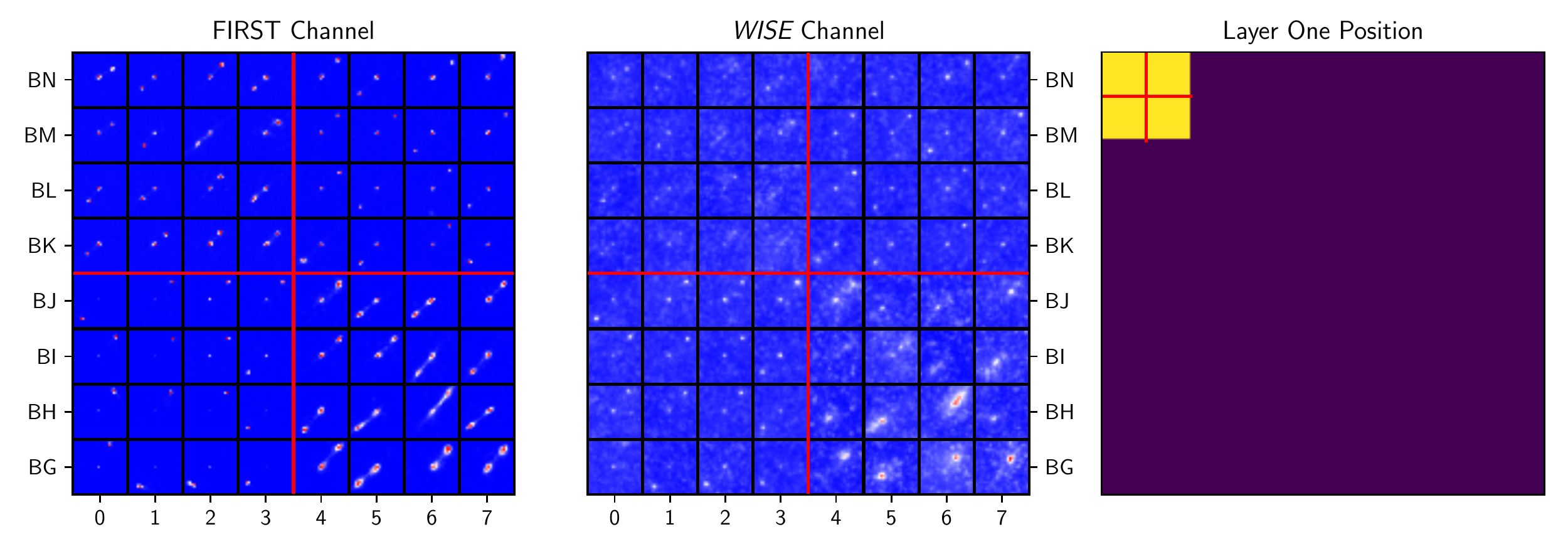}
	\includegraphics[width=0.85\linewidth]{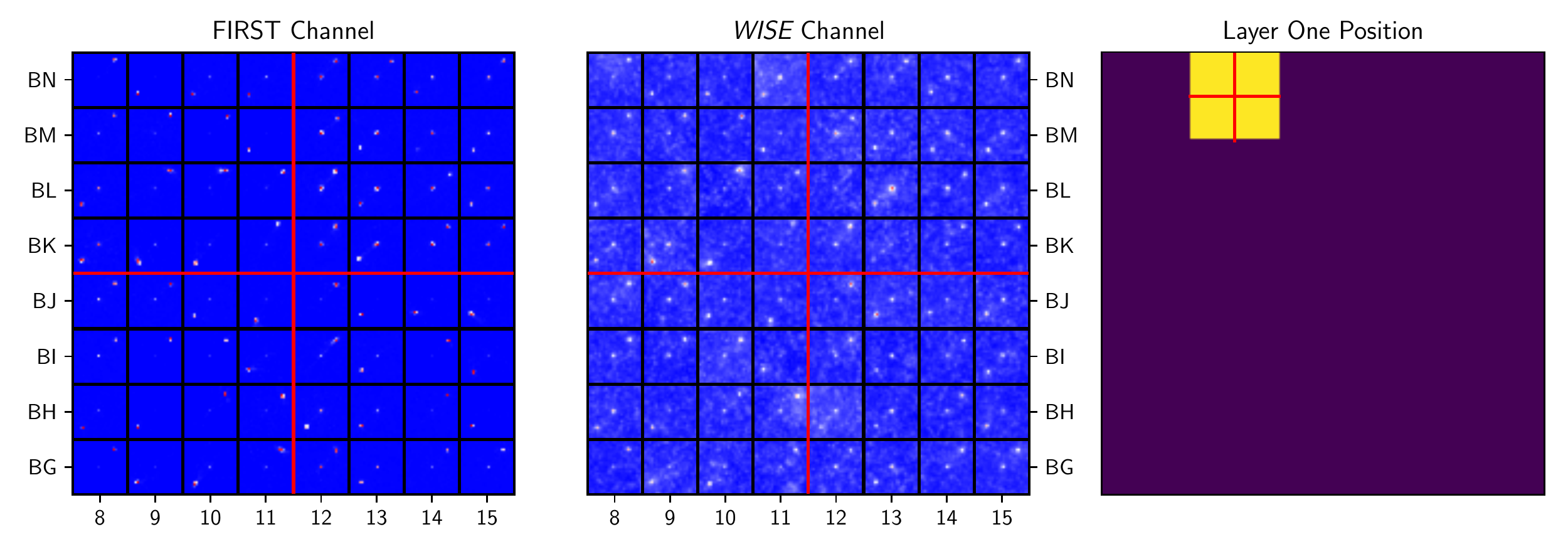}
	\includegraphics[width=0.85\linewidth]{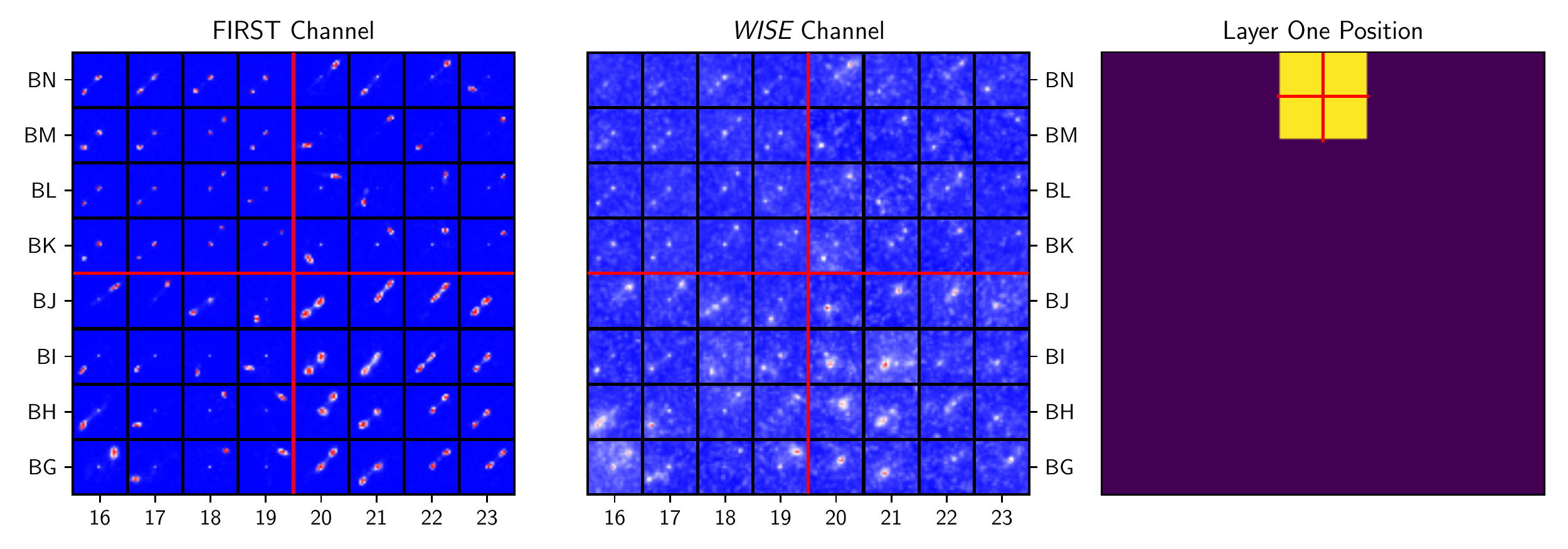}
	\includegraphics[width=0.85\linewidth]{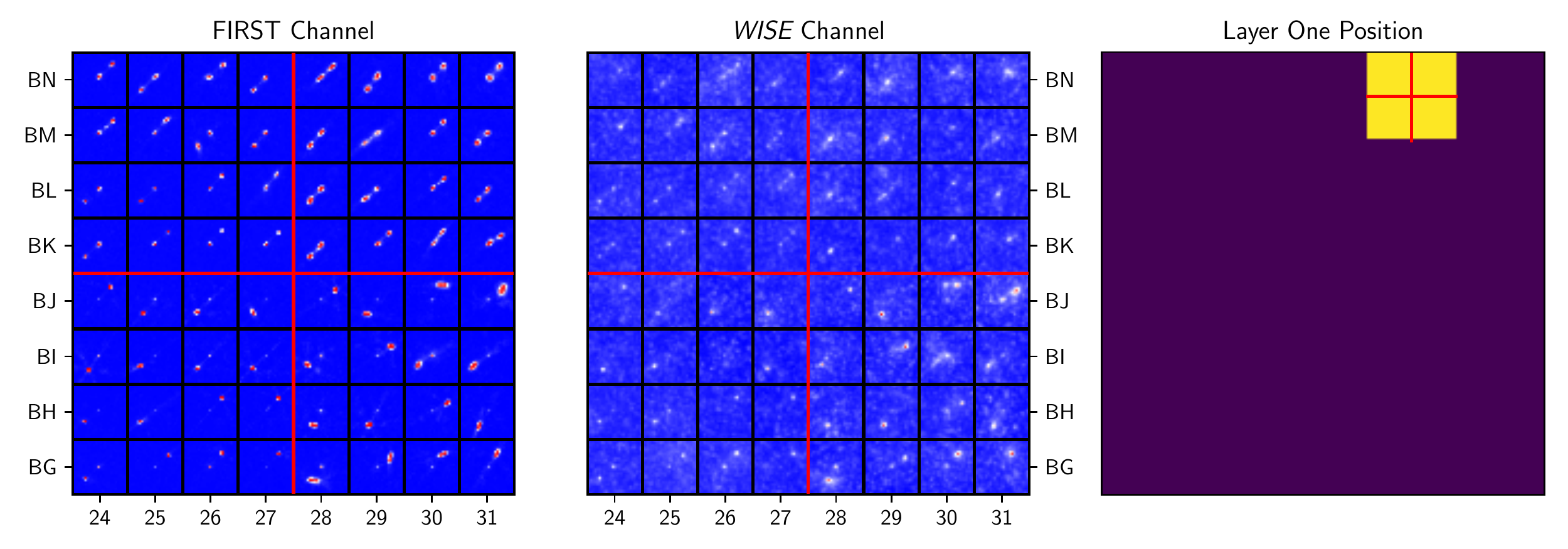}
	
	\caption{The \ac{FIRST} \textit{(left)} and \ac{WISE} \textit{(middle)} channels of the Layer Two \ac{SOM} constructed by concatenating 100 $4\times$ \acp{SOM}. The black vertical and horizontal lines represent the boundary between individual neurons on the lattice, while the red vertical and horizontal lines represent boundaries between individual $4\times4$ \acp{SOM}. Each of the \acp{SOM} were trained using the same data preprocessing and training stages as Layer One. The Layer One position \textit{(right)} shows the relative position of the four \acp{SOM} on the Layer One \ac{SOM} (Figure~\ref{fig:som}). }
\end{figure*}

\begin{figure*}
\ContinuedFloat
	\includegraphics[width=0.85\linewidth]{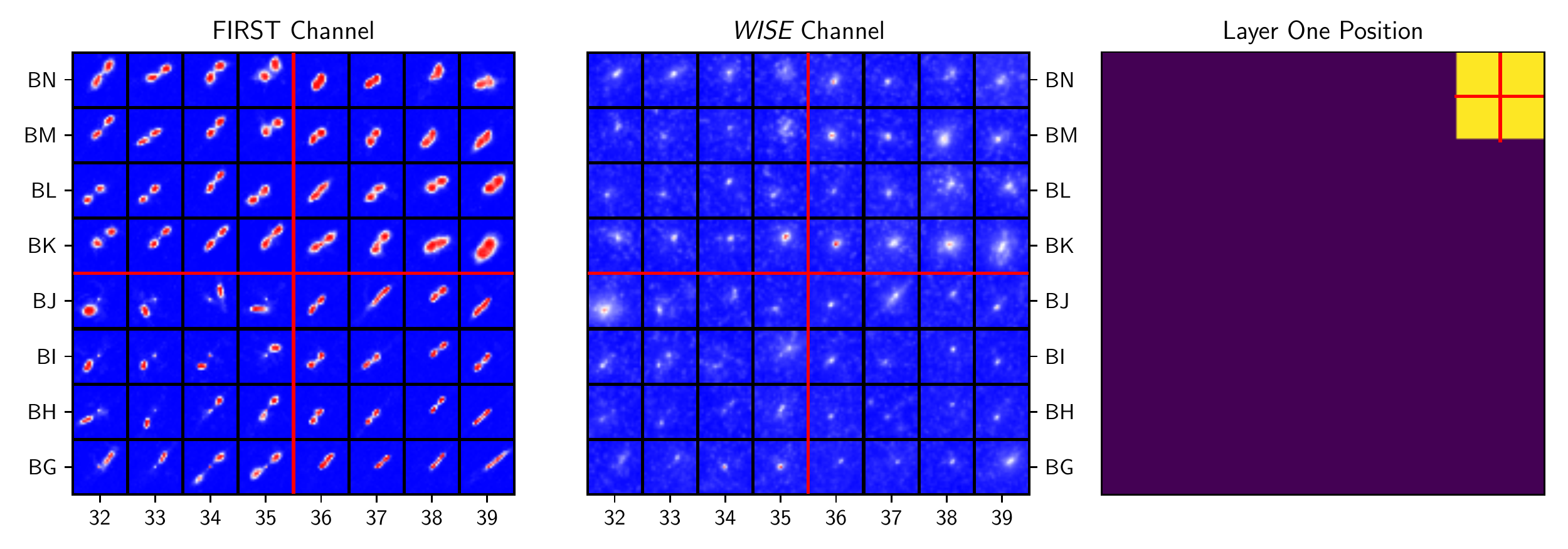}
	\includegraphics[width=0.85\linewidth]{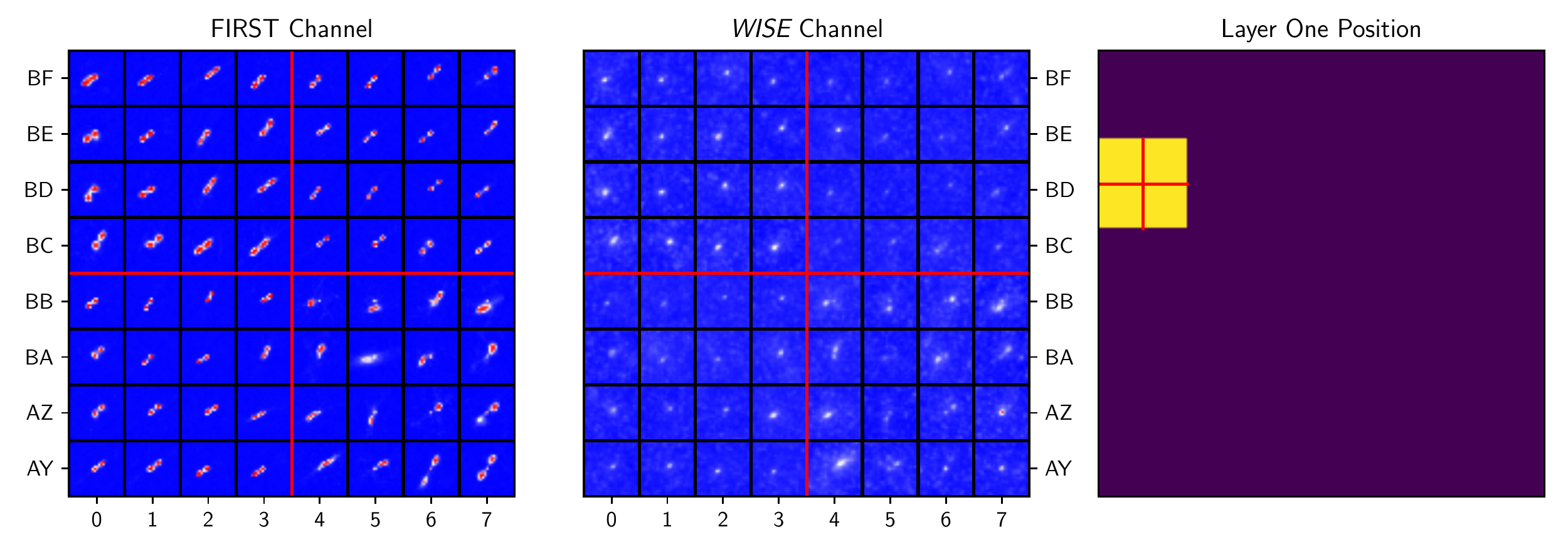}
	\includegraphics[width=0.85\linewidth]{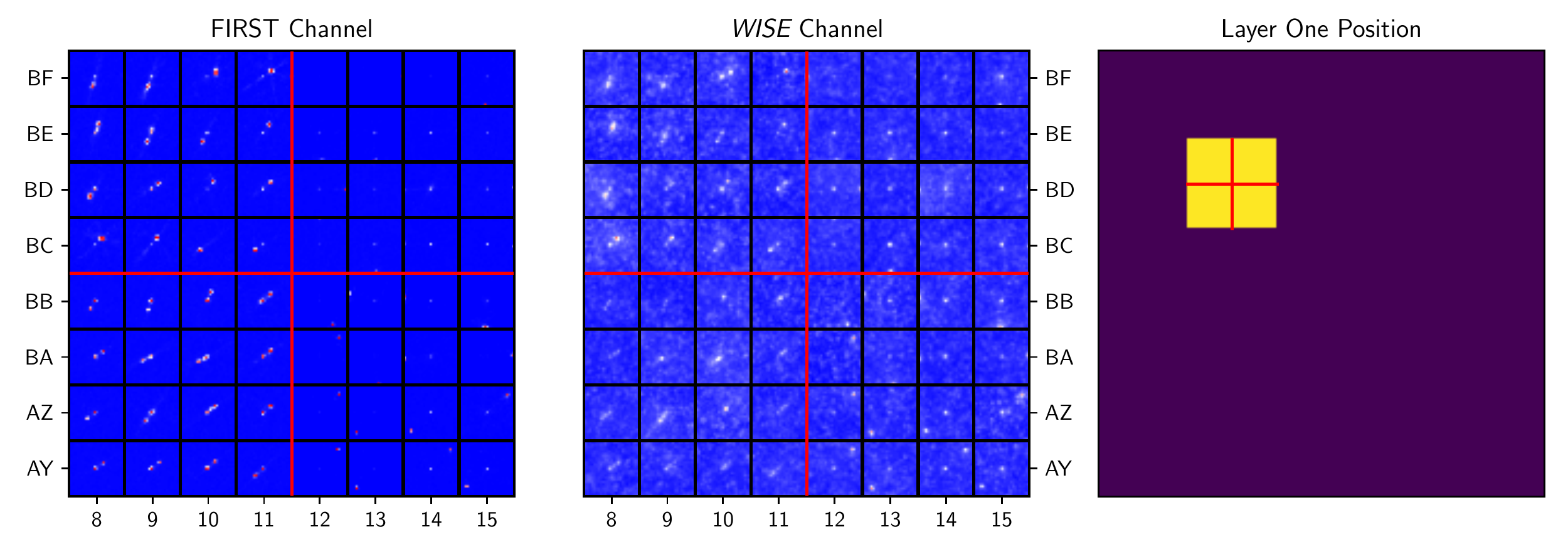}
	\includegraphics[width=0.85\linewidth]{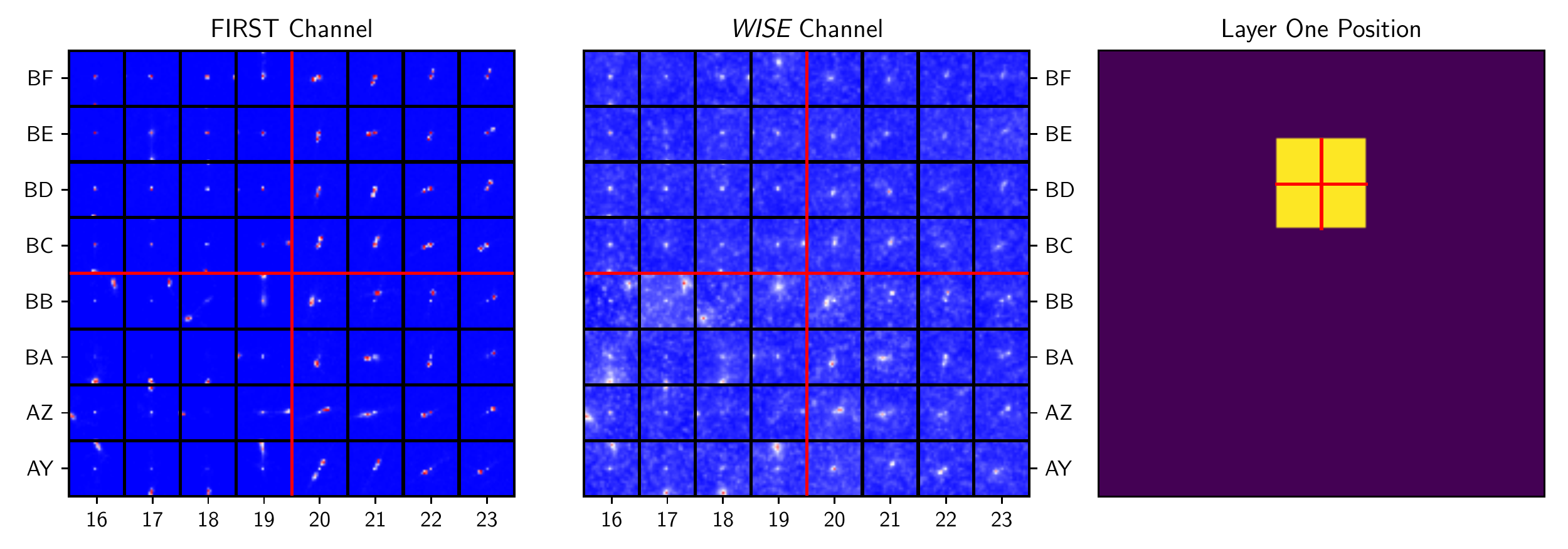}
	\caption{\textit{(Cont.)} The \ac{FIRST} \textit{(left)} and \ac{WISE} \textit{(middle)} channels of the Layer Two \ac{SOM} constructed by concatenating 100 $4\times$ \acp{SOM}. The black vertical and horizontal lines represent the boundary between individual neurons on the lattice, while the red vertical and horizontal lines represent boundaries between individual $4\times4$ \acp{SOM}. Each of the \acp{SOM} were trained using the same data preprocessing and training stages as Layer One. The Layer One position \textit{(right)} shows the relative position of the four \acp{SOM} on the Layer One \ac{SOM} (Figure~\ref{fig:som}). }
\end{figure*}

\begin{figure*}
\ContinuedFloat
	\includegraphics[width=0.85\linewidth]{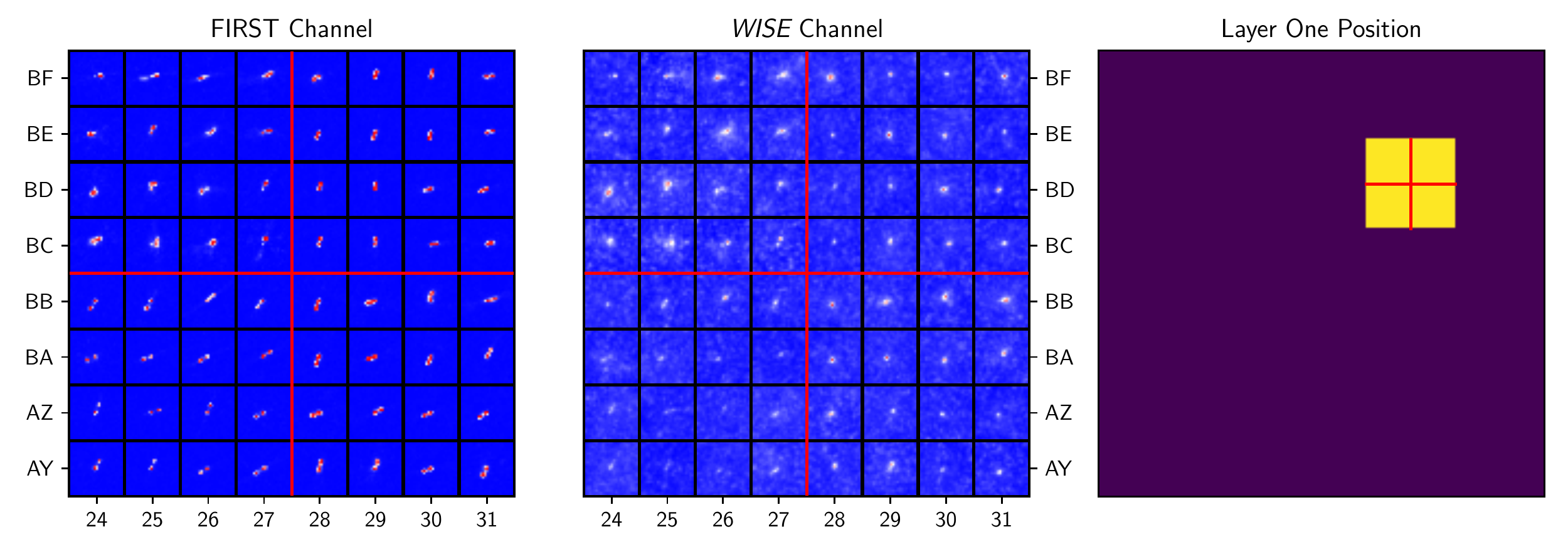}
	\includegraphics[width=0.85\linewidth]{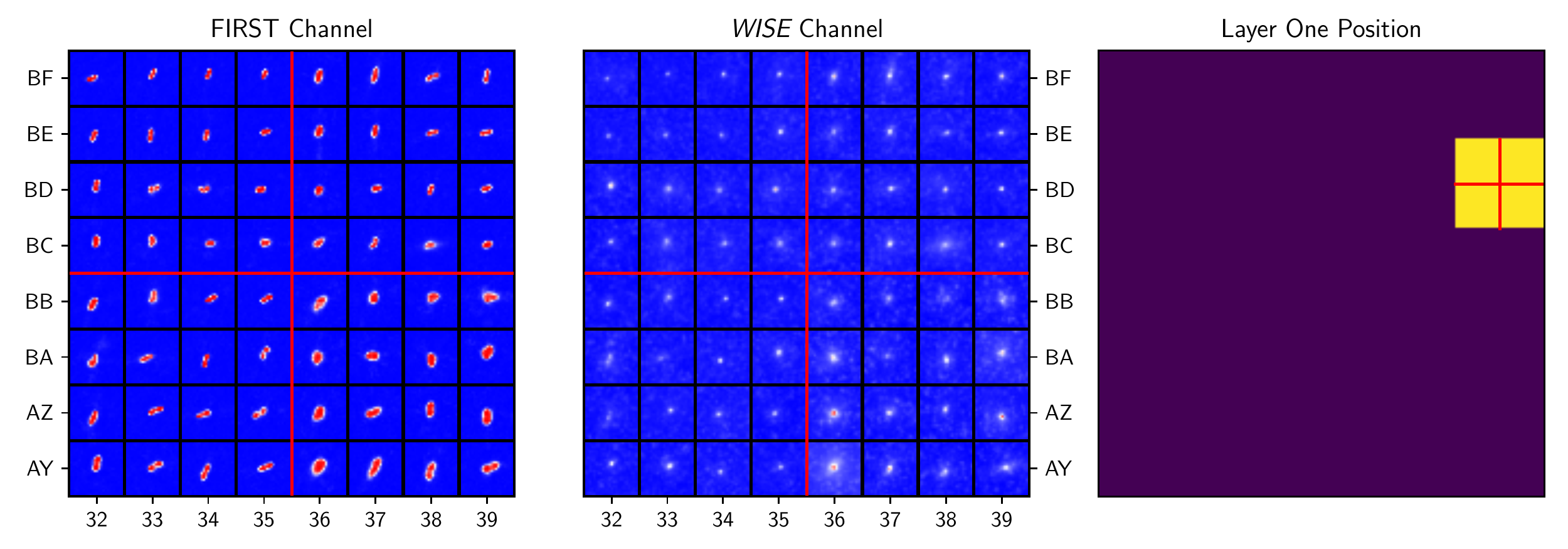}
	\includegraphics[width=0.85\linewidth]{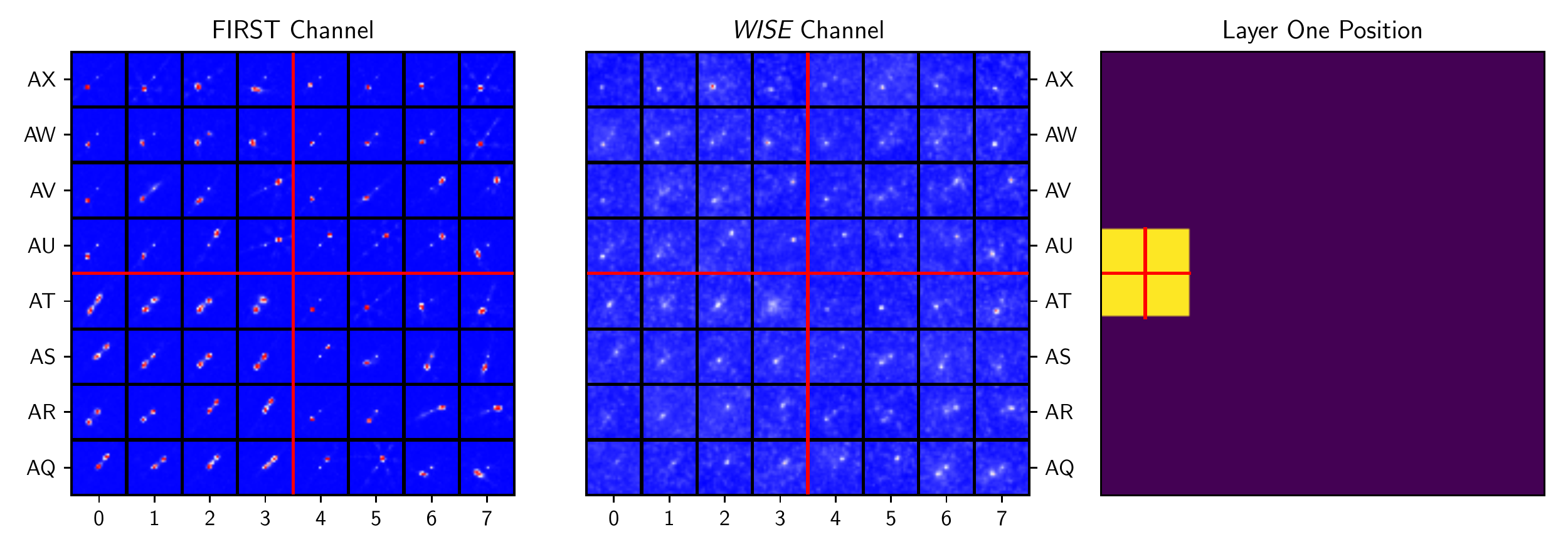}
	\includegraphics[width=0.85\linewidth]{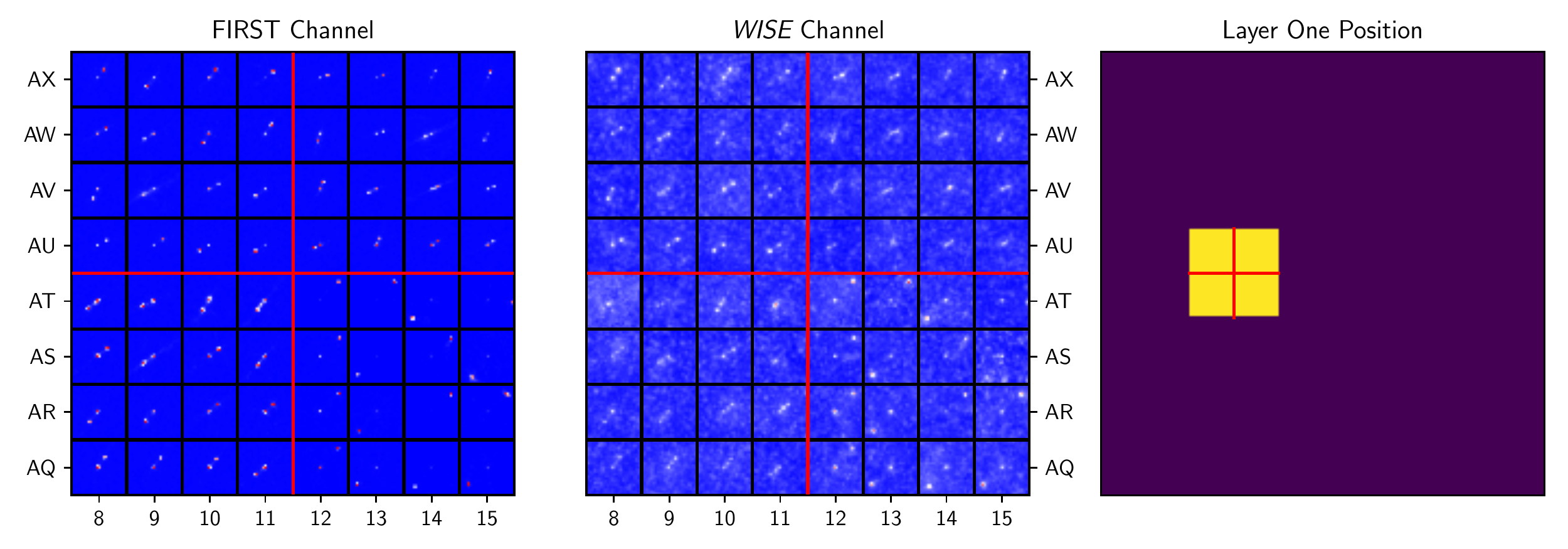}
	\caption{\textit{(Cont.)} The \ac{FIRST} \textit{(left)} and \ac{WISE} \textit{(middle)} channels of the Layer Two \ac{SOM} constructed by concatenating 100 $4\times$ \acp{SOM}. The black vertical and horizontal lines represent the boundary between individual neurons on the lattice, while the red vertical and horizontal lines represent boundaries between individual $4\times4$ \acp{SOM}. Each of the \acp{SOM} were trained using the same data preprocessing and training stages as Layer One. The Layer One position \textit{(right)} shows the relative position of the four \acp{SOM} on the Layer One \ac{SOM} (Figure~\ref{fig:som}). 
	}
\end{figure*}

\begin{figure*}
\ContinuedFloat
	\includegraphics[width=0.85\linewidth]{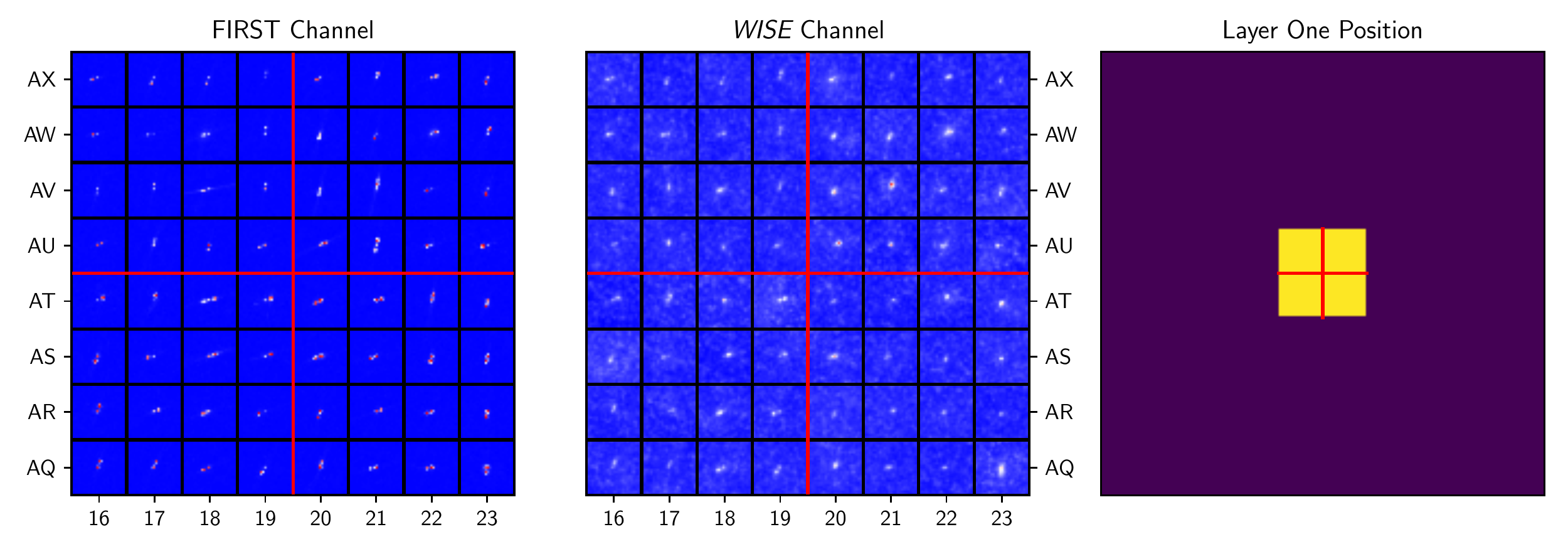}
	\includegraphics[width=0.85\linewidth]{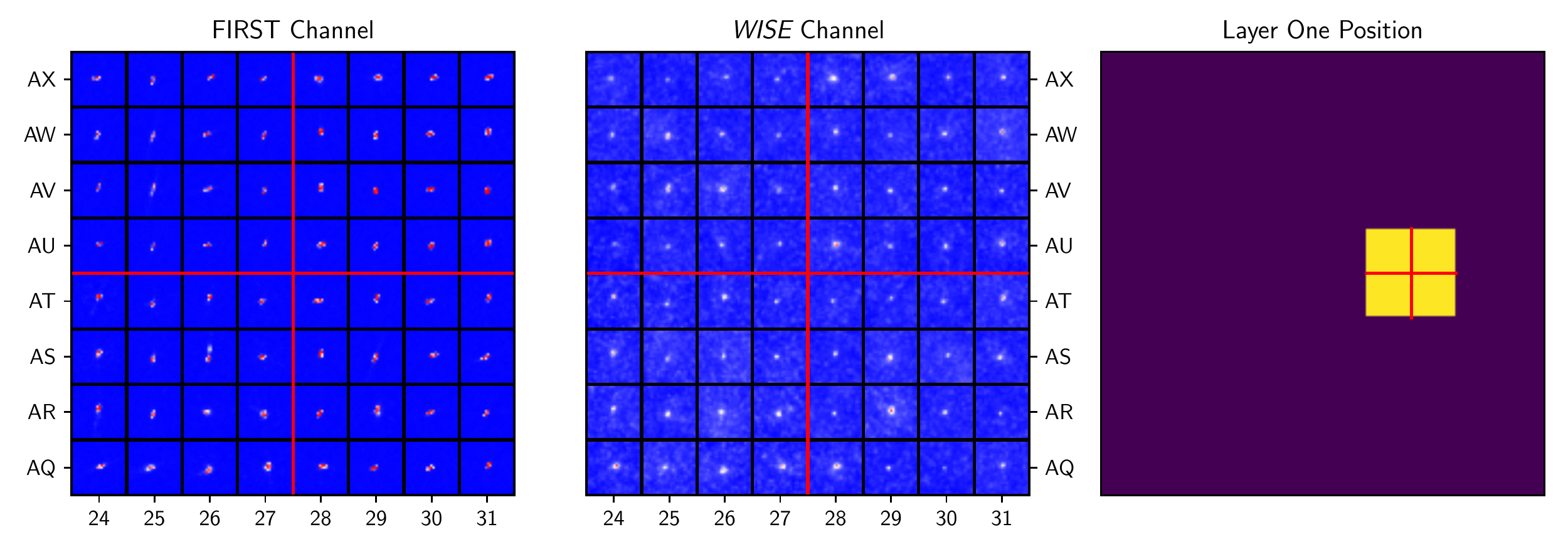}
	\includegraphics[width=0.85\linewidth]{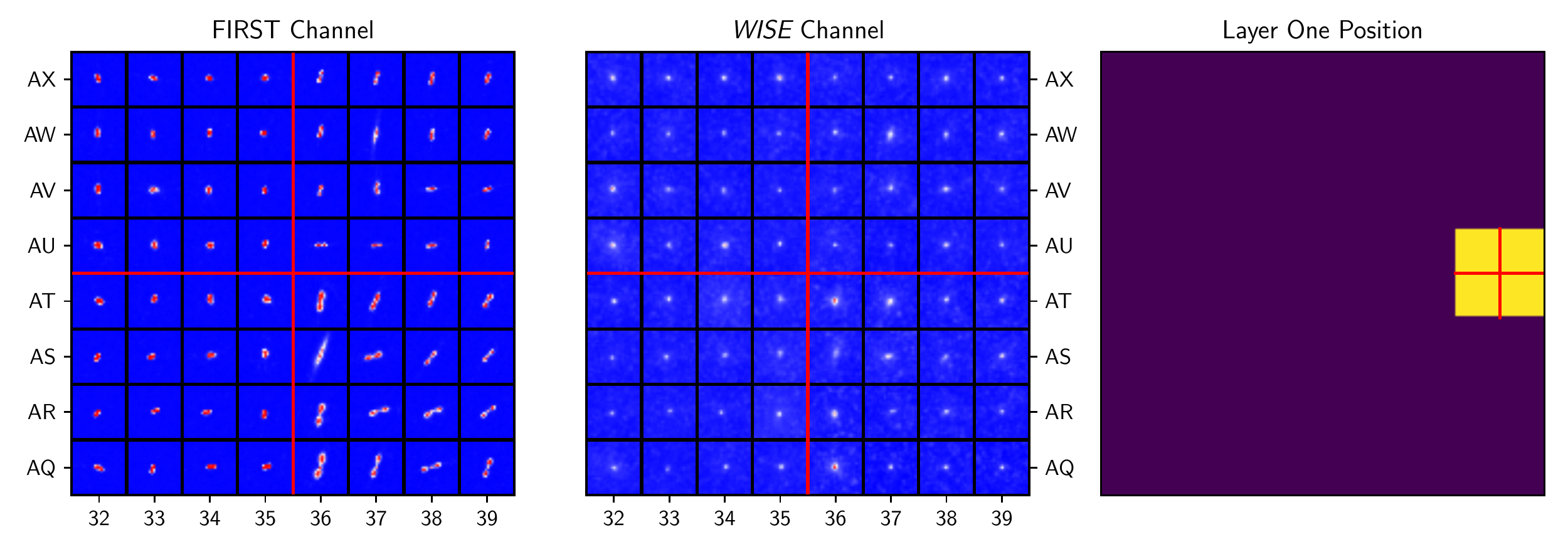}
	\includegraphics[width=0.85\linewidth]{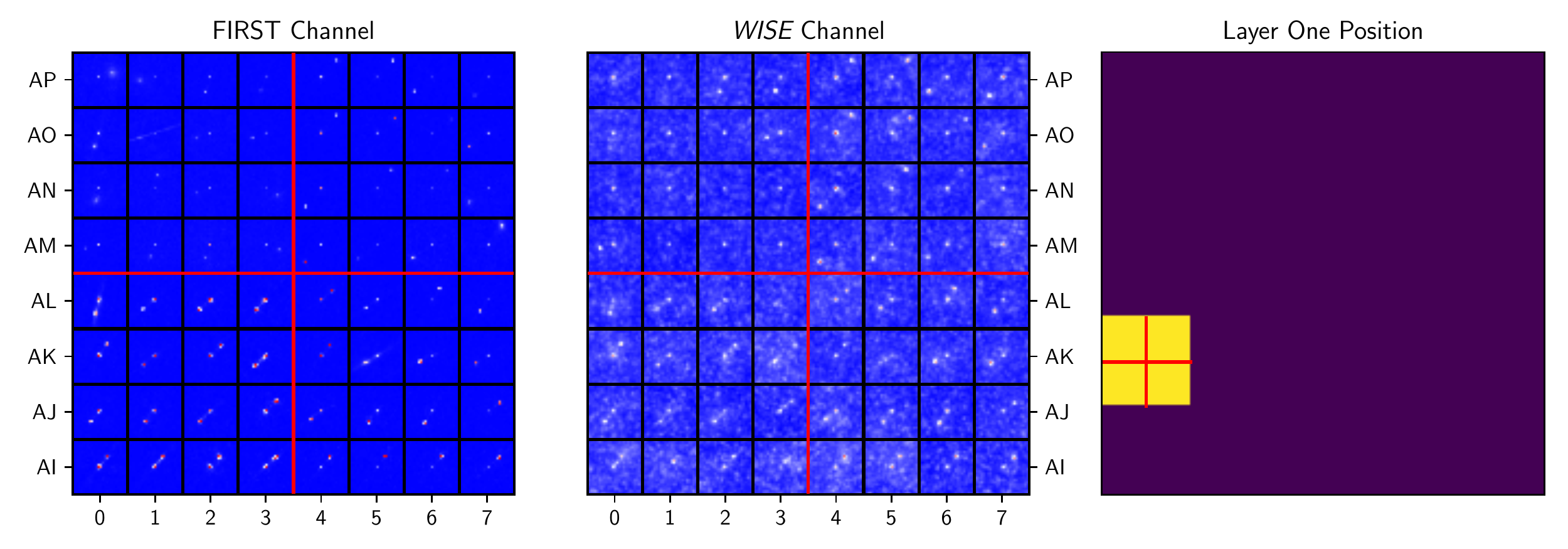}
	\caption{\textit{(Cont.)} The \ac{FIRST} \textit{(left)} and \ac{WISE} \textit{(middle)} channels of the Layer Two \ac{SOM} constructed by concatenating 100 $4\times$ \acp{SOM}. The black vertical and horizontal lines represent the boundary between individual neurons on the lattice, while the red vertical and horizontal lines represent boundaries between individual $4\times4$ \acp{SOM}. Each of the \acp{SOM} were trained using the same data preprocessing and training stages as Layer One. The Layer One position \textit{(right)} shows the relative position of the four \acp{SOM} on the Layer One \ac{SOM} (Figure~\ref{fig:som}). \label{fig:big_som}}
\end{figure*}

\begin{figure*}
\ContinuedFloat
	\includegraphics[width=0.85\linewidth]{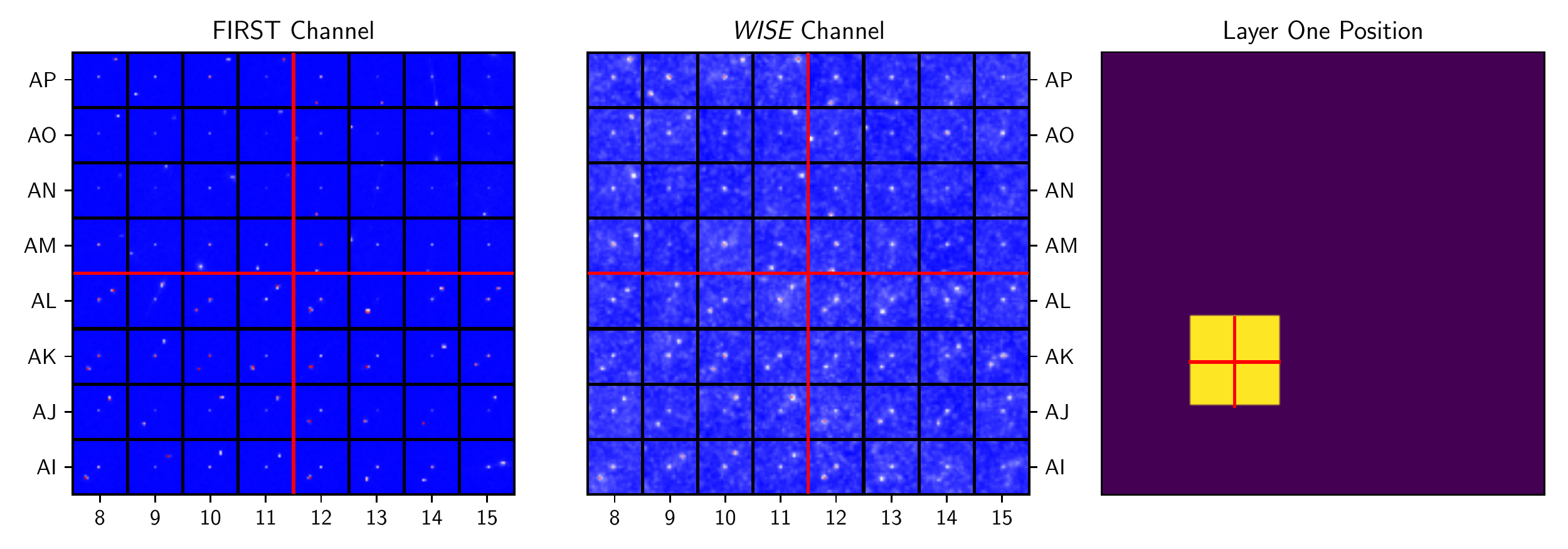}
	\includegraphics[width=0.85\linewidth]{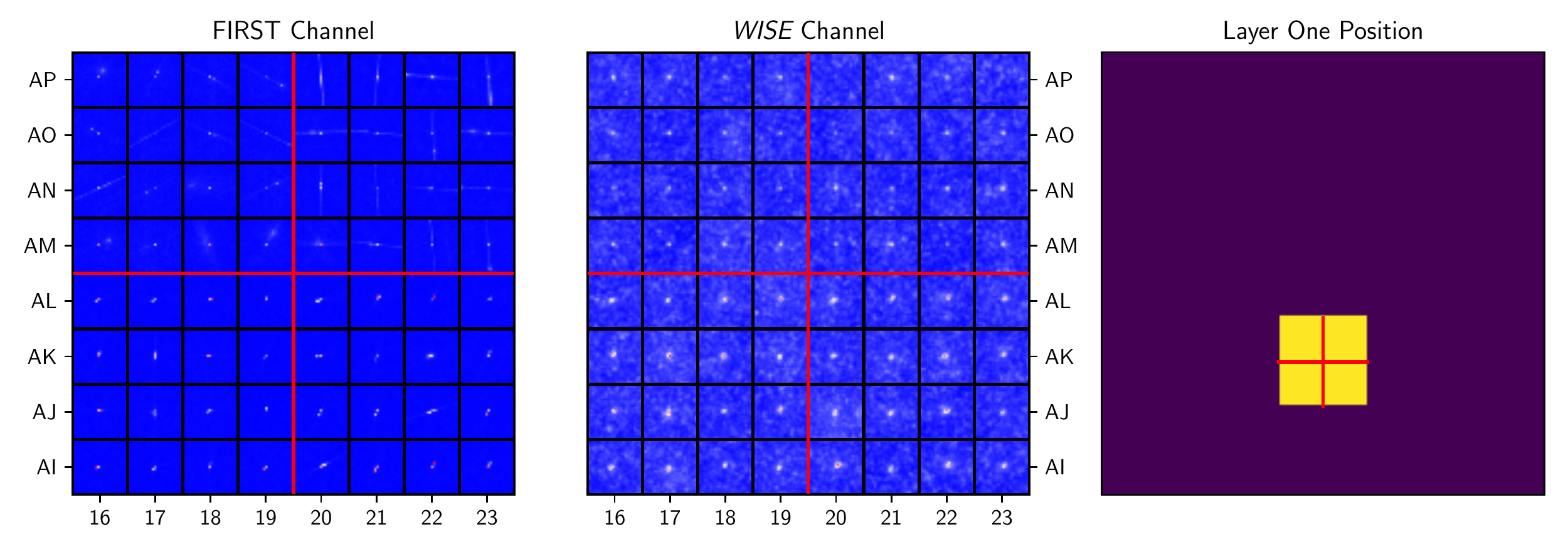}
	\includegraphics[width=0.85\linewidth]{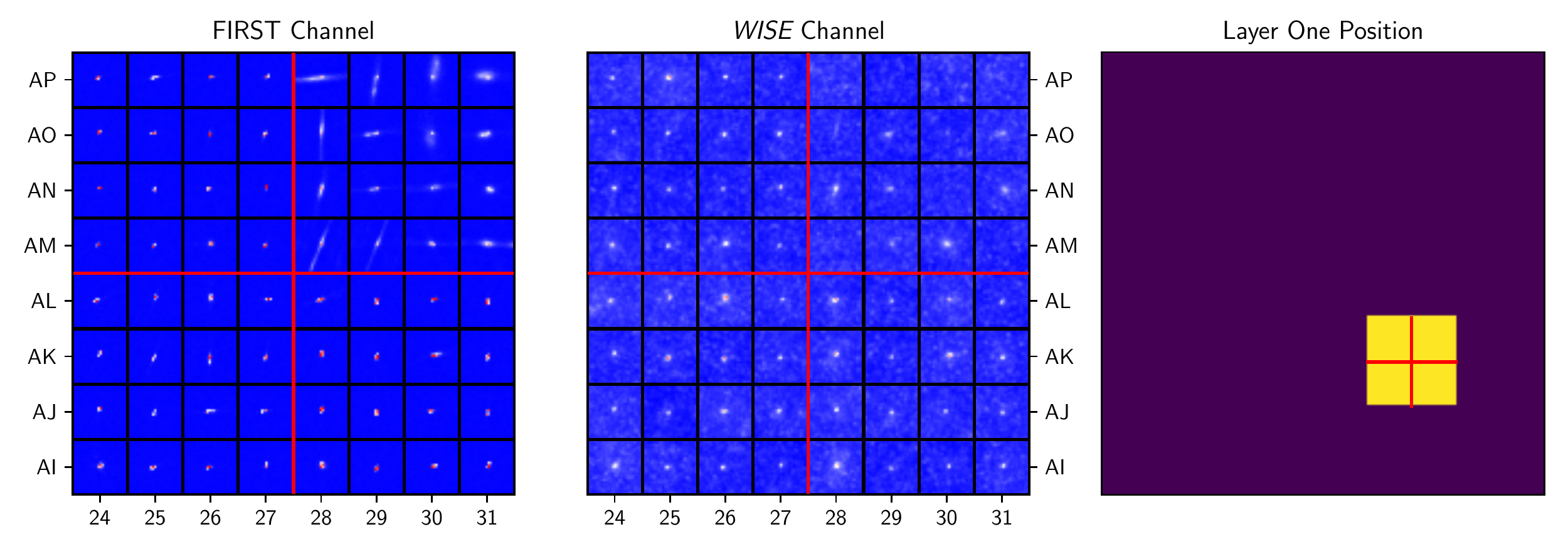}
	\includegraphics[width=0.85\linewidth]{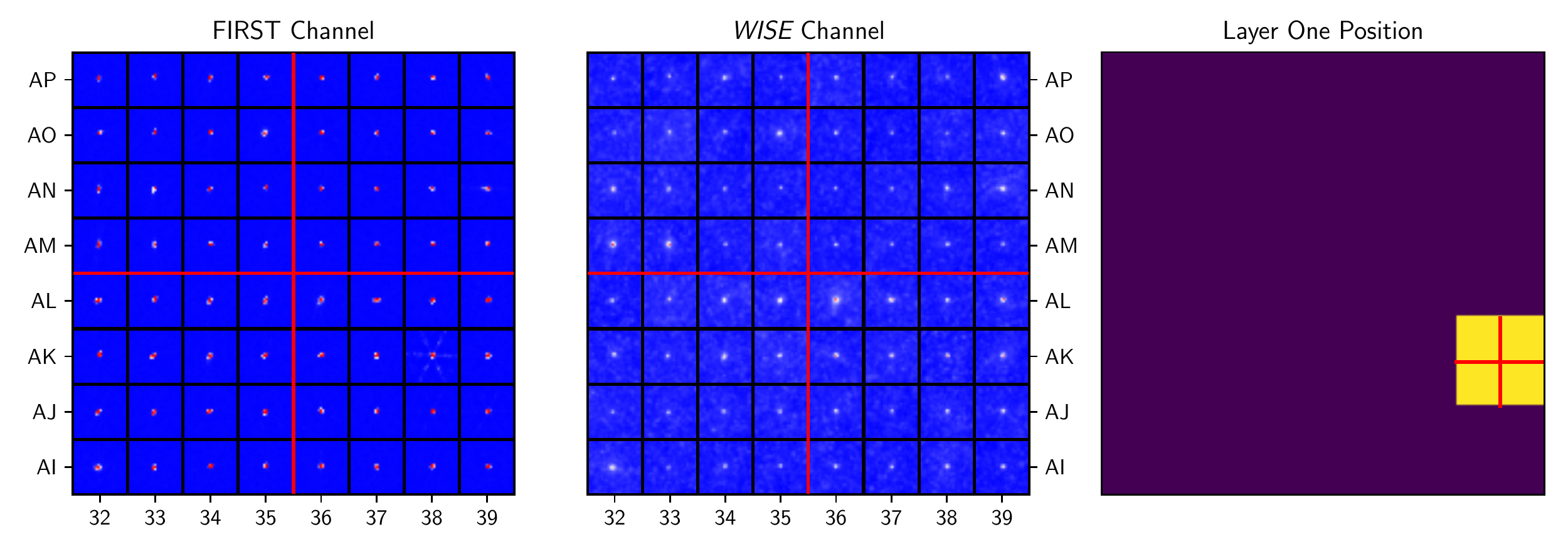}
	\caption{\textit{(Cont.)} The \ac{FIRST} \textit{(left)} and \ac{WISE} \textit{(middle)} channels of the Layer Two \ac{SOM} constructed by concatenating 100 $4\times$ \acp{SOM}. The black vertical and horizontal lines represent the boundary between individual neurons on the lattice, while the red vertical and horizontal lines represent boundaries between individual $4\times4$ \acp{SOM}. Each of the \acp{SOM} were trained using the same data preprocessing and training stages as Layer One. The Layer One position \textit{(right)} shows the relative position of the four \acp{SOM} on the Layer One \ac{SOM} (Figure~\ref{fig:som}). }
\end{figure*}

\begin{figure*}
\ContinuedFloat
	\includegraphics[width=0.85\linewidth]{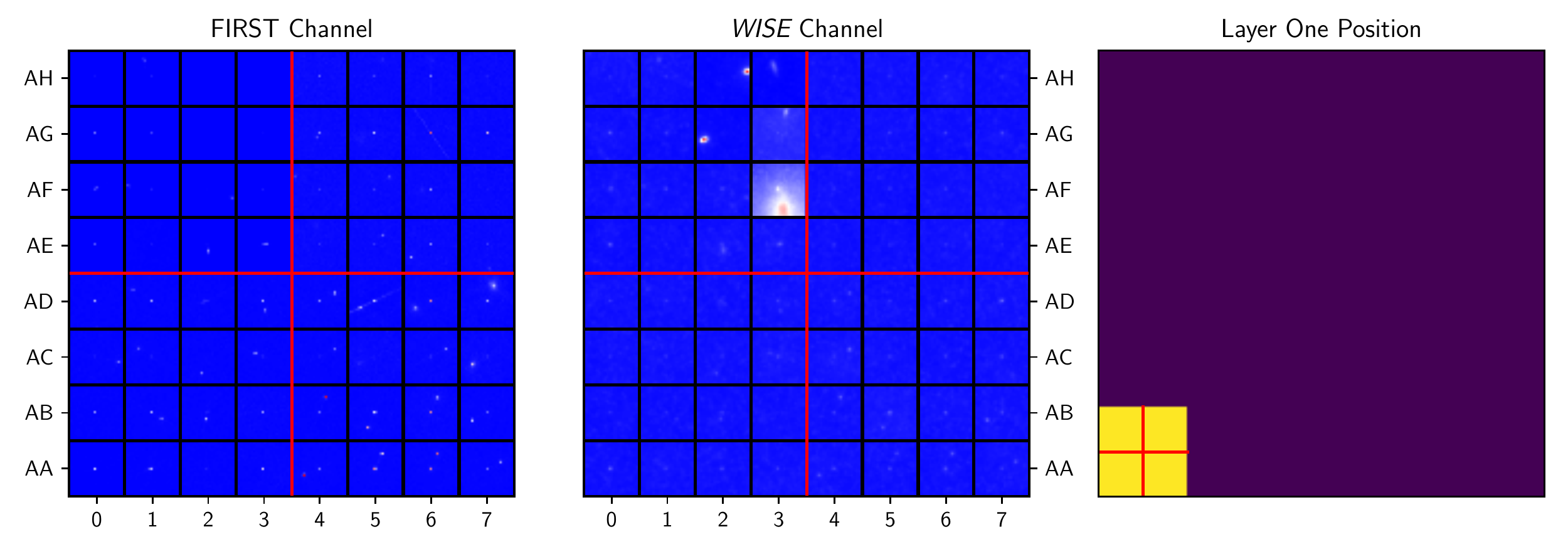}
	\includegraphics[width=0.85\linewidth]{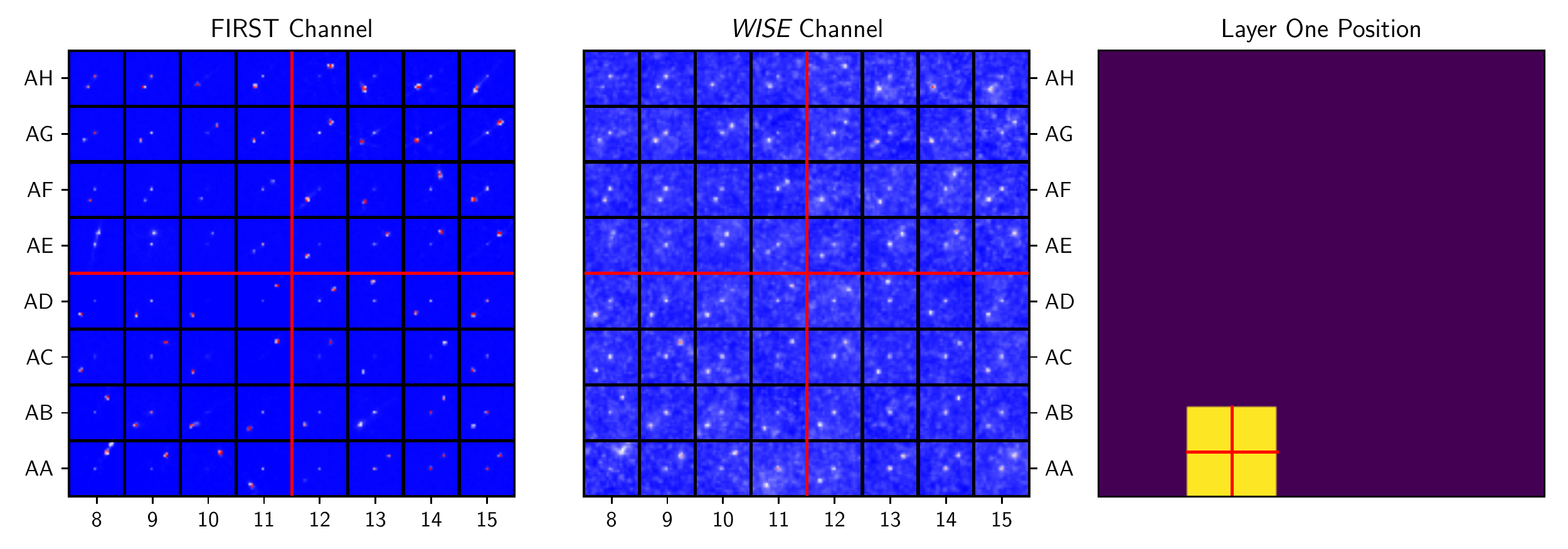}
	\includegraphics[width=0.85\linewidth]{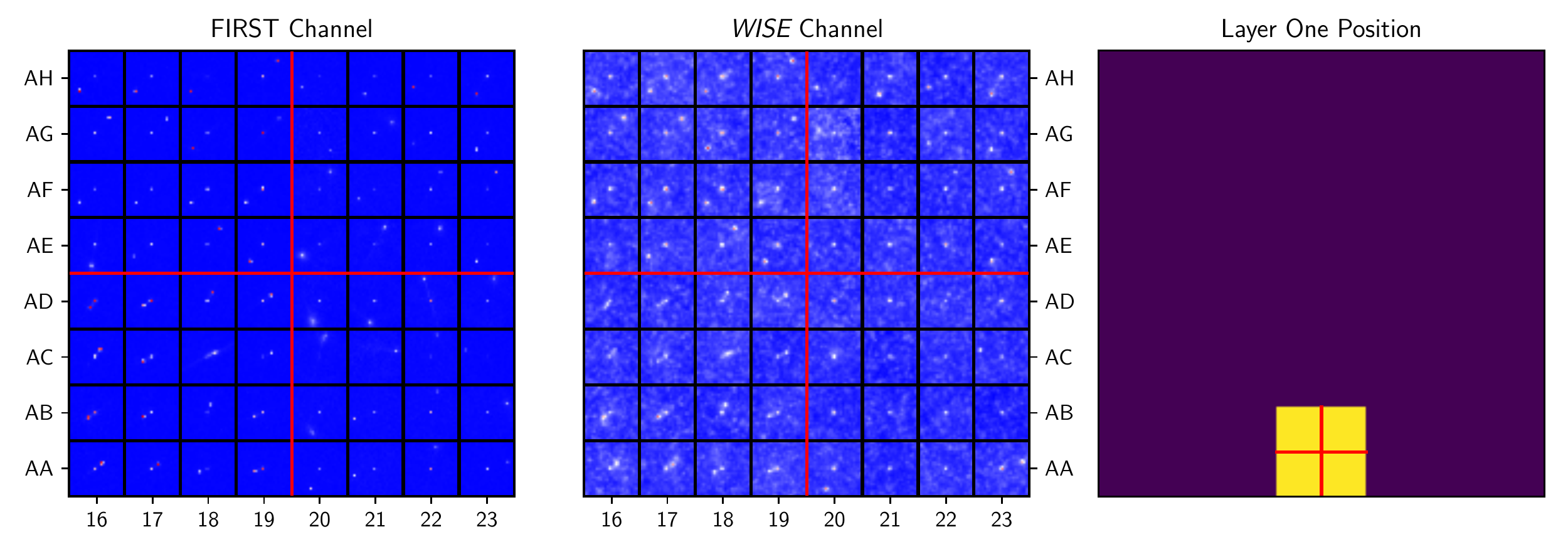}
	\includegraphics[width=0.85\linewidth]{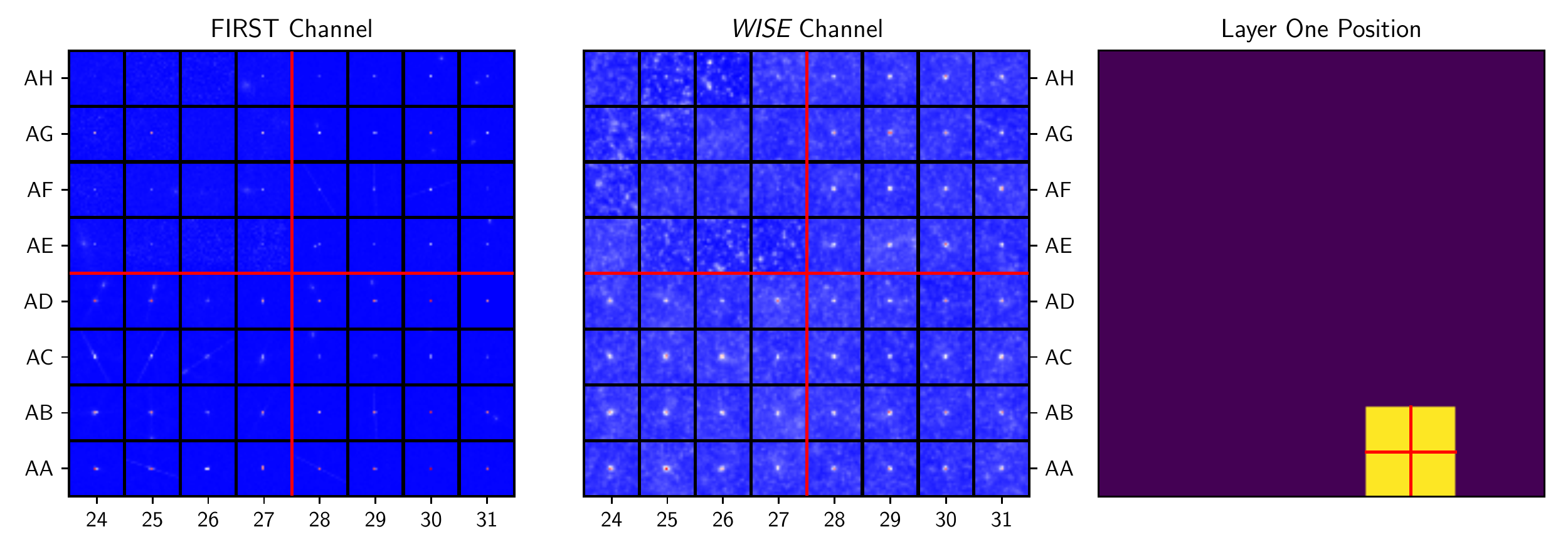}
	\caption{\textit{(Cont.)} The \ac{FIRST} \textit{(left)} and \ac{WISE} \textit{(middle)} channels of the Layer Two \ac{SOM} constructed by concatenating 100 $4\times$ \acp{SOM}. The black vertical and horizontal lines represent the boundary between individual neurons on the lattice, while the red vertical and horizontal lines represent boundaries between individual $4\times4$ \acp{SOM}. Each of the \acp{SOM} were trained using the same data preprocessing and training stages as Layer One. The Layer One position \textit{(right)} shows the relative position of the four \acp{SOM} on the Layer One \ac{SOM} (Figure~\ref{fig:som}). }
\end{figure*}

\begin{figure*}
\ContinuedFloat
	\includegraphics[width=0.85\linewidth]{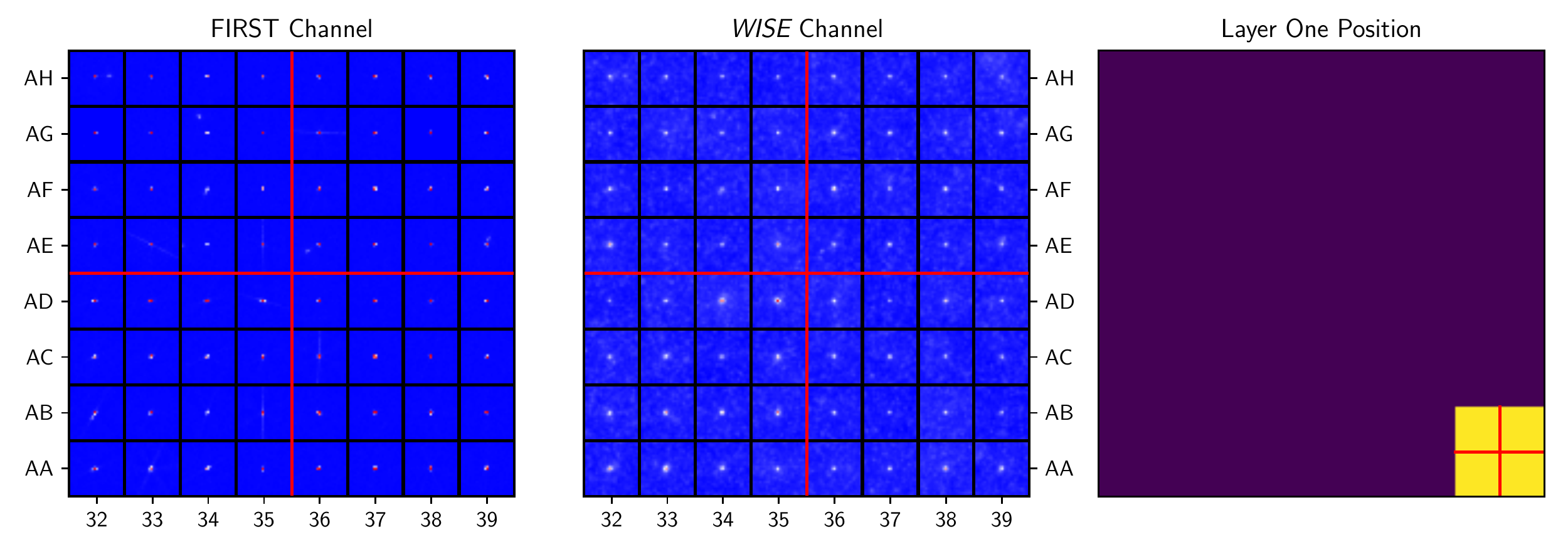}
	\caption{\textit{(Cont.)} The \ac{FIRST} \textit{(left)} and \ac{WISE} \textit{(middle)} channels of the Layer Two \ac{SOM} constructed by concatenating 100 $4\times$ \acp{SOM}. The black vertical and horizontal lines represent the boundary between individual neurons on the lattice, while the red vertical and horizontal lines represent boundaries between individual $4\times4$ \acp{SOM}. Each of the \acp{SOM} were trained using the same data preprocessing and training stages as Layer One. The Layer One position \textit{(right)} shows the relative position of the four \acp{SOM} on the Layer One \ac{SOM} (Figure~\ref{fig:som}). }
\end{figure*}

\newpage

\section{Example Tables}
\label{sec:cata_extract}

We include five row extracts of our \ac{FSC} and \ac{GRC} catalogues. 

\begin{table}
\begin{tabular}{rrlrrr}
\toprule
   \colstyle{GID} &  \colstyle{idx} & \colstyle{neuron\_index} & 
       \colstyle{ED} &  \colstyle{flip} & \colstyle{rotation} \\
\midrule
 279496 &    0 &  (29, 14, 0) & 27.301 &     1 &    0.297 \\
 215840 &    1 &   (17, 8, 0) & 23.787 &     1 &    1.239 \\
 717469 &    2 &  (35, 36, 0) & 19.560 &     0 &    0.279 \\
 215841 &    3 &   (17, 8, 0) & 22.714 &     1 &    4.381 \\
 474119 &    4 &  (36, 29, 0) & 12.071 &     1 &    2.356 \\
\bottomrule
\end{tabular}

\caption{Five example rows of our compiled \ac{FSC} catalogue. Columns correspond to those described by Table~\ref{tab:grp_cat}. Each row also carries with it the original information from \ac{FIRST} described in \citep{2015ApJ...801...26H}. A value of `1' in the `flip` column corresponds to an image being flipped as part of its transform. \label{tab:example_frc} }
\end{table}

\begin{table*}
\tiny
\begin{tabular}{rllrp{1cm}rp{1cm}rrrrrrr}
\toprule
 \colstyle{GID} & \colstyle{prob\_host} & \colstyle{comp\_host} & \colstyle{grp\_flag} &             \colstyle{D\_idx} &       \colstyle{D} & \colstyle{SP\_idx} & \colstyle{SP} & \colstyle{Q} & \colstyle{rel\_g} &  \colstyle{rel\_Q} & \colstyle{rel\_D} &  \colstyle{ir\_pp\_ra} &  \colstyle{ir\_pp\_dec} \\
 &&&&&$''$&&$''$&$''$&&&&$^{\circ}$&$^{\circ}$\\
\midrule
   0 &  J085632.98+595746.9     &  (J085632.98+595746.9) &         1 &    (28052, 28186) & 149.21 &         (28065, 28137, 28197, 28213, 28203) & 166.33 & 1.11 &   0.95 &   0.81 &   0.85 &    134.14 &      59.96 \\
   1 &  J085039.97+543753.4     &  (J085039.97+543753.4) &         1 &    (68434, 68412) & 114.90 &                (68458, 68418, 68446, 68408) & 129.04 & 1.12 &   0.53 &   0.75 &   0.83 &    132.67 &      54.63 \\
   2 &  J151347.88+530834.1     &  (J151347.88+530834.1) &         1 &    (81092, 81370) & 139.94 &         (81118, 81130, 81211, 81335, 81350) & 141.15 & 1.01 &   0.92 &   0.93 &   0.86 &    228.45 &      53.14 \\
   3 &  J140717.50+513213.4     &  (J140717.50+513213.4) &         1 &    (95447, 95258) & 151.76 &  (95461, 95421, 95450, 95371, 95337, 95286) & 171.89 & 1.13 &   0.91 &   0.84 &   0.85 &    211.82 &      51.54 \\
   4 &  J125142.02+503424.8     &  (J125142.02+503424.8) &         1 &  (103762, 103670) & 131.31 &                    (103758, 103703, 103688) & 139.48 & 1.06 &   0.90 &   0.89 &   0.83 &    192.92 &      50.57 \\
\bottomrule
\end{tabular}
\caption{Five example rows of our compiled \ac{GRC} catalogue. Columns correspond to those described in Table~\ref{tab:grp_cat}. \label{tab:example_grc} }
\end{table*}

\bsp	
\label{lastpage}
\end{document}